\documentclass[screen=true,review=false,authorversion=true,acmsmall]{acmart}
\renewcommand{\documentclass}[2][]{}
\newif\ifediting\editingfalse
\newif\iflong\longfalse
\newif\ifpreprint\preprinttrue
\documentclass[screen=true,review=true,acmsmall,anonymous]{acmart}

\makeatletter
\@ifundefined{ifediting}{
  \newif\ifediting%
  \editingfalse%
}{}
\makeatother

\makeatletter
\@ifundefined{iflong}{
  \newif\iflong%
  \longfalse%
}{}
\makeatother

\makeatletter
\@ifundefined{ifpreprint}{
  \newif\ifpreprint%
  \preprintfalse%
}{}
\makeatother

\makeatletter
\newif\ifanonymous
\if@ACM@anonymous
\anonymoustrue
\else
\anonymousfalse
\fi

\usepackage[utf8]{inputenc}
\usepackage{etoolbox}
\usepackage{multirow}
\usepackage{longtable}
\usepackage[capitalise,nameinlink,noabbrev]{cleveref}
\usepackage{paralist}
\usepackage{textcomp}
\usepackage[zi4]{inconsolata-slanted} 
\usepackage{needspace}
\usepackage{mathpartir}
\usepackage{mathtools}
\usepackage{etoolbox}

\usepackage{tikz}
\usetikzlibrary{calc}
\usetikzlibrary{patterns}
\usepackage{pgfplots}
\usepgfplotslibrary{colormaps,colorbrewer}
\pgfkeys{/pgf/number format/.cd,1000 sep={}}
\pgfplotsset{compat=1.14}

\makeatletter
\newcommand*\NewCodeLiteralCmd[2]{%
  \expandafter\newrobustcmd\expandafter*\csname\string#1@content\endcsname{\texttt{#2}}%
  \newcommand*#1{\texorpdfstring{\csname\string#1@content\endcsname}{#2}}%
}
\makeatother

\NewCodeLiteralCmd\LITSet{Set}
\NewCodeLiteralCmd\LITIntSet{IntSet}
\NewCodeLiteralCmd\LITcontainers{containers}
\NewCodeLiteralCmd\LIThstocoq{hs-to-coq}

\usepackage{xspace}
\newcommand*{\cn}[1][]{%
  \textcolor{red}{\textbf{%
    [?!%
    \ifx\relax#1\relax\else{} #1\fi
    ]%
  }}\xspace
}

\usepackage{draft-todo}
\ifediting
  \drafttrue

  \paperwidth=\dimexpr\paperwidth+6cm\relax%
  \oddsidemargin=\dimexpr\oddsidemargin+0cm\relax%
  \evensidemargin=\dimexpr\evensidemargin+6cm\relax%
  \marginparsep=\dimexpr\marginparsep+0cm\relax%
  \marginparwidth=\dimexpr\marginparwidth+6.0cm\relax%
\else
  \draftfalse
  \renewcommand{\todo}[2][]{}
\fi

\newtodo{asz}{violet!25}        
\newtodo{jb}{green!40}          
\newtodo{criz}{blue!67!cyan!20} 
\newtodo{scw}{red!20}           
\newtodo[Leo]{leo}{yellow!67}   
\newtodo{jp}{teal!50}           
\newtodo{jww}{purple!30}        
\newtodo{ly}{orange!50}         

\newcommand{\ensuretext}{\ifmmode\text\else\fi}

\newcounter{spaces}
\newcommand{\spacestart}{\setcounter{spaces}{0}}
\newcommand{\nextspace}{%
  \stepcounter{spaces}%
  \ifnum\value{spaces}=1\;\allowbreak\fi%
  \ifnum\value{spaces}=4\quad\quad\fi%
  }

\begingroup
\catcode`'=\active
\gdef\straighttick{
  \catcode`'=\active
  \def'{\textquotesingle}%
}
\endgroup

\makeatletter
\@ifundefined{lhs2tex.lhs2tex.sty.read}%
  {\@namedef{lhs2tex.lhs2tex.sty.read}{}%
   \newcommand\SkipToFmtEnd{}%
   \newcommand\EndFmtInput{}%
   \long\def\SkipToFmtEnd#1\EndFmtInput{}%
  }\SkipToFmtEnd

\newcommand\ReadOnlyOnce[1]{\@ifundefined{#1}{\@namedef{#1}{}}\SkipToFmtEnd}
\usepackage{amstext}
\usepackage{amssymb}
\usepackage{stmaryrd}
\DeclareFontFamily{OT1}{cmtex}{}
\DeclareFontShape{OT1}{cmtex}{m}{n}
  {<5><6><7><8>cmtex8
   <9>cmtex9
   <10><10.95><12><14.4><17.28><20.74><24.88>cmtex10}{}
\DeclareFontShape{OT1}{cmtex}{m}{it}
  {<-> ssub * cmtt/m/it}{}

\DeclareFontShape{OT1}{cmtt}{bx}{n}
  {<5><6><7><8>cmtt8
   <9>cmbtt9
   <10><10.95><12><14.4><17.28><20.74><24.88>cmbtt10}{}
\DeclareFontShape{OT1}{cmtex}{bx}{n}
  {<-> ssub * cmtt/bx/n}{}

\newcommand{\anonymous}{\kern0.06em \vbox{\hrule\@width.5em}}

\usepackage{polytable}

\@ifundefined{mathindent}%
  {\newdimen\mathindent\mathindent\leftmargini}%
  {}%

\def\resethooks{%
  \global\let\SaveRestoreHook\empty
  \global\let\ColumnHook\empty}
\newcommand*{\savecolumns}[1][default]%
  {\g@addto@macro\SaveRestoreHook{\savecolumns[#1]}}
\newcommand*{\restorecolumns}[1][default]%
  {\g@addto@macro\SaveRestoreHook{\restorecolumns[#1]}}
\newcommand*{\aligncolumn}[2]%
  {\g@addto@macro\ColumnHook{\column{#1}{#2}}}

\resethooks

\newcommand{\onelinecommentchars}{\quad-{}- }
\newcommand{\commentbeginchars}{\enskip\{-}
\newcommand{\commentendchars}{-\}\enskip}

\newcommand{\visiblecomments}{%
  \let\onelinecomment=\onelinecommentchars
  \let\commentbegin=\commentbeginchars
  \let\commentend=\commentendchars}

\newcommand{\invisiblecomments}{%
  \let\onelinecomment=\empty
  \let\commentbegin=\empty
  \let\commentend=\empty}

\visiblecomments

\newlength{\blanklineskip}
\newcommand{\hsindent}[1]{\quad}
\let\hspre\empty
\let\hspost\empty

\EndFmtInput
\makeatother
\makeatletter

\newcommand{\hsnewpar}[1]%
  {{\parskip=0pt\parindent=0pt\par\vskip #1\noindent}}

\newcommand{\hscodestyle}{\ttfamily}

\newcommand{\sethscode}[1]%
  {\expandafter\let\expandafter\hscode\csname #1\endcsname
   \expandafter\let\expandafter\endhscode\csname end#1\endcsname}

  {\hsnewpar\abovedisplayskip
   \advance\leftskip\mathindent
   \hscodestyle
   \def\hspre{}%
   \def\hspost{}%
   \pboxed}%
  {\endpboxed%
   \hsnewpar\belowdisplayskip
   \ignorespacesafterend}

\newcommand{\plainhs}{\sethscode{plainhscode}}
\plainhs

\makeatother

\renewcommand{\onelinecomment}{--\hspace{.5pt}\itshape }
\renewcommand{\hsindent}[1]{{\quad\quad}}
\newcommand{\summarytallytable}{\begin{tabular}{l|rrr|rrr|rrr}
 &  & \multicolumn{1}{c}{Haskell} &  &  & \multicolumn{1}{c}{Gallina} &  &  & \multicolumn{1}{c}{Proofs} & \\
 & Set & IntSet & Set & IntSet & Set & IntSet\\
\noalign{\hrule height 0.4pt}
Functions, &  & 619 &  &  & 923 &  &  & 5434 & \\
\quad  verified & 316 & 303 & 448 & 475 & 2241 & 3193\\
\noalign{\hrule height 0.4pt}
Functions, &  & 265 &  &  & 466 &  &  &  & \\
\quad  unverified & 142 & 123 & 207 & 259 &  & \\
\noalign{\hrule height 0.4pt}
Functions, &  & 95 &  &  &  &  &  &  & \\
\quad  untranslated & 46 & 49 &  &  &  & \\
\noalign{\hrule height 0.4pt}
Functions, &  & 247 &  &  & 341 &  &  & 1402 & \\
\quad  verified & 134 & 113 & 223 & 118 & 721 & 681\\
\noalign{\hrule height 0.4pt}
Functions, unverified &  & 134 &  & 157 &  & \\
\noalign{\hrule height 0.4pt}
Functions, &  & 97 &  &  &  &  &  &  & \\
\quad  untranslated & 45 & 52 &  &  &  & \\
\noalign{\hrule height 0.4pt}
Type classes, &  & 38 &  &  & 147 &  &  & 822 & \\
\quad  verified & 21 & 17 & 101 & 46 & 370 & 452\\
\noalign{\hrule height 0.4pt}
Type classes, unverified & 28 &  & 79 &  &  & \\
\noalign{\hrule height 0.4pt}
Type classes, &  & 78 &  &  &  &  &  &  & \\
\quad  untranslated & 43 & 35 &  &  &  & \\
\noalign{\hrule height 0.4pt}
Headers and types &  & 298 &  &  & 132 &  &  & 24 & \\
\quad  & 153 & 145 & 51 & 81 & 14 & 10\\
\noalign{\hrule height 0.4pt}
Tactics &  &  &  &  & 7 &  &  & 222 & \\
\quad  &  &  & 3 & 4 & 149 & 73\\
\noalign{\hrule height 0.4pt}
Well-formedness &  &  &  &  &  &  &  & 505 & \\
\quad  &  &  &  &  & 302 & 203\\
\noalign{\hrule height 0.4pt}
\texttt{FSetInterface} &  &  &  &  &  &  &  & 1653 & \\
\quad  &  &  &  &  & 674 & 777\\
\noalign{\hrule height 0.4pt}
Tests &  & 308 &  & 304 &  & 784\\
\noalign{\hrule height 0.4pt}
Arithmetic &  &  &  &  & 153 &  &  & 1265 & \\
\quad  &  &  &  & 153 &  & 586\\
\noalign{\hrule height 0.4pt}
Lists &  &  &  &  &  &  &  & 1489 & \\
\quad  &  &  &  &  &  & 183\\
\noalign{\hrule height 0.4pt}
Dyadic &  &  &  &  &  & \\
\noalign{\hrule height 0.4pt}
Order &  &  &  &  &  &  &  & 453 & \\
\quad  &  &  &  &  & 30 & \\
\end{tabular}
}
\newcommand{\translationcoordinates}{(1,12) [IntSet] 
(1,2) [IntSet] 
(1,6) [IntSet] 
(1,8) [IntSet] 
(10,3) [IntSet] 
(10,5) [IntSet] 
(10,7) [IntSet] 
(10,7) [IntSet] 
(10,8) [Set] 
(11,14) [Set] 
(11,2) [Set] 
(11,7) [IntSet] 
(12,10) [IntSet] 
(12,11) [Set] 
(12,6) [Set] 
(12,6) [Set] 
(12,6) [Set] 
(12,9) [Set] 
(13,7) [Set] 
(13,7) [Set] 
(13,7) [Set] 
(13,9) [IntSet] 
(14,10) [Set] 
(15,11) [Set] 
(15,4) [Set] 
(15,8) [Set] 
(15,9) [IntSet] 
(15,9) [IntSet] 
(15,9) [Set] 
(16,13) [Set] 
(16,14) [Set] 
(16,9) [IntSet] 
(17,13) [IntSet] 
(17,16) [Set] 
(18,12) [Set] 
(18,12) [Set] 
(18,13) [Set] 
(19,13) [Set] 
(19,15) [IntSet] 
(2,10) [IntSet] 
(2,10) [IntSet] 
(2,13) [IntSet] 
(2,2) [IntSet] 
(2,2) [IntSet] 
(2,2) [IntSet] 
(2,2) [IntSet] 
(2,2) [IntSet] 
(2,2) [IntSet] 
(2,2) [IntSet] 
(2,2) [IntSet] 
(2,2) [IntSet] 
(2,2) [IntSet] 
(2,2) [IntSet] 
(2,2) [Set] 
(2,2) [Set] 
(2,2) [Set] 
(2,2) [Set] 
(2,3) [IntSet] 
(2,3) [IntSet] 
(2,3) [IntSet] 
(2,3) [IntSet] 
(2,3) [IntSet] 
(2,3) [IntSet] 
(2,3) [IntSet] 
(2,3) [IntSet] 
(2,3) [IntSet] 
(2,3) [IntSet] 
(2,3) [Set] 
(2,3) [Set] 
(2,3) [Set] 
(2,3) [Set] 
(2,4) [IntSet] 
(2,4) [Set] 
(2,4) [Set] 
(2,4) [Set] 
(2,4) [Set] 
(2,5) [IntSet] 
(2,5) [IntSet] 
(2,5) [Set] 
(2,5) [Set] 
(21,16) [Set] 
(21,9) [IntSet] 
(21,9) [IntSet] 
(22,14) [Set] 
(22,14) [Set] 
(22,16) [Set] 
(23,15) [Set] 
(23,2) [IntSet] 
(25,18) [Set] 
(25,8) [IntSet] 
(26,15) [IntSet] 
(27,12) [Set] 
(27,12) [Set] 
(28,14) [IntSet] 
(28,7) [Set] 
(30,15) [Set] 
(30,15) [Set] 
(30,17) [IntSet] 
(30,2) [Set] 
(31,13) [IntSet] 
(34,13) [IntSet] 
(34,13) [IntSet] 
(35,13) [IntSet] 
(39,19) [Set] 
(39,27) [Set] 
(4,2) [Set] 
(4,3) [IntSet] 
(4,5) [IntSet] 
(4,5) [IntSet] 
(4,7) [IntSet] 
(40,19) [Set] 
(43,27) [IntSet] 
(45,30) [IntSet] 
(48,27) [IntSet] 
(52,28) [Set] 
(55,30) [IntSet] 
(56,35) [IntSet] 
(57,55) [IntSet] 
(58,35) [IntSet] 
(59,29) [IntSet] 
(6,3) [Set] 
(6,3) [Set] 
(6,3) [Set] 
(6,3) [Set] 
(6,5) [Set] 
(6,5) [Set] 
(6,6) [Set] 
(63,29) [IntSet] 
(7,4) [IntSet] 
(7,4) [IntSet] 
(7,4) [Set] 
(7,4) [Set] 
(79,28) [Set] 
(8,3) [IntSet] 
(8,3) [Set] 
(8,5) [Set] 
(8,5) [Set] 
(9,15) [Set] 
(9,3) [Set] 
(9,5) [Set] 
(9,5) [Set] 
(9,6) [IntSet] 
(9,7) [Set] 
(9,7) [Set] 
(9,7) [Set] 
}
\newcommand{\provingcoordinates}{(1,2) [IntSet] 
(1,2) [IntSet] 
(1,2) [Set] 
(1,2) [Set] 
(102,11) [Set] 
(107,27) [Set] 
(109,27) [Set] 
(12,2) [IntSet] 
(12,2) [Set] 
(12,2) [Set] 
(12,4) [IntSet] 
(125,19) [IntSet] 
(128,2) [IntSet] 
(13,2) [IntSet] 
(133,39) [Set] 
(134,11) [Set] 
(139,12) [IntSet] 
(14,10) [IntSet] 
(14,2) [IntSet] 
(14,2) [IntSet] 
(14,2) [Set] 
(14,2) [Set] 
(15,23) [Set] 
(154,30) [IntSet] 
(16,10) [IntSet] 
(16,10) [Set] 
(17,2) [IntSet] 
(17,6) [Set] 
(17,6) [Set] 
(173,57) [IntSet] 
(187,48) [IntSet] 
(20,2) [IntSet] 
(20,4) [IntSet] 
(20,8) [Set] 
(203,15) [Set] 
(209,2) [IntSet] 
(22,12) [Set] 
(22,12) [Set] 
(22,12) [Set] 
(222,58) [IntSet] 
(224,63) [IntSet] 
(23,6) [Set] 
(23,6) [Set] 
(23,9) [Set] 
(245,59) [IntSet] 
(251,26) [IntSet] 
(255,28) [IntSet] 
(26,9) [Set] 
(261,2) [IntSet] 
(27,16) [Set] 
(271,43) [IntSet] 
(275,2) [Set] 
(299,8) [IntSet] 
(3,2) [Set] 
(30,2) [Set] 
(30,4) [IntSet] 
(31,18) [Set] 
(31,21) [Set] 
(31,7) [Set] 
(312,45) [IntSet] 
(32,2) [IntSet] 
(32,9) [Set] 
(32,9) [Set] 
(33,15) [Set] 
(33,15) [Set] 
(331,30) [Set] 
(35,14) [Set] 
(350,30) [Set] 
(4,8) [Set] 
(4,8) [Set] 
(42,2) [IntSet] 
(43,6) [Set] 
(44,39) [Set] 
(44,40) [Set] 
(46,11) [IntSet] 
(46,2) [Set] 
(50,13) [Set] 
(52,16) [IntSet] 
(54,19) [Set] 
(56,13) [Set] 
(57,12) [Set] 
(57,7) [Set] 
(58,12) [Set] 
(6,2) [Set] 
(66,16) [Set] 
(66,2) [IntSet] 
(69,17) [IntSet] 
(70,17) [Set] 
(71,22) [Set] 
(72,25) [Set] 
(78,30) [Set] 
(8,2) [Set] 
(82,28) [Set] 
(87,23) [IntSet] 
(88,1) [IntSet] 
(88,9) [IntSet] 
(9,2) [IntSet] 
(9,2) [IntSet] 
(99,2) [IntSet] 
}

\newcommand{\funcs}{149}
\newcommand{\tycls}{22}
\newcommand{\locHaskell}{2207}
\newcommand{\funcsUntranslated}{15}
\newcommand{\tyclsUntranslated}{11}
\newcommand{\locHaskellUntranslated}{270}
\newcommand{\locGallina}{2709}

\newcommand{\locHaskellVerified}{1202}

\newcommand{\percVerifiedSet}{66}

\newcommand{\percVerifiedIntSet}{49}

\newcommand{\locArith}{1265}
\newcommand{\locDyadic}{1169}

\newcommand{\locOrder}{453}

\newcommand{\locLists}{1489}

\newcommand{\locProofPerHaskell}{8.5}
\newcommand{\locProofPerHaskellSet}{7.1}
\newcommand{\locProofPerHaskellIntSet}{10.0}
\newcommand{\plotHaskellTests}{308}
\newcommand{\plotHaskellConstantSet}{153}
\newcommand{\plotHaskellConstantIntSet}{145}
\newcommand{\plotHaskellVerifiedSet}{471}
\newcommand{\plotHaskellVerifiedIntSet}{433}
\newcommand{\plotHaskellUnverifiedSet}{170}
\newcommand{\plotHaskellUnverifiedIntSet}{257}
\newcommand{\plotHaskellUntranslatedSet}{134}
\newcommand{\plotHaskellUntranslatedIntSet}{136}
\newcommand{\plotGallinaTests}{304}
\newcommand{\plotGallinaConstantSet}{54}
\newcommand{\plotGallinaConstantIntSet}{85}
\newcommand{\plotGallinaVerifiedSet}{772}
\newcommand{\plotGallinaVerifiedIntSet}{639}
\newcommand{\plotGallinaUnverifiedSet}{286}
\newcommand{\plotGallinaUnverifiedIntSet}{416}
\newcommand{\plotGallinaUntranslatedSet}{0}
\newcommand{\plotGallinaUntranslatedIntSet}{0}
\newcommand{\plotProofsTests}{784}
\newcommand{\plotProofsConstantSet}{465}
\newcommand{\plotProofsConstantIntSet}{286}
\newcommand{\plotProofsVerifiedSet}{3332}
\newcommand{\plotProofsVerifiedIntSet}{4326}
\newcommand{\plotProofsUnverifiedSet}{0}
\newcommand{\plotProofsUnverifiedIntSet}{0}
\newcommand{\plotProofsUntranslatedSet}{0}
\newcommand{\plotProofsUntranslatedIntSet}{0}
\newcommand{\plotOrder}{453}
\newcommand{\plotDyadic}{1169}
\newcommand{\plotLists}{1489}
\newcommand{\plotArith}{1418}
\newcommand{\plotCoqInterface}{1653}
\newcommand{\listVerifSet}{\texttt{delete}, \texttt{deleteMax}, \texttt{deleteMin}, \texttt{difference}, \texttt{disjoint}, \texttt{drop}, \texttt{elems}, \texttt{empty}, \texttt{filter}, \texttt{foldl}, \texttt{foldl'}, \texttt{foldr}, \texttt{foldr'}, \texttt{fromAscList}, \texttt{fromDescList}, \texttt{fromDistinctAscList}, \texttt{fromDistinctDescList}, \texttt{insert}, \texttt{intersection}, \texttt{isSubsetOf}, \texttt{lookupMax}, \texttt{lookupMin}, \texttt{mapMonotonic}, \texttt{maxView}, \texttt{member}, \texttt{minView}, \texttt{notMember}, \texttt{null}, \texttt{partition}, \texttt{singleton}, \texttt{size}, \texttt{split}, \texttt{splitAt}, \texttt{splitMember}, \texttt{take}, \texttt{toAscList}, \texttt{toDescList}, \texttt{toList}, \texttt{union}, \texttt{unions}\\\textbf{Instances:} \texttt{Eq}, \texttt{Eq1}, \texttt{Monoid}, \texttt{Ord}, \texttt{Ord1}, \texttt{Semigroup}\\\textbf{Internal functions:} \texttt{balanceL}, \texttt{balanceR}, \texttt{combineEq}, \texttt{glue}, \texttt{insertMax}, \texttt{insertMin}, \texttt{insertR}, \texttt{link}, \texttt{maxViewSure}, \texttt{merge}, \texttt{minViewSure}, \texttt{valid}}
\newcommand{\listUnverifSet}{\texttt{cartesianProduct}, \texttt{disjointUnion}, \texttt{dropWhileAntitone}, \texttt{fromList}, \texttt{isProperSubsetOf}, \texttt{lookupGE}, \texttt{lookupGT}, \texttt{lookupIndex}, \texttt{lookupLE}, \texttt{lookupLT}, \texttt{map}, \texttt{powerSet}, \texttt{spanAntitone}, \texttt{splitRoot}, \texttt{takeWhileAntitone}\\\textbf{Instances:} \texttt{Foldable}\\\textbf{Internal functions:} }
\newcommand{\listUntransSet}{\texttt{deleteAt}, \texttt{deleteFindMax}, \texttt{deleteFindMin}, \texttt{elemAt}, \texttt{findIndex}, \texttt{findMax}, \texttt{findMin}\\\textbf{Instances:} \texttt{Data}, \texttt{IsList}, \texttt{NFData}, \texttt{Read}, \texttt{Show}, \texttt{Show1}\\\textbf{Internal functions:} \texttt{showTree}}
\newcommand{\listVerifIntSet}{\texttt{delete}, \texttt{difference}, \texttt{disjoint}, \texttt{elems}, \texttt{empty}, \texttt{filter}, \texttt{foldl}, \texttt{foldr}, \texttt{fromList}, \texttt{insert}, \texttt{intersection}, \texttt{isProperSubsetOf}, \texttt{isSubsetOf}, \texttt{map}, \texttt{member}, \texttt{notMember}, \texttt{null}, \texttt{partition}, \texttt{singleton}, \texttt{size}, \texttt{split}, \texttt{splitMember}, \texttt{toAscList}, \texttt{toDescList}, \texttt{toList}, \texttt{union}, \texttt{unions}\\\textbf{Instances:} \texttt{Eq}, \texttt{Monoid}, \texttt{Ord}, \texttt{Semigroup}\\\textbf{Internal functions:} \texttt{branchMask}, \texttt{equal}, \texttt{highestBitMask}, \texttt{mask}, \texttt{nequal}, \texttt{nomatch}, \texttt{revNat}, \texttt{shorter}, \texttt{valid}, \texttt{zero}}
\newcommand{\listUnverifIntSet}{\texttt{deleteMax}, \texttt{deleteMin}, \texttt{foldl'}, \texttt{foldr'}, \texttt{lookupGE}, \texttt{lookupGT}, \texttt{lookupLE}, \texttt{lookupLT}, \texttt{maxView}, \texttt{minView}, \texttt{splitRoot}\\\textbf{Instances:} \\\textbf{Internal functions:} \texttt{bitcount}, \texttt{bitmapOf}, \texttt{bitmapOfSuffix}, \texttt{deleteBM}, \texttt{highestBitSet}, \texttt{indexOfTheOnlyBit}, \texttt{insertBM}, \texttt{link}, \texttt{lowestBitMask}, \texttt{lowestBitSet}, \texttt{maskW}, \texttt{prefixOf}, \texttt{shiftRL}, \texttt{subsetCmp}, \texttt{suffixBitMask}, \texttt{suffixOf}, \texttt{tip}, \texttt{unsafeFindMax}, \texttt{unsafeFindMin}}
\newcommand{\listUntransIntSet}{\texttt{deleteFindMax}, \texttt{deleteFindMin}, \texttt{findMax}, \texttt{findMin}, \texttt{fromAscList}, \texttt{fromDistinctAscList}\\\textbf{Instances:} \texttt{Data}, \texttt{IsList}, \texttt{NFData}, \texttt{Read}, \texttt{Show}\\\textbf{Internal functions:} \texttt{showTree}}

\newcommand{\verifiedAPIfigure}{%
\begin{figure}
\raggedright
\begin{description}
\item[\ensuretext{\straighttick\ttfamily{}Set}:]  \listVerifSet
\item[\ensuretext{\straighttick\ttfamily{}IntSet}:]  \listVerifIntSet
\end{description}
\caption{The verified subset of functions and type classes in \ensuretext{\straighttick\ttfamily{}Data.Set} and \ensuretext{\straighttick\ttfamily{}Data.IntSet}}
\label{fig:verifiedAPI}
\end{figure}
}

\newcommand{\untranslatedAPIfigure}{%
\begin{figure}
\raggedright
\begin{description}
\item[\ensuretext{\straighttick\ttfamily{}Set}:]  \listUntransSet
\item[\ensuretext{\straighttick\ttfamily{}IntSet}:]  \listUntransIntSet
\end{description}
\caption{Untranslated functions and type classes in \ensuretext{\straighttick\ttfamily{}Data.Set} and \ensuretext{\straighttick\ttfamily{}Data.IntSet}}
\label{fig:untranslatedAPI}
\vspace*{-1ex}
\end{figure}
}

\def\Vhrulefill{\leavevmode\leaders\hrule height 1.0ex depth \dimexpr0.4pt-1.0ex\hfill\kern0pt}
\newcommand\dito[1]{%
  \parbox{\widthof{#1}}{\Vhrulefill\textquotedbl\Vhrulefill}%
}

\newcommand{\locfigure}{%
\begin{figure}
\begin{tikzpicture}
\pgfplotscreateplotcyclelist{mylist}{%
  {index of colormap=1,draw=black,fill=.},
  {index of colormap=2,draw=black,fill=.},
  {index of colormap=2,draw=black,fill=.,postaction={pattern=north east lines}},
  {index of colormap=2,draw=black,fill=.,postaction={pattern=crosshatch}},
  {index of colormap=4,draw=black,fill=.!80!black},
}
\begin{axis}[
  colormap/Set3-5,
  height=9cm,
  width=\linewidth,
  xbar stacked,
  xmin=0,
  xmax=5150,
  cycle list name = mylist,
  yticklabel style={rotate=90},
  xlabel={Lines of code},
  legend style={%
    legend pos= north east,
    legend cell align=left,
    legend columns=1
  },
  symbolic y coords={%
    TestsProofs,
    CoqInterface,
    IntSetProofs,
    SetProofs,
    Order,
    Dyadic,
    Lists,
    Arith,
    ,
    TestsGallina,
    IntSetGallina,
    SetGallina,
    ,
    TestsHS,
    IntSetHS,
    SetHS
  },
  ytick={IntSetHS,IntSetGallina,[normalized]3.5},
  yticklabels={Haskell,Gallina,Proofs},
  ytick style={draw=none},
  ]
\addplot coordinates {%
  (0,TestsProofs)
  (0,CoqInterface)
  (\plotProofsConstantIntSet,IntSetProofs)
  (\plotProofsConstantSet,SetProofs)
  (\plotOrder,Order)
  (\plotDyadic,Dyadic)
  (\plotLists,Lists)
  (\plotArith,Arith)
  (0,TestsGallina)
  (\plotGallinaConstantIntSet,IntSetGallina)
  (\plotGallinaConstantSet,SetGallina)
  (0,TestsHS)
  (\plotHaskellConstantIntSet,IntSetHS)
  (\plotHaskellConstantSet,SetHS)
};
\addlegendentry{Headers, types and proof preparations};
\addplot coordinates {%
  (0,TestsProofs)
  (0,CoqInterface)
  (\plotProofsVerifiedIntSet,IntSetProofs)
  (\plotProofsVerifiedSet,SetProofs)
  (0,Order)
  (0,Dyadic)
  (0,Lists)
  (0,Arith)
  (0,TestsGallina)
  (\plotGallinaVerifiedIntSet,IntSetGallina)
  (\plotGallinaVerifiedSet,SetGallina)
  (0,TestsHS)
  (\plotHaskellVerifiedIntSet,IntSetHS)
  (\plotHaskellVerifiedSet,SetHS)
};
\addlegendentry{Functions and type class instances (verified)};
\addplot coordinates {%
  (0,TestsProofs)
  (0,CoqInterface)
  (\plotProofsUnverifiedIntSet,IntSetProofs)
  (\plotProofsUnverifiedSet,SetProofs)
  (0,Order)
  (0,Dyadic)
  (0,Lists)
  (0,Arith)
  (0,TestsGallina)
  (\plotGallinaUnverifiedIntSet,IntSetGallina)
  (\plotGallinaUnverifiedSet,SetGallina)
  (0,TestsHS)
  (\plotHaskellUnverifiedIntSet,IntSetHS)
  (\plotHaskellUnverifiedSet,SetHS)
};
\addlegendentry{\dito{Functions and type class instances} (unverified)};
\addplot coordinates {%
  (0,TestsProofs)
  (0,CoqInterface)
  (\plotProofsUntranslatedIntSet,IntSetProofs)
  (\plotProofsUntranslatedSet,SetProofs)
  (0,Order)
  (0,Dyadic)
  (0,Lists)
  (0,Arith)
  (0,TestsGallina)
  (\plotGallinaUntranslatedIntSet,IntSetGallina)
  (\plotGallinaUntranslatedSet,SetGallina)
  (0,TestsHS)
  (\plotHaskellUntranslatedIntSet,IntSetHS)
  (\plotHaskellUntranslatedSet,SetHS)
};
\addlegendentry{\dito{Functions and type class instances} (untranslated)};
\addplot coordinates {%
  (\plotProofsTests,TestsProofs)
  (\plotCoqInterface,CoqInterface)
  (0,IntSetProofs)
  (0,SetProofs)
  (0,Order)
  (0,Dyadic)
  (0,Lists)
  (0,Arith)
  (\plotGallinaTests,TestsGallina)
  (0,IntSetGallina)
  (0,SetGallina)
  (\plotHaskellTests,TestsHS)
  (0,IntSetHS)
  (0,SetHS)
 };
\addlegendentry{Tests and specifications};

\addplot [
  nodes near coords,
  stacked ignores zero=false,
  point meta=explicit symbolic,
  nodes near coords align={anchor=west},
  ] coordinates {%
  (0,TestsProofs)   [Tests]
  (0,CoqInterface)  [Instantiating Coq’s \ensuretext{\straighttick\ttfamily{}FSetInterface}]
  (0,IntSetProofs)  [\ensuretext{\straighttick\ttfamily{}IntSet}]
  (0,SetProofs)     [\ensuretext{\straighttick\ttfamily{}Set}]
  (0,Order)         [Automation for reasoning about \ensuretext{\straighttick\ttfamily{}Ord}]
  (0,Dyadic)        [A theory of dyadic intervals]
  (0,Lists)         [Lemmas about lists and sortedness]
  (0,Arith)         [Lemmas about integers and bit-wise operations]
  (0,TestsGallina)  [Tests]
  (0,IntSetGallina) [\ensuretext{\straighttick\ttfamily{}IntSet}]
  (0,SetGallina)    [\ensuretext{\straighttick\ttfamily{}Set}]
  (0,TestsHS)       [Tests]
  (0,IntSetHS)      [\ensuretext{\straighttick\ttfamily{}IntSet}]
  (0,SetHS)         [\ensuretext{\straighttick\ttfamily{}Set}]
};

\end{axis}
\end{tikzpicture}
\caption{A quantitative overview of the Haskell code, its translation into Coq and our proofs}

\label{fig:locfigure}
\end{figure}
}

\newcommand{\repohash}{67aaf28b5}
\newcommand{\fileContainers}[1]{\href{https://github.com/antalsz/hs-to-coq/blob/67aaf28b5/examples/containers/#1}{examples/containers/#1}}
\newcommand{\fileHsToCoq}[1]{\href{https://github.com/antalsz/hs-to-coq/blob/67aaf28b5/#1}{#1}}
\newcommand{\containerCommit}[1]{\footnote{\url{https://github.com/haskell/containers/commit/#1}}}
\newcommand{\containerPull}[1]{\footnote{\url{https://github.com/haskell/containers/pull/#1}}}
\newcommand{\containerIssue}[1]{\footnote{\url{https://github.com/haskell/containers/issues/#1}}}

\ifanonymous
\setcopyright{none}
\renewcommand\footnotetextcopyrightpermission[1]{} 
\else\ifpreprint
\setcopyright{none}
\renewcommand\footnotetextcopyrightpermission[1]{} 
\makeatletter
\fancypagestyle{standardpagestyle}{%
\fancyhf{}%
\fancyhead[LE]{\@headfootfont\thepage}%
\fancyhead[RO]{\@headfootfont\thepage}%
\fancyhead[RE]{\@headfootfont\@shortauthors}%
\fancyhead[LO]{\@headfootfont\shorttitle}%
\fancyfoot[RO,LE]{}%
}
\fancypagestyle{firstpagestyle}{%
\fancyhf{}%
\fancyhead[LE]{}%
\fancyhead[RO]{}%
\fancyhead[RE]{}%
\fancyhead[LO]{}%
\fancyfoot[RO,LE]{}%
}
\makeatother

\setcopyright{acmlicensed}
\acmPrice{15.00}
\acmYear{2018}
\copyrightyear{2018}
\acmConference[ICFP'18]{ICFP 2018}{September 23--29, 2018}{St. Louis, MO, USA}
\fi\fi



\begin{document}

\title{Ready, \LITSet, Verify!}
\subtitle{Applying \LIThstocoq{} to real-world Haskell code}

\newcommand{\Penn}{%
\affiliation{%
  \institution{University of Pennsylvania}
  \streetaddress{3330 Walnut St}
  \city{Philadelphia}
  \state{PA}
  \postcode{19104}
  \country{USA}
}}
\newcommand{\unsw}{%
\affiliation{%
  \institution{University of New South Wales}
  \city{Sydney}
  \state{NSW}
  \country{Australia}
}}

\author{Joachim Breitner} \email{joachim@cis.upenn.edu} \Penn
\author{Antal Spector-Zabusky} \email{antals@cis.upenn.edu} \Penn
\author{Yao Li} \email{liyao@cis.upenn.edu} \Penn
\author{Christine Rizkallah} \email{criz@cis.upenn.edu} \unsw
\author{John Wiegley} \email{john.wiegley@baesystems.com} \affiliation{\institution{BAE Systems}\country{USA}}
\author{Stephanie Weirich} \email{sweirich@cis.upenn.edu} \Penn

\begin{abstract}
  Good tools can bring mechanical verification to programs written in mainstream
  functional languages.
  We use \texttt{hs\char45{}to\char45{}coq} to translate significant portions of Haskell's \texttt{containers}
  library into Coq,
  and verify it against specifications that we derive from a variety of sources
  including type class laws, the library's test suite, and interfaces from Coq's
  standard library.
  Our work shows that it is feasible to verify mature, widely-used, highly
  optimized, and unmodified Haskell code.
  We also learn more about the theory of weight-balanced trees, extend
  \texttt{hs\char45{}to\char45{}coq} to handle partiality, and -- since we found no bugs -- attest to
  the superb quality of well-tested functional code.
\end{abstract}

\maketitle

\renewcommand{\shortauthors}{Breitner, Spector-Zabusky, Li, Rizkallah, Wiegley and Weirich}

\section{Introduction}
\label{sec:introduction}
%

What would it take to tempt functional programmers to verify their code?

Certainly, better tools would help.
We see that functional programmers who use dependently-typed languages
or proof assistants, such as
Coq~\citep{Coq:manual}, Agda~\citep{agda}, Idris~\citep{idris}, and
Isabelle~\citep{isabelle}, do verify their code, since their tools allow it.
%
However, adopting these platforms means rewriting everything from scratch.
What about the verification of \emph{existing} code, such as libraries written
in mature languages like Haskell?

Haskell programmers can reach for LiquidHaskell \citep{Vazou:2014:RTH:2628136.2628161} which smoothly integrates the expressive power of refinement types with Haskell, using SMT solvers for fully automatic verification.
But some verification endeavors require the full strength of a mature interactive proof assistant like Coq.
The \texttt{hs\char45{}to\char45{}coq} tool, developed by \citet*{hs-to-coq-cpp}, translates Haskell types, functions and type classes into equivalent Coq code -- a form of shallow embedding -- which can be verified just like normal Coq definitions.

But can this approach be used for more than the small, textbook-sized examples it has been applied to so far? Yes, it can! In this work, we use \texttt{hs\char45{}to\char45{}coq} to translate and
verify the two set data structures from Haskell's
\texttt{containers} package.\footnote{Specifically, we target version 0.5.11.0, which
  was released on January 22, 2018 and was the most recent release of this
  library at the time of publication; it is available at
  \url{https://github.com/haskell/containers/tree/v0.5.11.0}.}  This codebase is
not a toy.  It is decades old, highly tuned for performance, type-class
polymorphic, and implemented in terms of low-level features like bit
manipulation operators and raw pointer equality.  It is also an integral part of
the Haskell ecosystem.
%
%
We make the following contributions:

\begin{itemize}
\item We demonstrate that \texttt{hs\char45{}to\char45{}coq} is suitable for the
  verification of unmodified, real-world Haskell libraries. By
  “real-world”, we mean code that is up-to-date, in common use, and
  optimized for performance.  In \cref{sec:containers} we describe the
  \texttt{containers} library in more detail and discuss why it fits this
  description.

\item We present a case study not just of verifying a popular Haskell library,
  but also of developing a good \emph{specification} of that library. This
  process is worth consideration because it is not at all obvious what we mean
  when we say that we have ``verified'' a library. \Cref{sec:specs} discusses
  the techniques that we have used to construct a rich, two-sided
  specification; one that draws from diverse, cross-validated sources and yet
  is suitable for verification.

\item We extend \texttt{hs\char45{}to\char45{}coq} and its associated standard
  libraries to support our verification goal. In particular, in
  \cref{sec:translating} we describe the challenges that arise when
  translating the \ensuretext{\straighttick\ttfamily{}Data.Set} and \ensuretext{\straighttick\ttfamily{}Data.IntSet}
  modules, and our solutions.  Notably, we drop the restriction in previous
  work~\citep{hs-to-coq-cpp} that the input of the translation must be
  intrinsically total. Instead, we show how to safely defer reasoning about
  incomplete pattern matching and potential nontermination to later stages of
  the verification process.

\item
  We increase confidence in the translation done of \texttt{hs\char45{}to\char45{}coq}.
  In one direction, properties of the Haskell test suite turn into Coq theorems that we prove.
  In the other direction, the translated code, when extracted back to Haskell, passes the original test suite.

\item We provide new implementation-agnostic insight into the verification of
  the weight-balanced tree data structure, as we describe in
  \cref{sec:additional}.
  In particular, we find the right precondition for the central balancing
  operations needed to verify the particular variant used in \ensuretext{\straighttick\ttfamily{}Data.Set}.
\end{itemize}

Our work provides a rich specification for Haskell's finite set
libraries that is directly and mechanically connected to the current
implementation. As a result, Haskell programmers can be assured that these
libraries behave as expected. Of course, there is a limit to the assurances
that we can provide through this sort of effort. We discuss the verification
gap and other limitations of our approach in \cref{sec:limitations}.

We would like to have been able to claim the contribution of findings bugs in \texttt{containers}, but there simply were none.
Still, our efforts resulted in improvements to the \texttt{containers} library.
First, an insight during the
verification process led to an optimization that makes the \ensuretext{\straighttick\ttfamily{}Data.Set.union}
function~4\% faster. Second, we discovered an incompleteness in the
specification of the validity checker used in the test suite.

The tangible artifacts of this work have been incorporated into the
\texttt{hs\char45{}to\char45{}coq} distribution and are available as open source tools and
libraries.\footnote{\url{https://github.com/antalsz/hs-to-coq}.\ifanonymous{}
  Reviewers should note that links to commits and files can deanonymize the authors
  and may use the archive included with the submission, which contains a snapshot of the repository
  at commit \repohash{}.\fi}%
\scw{Don't forget to submit the tarball}


\section{The \LITcontainers{} library}
\label{sec:containers}

We select the \texttt{containers} library for our verification efforts because it is
a critical component of the Haskell ecosystem. With over 4000 publicly available Haskell packages using on \ensuretext{\straighttick\ttfamily{}containers}, it is the third-most dependent-up package on
the Haskell package repository Hackage, after \ensuretext{\straighttick\ttfamily{}base} and \ensuretext{\straighttick\ttfamily{}bytestring}.%
\footnote{\url{http://packdeps.haskellers.com/reverse}}

The \texttt{containers} library is both mature and highly optimized. It has existed for
over a decade and has undergone many significant revisions in order to improve
its performance.  It contains seven container data structures,
covering support for finite sets (\ensuretext{\straighttick\ttfamily{}Data.Set} and
\ensuretext{\straighttick\ttfamily{}Data.IntSet}), finite maps (\ensuretext{\straighttick\ttfamily{}Data.Map} and \ensuretext{\straighttick\ttfamily{}Data.IntMap}), sequences
(\ensuretext{\straighttick\ttfamily{}Data.Sequence}), graphs (\ensuretext{\straighttick\ttfamily{}Data.Graph}), and trees (\ensuretext{\straighttick\ttfamily{}Data.Tree}). However most users
of the \texttt{containers} library only use the map and set modules;%
\footnote{We calculated that 78\% of the packages on Hackage that depend on \texttt{containers} use only sets and maps.}%
moreover, the map modules are essentially analogues of the set
modules.  Therefore, we focus on \ensuretext{\straighttick\ttfamily{}Data.Set} and \ensuretext{\straighttick\ttfamily{}Data.IntSet} in this work.

\subsection{Weight-balanced trees and big-endian Patricia trees}\label{sec:tree-structures}

\begin{figure}[t]
\abovedisplayskip=0pt
\belowdisplayskip=0pt
\raggedright
\begingroup\straighttick\begin{hscode}\SaveRestoreHook
\column{B}{@{}>{\hspre}l<{\hspost}@{}}%
\column{6}{@{}>{\hspre}l<{\hspost}@{}}%
\column{13}{@{}>{\hspre}l<{\hspost}@{}}%
\column{14}{@{}>{\hspre}l<{\hspost}@{}}%
\column{16}{@{}>{\hspre}l<{\hspost}@{}}%
\column{36}{@{}>{\hspre}l<{\hspost}@{}}%
\column{41}{@{}>{\hspre}l<{\hspost}@{}}%
\column{45}{@{}>{\hspre}l<{\hspost}@{}}%
\column{E}{@{}>{\hspre}l<{\hspost}@{}}%
\>[B]{}\mbox{\onelinecomment  ------------------------------------------------------------------}{\spacestart}{\nextspace}{}\<[E]%
\\[-0.3ex]%
\>[B]{}\mbox{\onelinecomment   Sets are size balanced trees}{\spacestart}{\nextspace}{}\<[E]%
\\[-0.3ex]%
\>[B]{}\mbox{\onelinecomment  ------------------------------------------------------------------}{\spacestart}{\nextspace}{}\<[E]%
\\[-0.3ex]%
\>[B]{}\textbf{data}{\spacestart}{\nextspace}Set{\spacestart}{\nextspace}a{\spacestart}{\nextspace}{\nextspace}{}\<[13]%
\>[13]{}={\spacestart}{\nextspace}{\nextspace}{}\<[16]%
\>[16]{}Bin{\spacestart}{\nextspace}\mbox{\enskip\{-\# UNPACK  \#-\}\enskip}{\spacestart}{\nextspace}!Size{\spacestart}{\nextspace}!a{\spacestart}{\nextspace}!(Set{\spacestart}{\nextspace}a){\spacestart}{\nextspace}!(Set{\spacestart}{\nextspace}a){\spacestart}{\nextspace}{}\<[E]%
\\[-0.3ex]%
\>[B]{}{\spacestart}{\nextspace}{\nextspace}{\nextspace}{\nextspace}{\nextspace}{\nextspace}{\nextspace}{\nextspace}{\nextspace}{\nextspace}{\nextspace}{\nextspace}{}\<[13]%
\>[13]{}|{\spacestart}{\nextspace}{\nextspace}{}\<[16]%
\>[16]{}Tip{\spacestart}{\nextspace}{}\<[E]%
\\[\blanklineskip]%
\>[B]{}\textbf{type}{\spacestart}{\nextspace}Size{\spacestart}{\nextspace}{\nextspace}{\nextspace}{}\<[13]%
\>[13]{}={\spacestart}{\nextspace}Int{\spacestart}{\nextspace}{}\<[E]%
\\[\blanklineskip]%
\>[B]{}\mbox{\onelinecomment  | $O(\log n)$. Is the element in the set?}{\spacestart}{\nextspace}{}\<[E]%
\\[-0.3ex]%
\>[B]{}member{\spacestart}{\nextspace}::{\spacestart}{\nextspace}Ord{\spacestart}{\nextspace}a{\spacestart}{\nextspace}=>{\spacestart}{\nextspace}a{\spacestart}{\nextspace}->{\spacestart}{\nextspace}Set{\spacestart}{\nextspace}a{\spacestart}{\nextspace}->{\spacestart}{\nextspace}Bool{\spacestart}{\nextspace}{}\<[E]%
\\[-0.3ex]%
\>[B]{}member{\spacestart}{\nextspace}={\spacestart}{\nextspace}go{\spacestart}{\nextspace}{}\<[E]%
\\[-0.3ex]%
\>[B]{}~~{\spacestart}{\nextspace}{\nextspace}{}\<[6]%
\>[6]{}\textbf{where}{\spacestart}{\nextspace}{\nextspace}{\nextspace}{}\<[14]%
\>[14]{}go{\spacestart}{\nextspace}!\char95 {\spacestart}{\nextspace}Tip{\spacestart}{\nextspace}={\spacestart}{\nextspace}False{\spacestart}{\nextspace}{}\<[E]%
\\[-0.3ex]%
\>[B]{}{\spacestart}{\nextspace}{\nextspace}{\nextspace}{\nextspace}{\nextspace}{\nextspace}{\nextspace}{\nextspace}{\nextspace}{\nextspace}{\nextspace}{\nextspace}{\nextspace}{}\<[14]%
\>[14]{}go{\spacestart}{\nextspace}x{\spacestart}{\nextspace}(Bin{\spacestart}{\nextspace}\char95 {\spacestart}{\nextspace}y{\spacestart}{\nextspace}l{\spacestart}{\nextspace}r){\spacestart}{\nextspace}={\spacestart}{\nextspace}{\nextspace}{}\<[36]%
\>[36]{}\textbf{case}{\spacestart}{\nextspace}compare{\spacestart}{\nextspace}x{\spacestart}{\nextspace}y{\spacestart}{\nextspace}\textbf{of}{\spacestart}{\nextspace}{}\<[E]%
\\[-0.3ex]%
\>[B]{}{\spacestart}{\nextspace}{\nextspace}{\nextspace}{\nextspace}{\nextspace}{\nextspace}{\nextspace}{\nextspace}{\nextspace}{\nextspace}{\nextspace}{\nextspace}{\nextspace}{\nextspace}{\nextspace}{\nextspace}{\nextspace}{\nextspace}{\nextspace}{\nextspace}{\nextspace}{\nextspace}{\nextspace}{\nextspace}{\nextspace}{\nextspace}{\nextspace}{\nextspace}{\nextspace}{\nextspace}{\nextspace}{\nextspace}{\nextspace}{\nextspace}{\nextspace}{}\<[36]%
\>[36]{}~~{\spacestart}{\nextspace}{\nextspace}{}\<[41]%
\>[41]{}LT{\spacestart}{\nextspace}{\nextspace}{}\<[45]%
\>[45]{}->{\spacestart}{\nextspace}go{\spacestart}{\nextspace}x{\spacestart}{\nextspace}l{\spacestart}{\nextspace}{}\<[E]%
\\[-0.3ex]%
\>[B]{}{\spacestart}{\nextspace}{\nextspace}{\nextspace}{\nextspace}{\nextspace}{\nextspace}{\nextspace}{\nextspace}{\nextspace}{\nextspace}{\nextspace}{\nextspace}{\nextspace}{\nextspace}{\nextspace}{\nextspace}{\nextspace}{\nextspace}{\nextspace}{\nextspace}{\nextspace}{\nextspace}{\nextspace}{\nextspace}{\nextspace}{\nextspace}{\nextspace}{\nextspace}{\nextspace}{\nextspace}{\nextspace}{\nextspace}{\nextspace}{\nextspace}{\nextspace}{\nextspace}{\nextspace}{\nextspace}{\nextspace}{\nextspace}{}\<[41]%
\>[41]{}GT{\spacestart}{\nextspace}{\nextspace}{}\<[45]%
\>[45]{}->{\spacestart}{\nextspace}go{\spacestart}{\nextspace}x{\spacestart}{\nextspace}r{\spacestart}{\nextspace}{}\<[E]%
\\[-0.3ex]%
\>[B]{}{\spacestart}{\nextspace}{\nextspace}{\nextspace}{\nextspace}{\nextspace}{\nextspace}{\nextspace}{\nextspace}{\nextspace}{\nextspace}{\nextspace}{\nextspace}{\nextspace}{\nextspace}{\nextspace}{\nextspace}{\nextspace}{\nextspace}{\nextspace}{\nextspace}{\nextspace}{\nextspace}{\nextspace}{\nextspace}{\nextspace}{\nextspace}{\nextspace}{\nextspace}{\nextspace}{\nextspace}{\nextspace}{\nextspace}{\nextspace}{\nextspace}{\nextspace}{\nextspace}{\nextspace}{\nextspace}{\nextspace}{\nextspace}{}\<[41]%
\>[41]{}EQ{\spacestart}{\nextspace}{\nextspace}{}\<[45]%
\>[45]{}->{\spacestart}{\nextspace}True{}\<[E]%
\ColumnHook
\end{hscode}\resethooks\endgroup
\caption[The \ensuretext{\straighttick\ttfamily{}Set} data type and its membership function]{%
The \ensuretext{\straighttick\ttfamily{}Set} data type and its membership function\footnotemark%
}
\label{fig:setdef}
\end{figure}
\footnotetext{%
  From \url{http://hackage.haskell.org/package/containers-0.5.11.0/docs/src/Data.Set.Internal.html\#Set}.\\
  All code listings in this paper are manually reformatted and may omit module names from fully qualified names.}%

The \ensuretext{\straighttick\ttfamily{}Data.Set} module implements finite sets using weight-balanced binary
search trees.  The definition of the \ensuretext{\straighttick\ttfamily{}Set} datatype in this module, along with
its membership function, is given in \cref{fig:setdef}.
These sets and operations are polymorphic over the element type and require
only that this type is linearly ordered, as expressed by the
\ensuretext{\straighttick\ttfamily{}Ord} constraint on the \ensuretext{\straighttick\ttfamily{}member} function. The \ensuretext{\straighttick\ttfamily{}member} function 
descends the ordered search tree to determine whether it contains a
particular element.

The \ensuretext{\straighttick\ttfamily{}Size} component stored with the \ensuretext{\straighttick\ttfamily{}Bin} constructor is used by the
operations in the library to ensure that the tree stays balanced. The
implementation maintains the balancing
invariant
\[
s_1 + s_2 \le 1 \vee (s_1 \le 3 s_2 \wedge s_2 \le 3 s_1),
\]
where $s_1$ and $s_2$ are the sizes of the left and right subtrees of a \ensuretext{\straighttick\ttfamily{}Bin}
constructor. This definition is based on the description by
\citet{adams-tr}, who modified the
original weight-balanced tree proposed by~\citet{nievergelt}.
Thanks to this balancing, operations such as insertion, membership
testing, and deletion take time logarithmic in the size of the tree.

This type definition has been tweaked to improve the performance of the
library. The annotations on the \ensuretext{\straighttick\ttfamily{}Bin} data constructor instruct the compiler
to unpack the size component, removing a level of indirection. The \ensuretext{\straighttick\ttfamily{}!}
annotations indicate that all components should be strictly evaluated.

\medskip 

The \ensuretext{\straighttick\ttfamily{}Data.IntSet} module also provides search trees, specialized to
values of type \ensuretext{\straighttick\ttfamily{}Int} to provide more efficient operations, especially \ensuretext{\straighttick\ttfamily{}union}.
This implementation is based on big-endian
Patricia trees, as
proposed in \citeauthor{Morrison:1968}’s work on
PATRICIA~\citeyearpar{Morrison:1968} and described in a pure functional setting by
\citet{okasakigill}.

The definition of this data structure is shown in \cref{fig:comments}.
The core idea is to use the bits of the stored values to decide in which
subtree of a node they should be placed. In a node \mbox{\ensuretext{\straighttick\ttfamily{}Bin{\spacestart}{\nextspace}p{\spacestart}{\nextspace}m{\spacestart}{\nextspace}s1{\spacestart}{\nextspace}s2}}, the
mask \ensuretext{\straighttick\ttfamily{}m} has exactly one bit set. All bits higher than the mask bit are equal
in all elements of that node; they form the prefix \ensuretext{\straighttick\ttfamily{}p}. The mask bit is the
highest bit that is not shared by all elements. In particular, all elements in
\ensuretext{\straighttick\ttfamily{}s1} have this bit cleared, while all elements in \ensuretext{\straighttick\ttfamily{}s2} have it set. When looking
up a value \ensuretext{\straighttick\ttfamily{}x}, the mask bit of \ensuretext{\straighttick\ttfamily{}x} tells us into which branch to
descend.

\begin{figure}
\abovedisplayskip=0pt
\belowdisplayskip=0pt
\raggedright
\begingroup\straighttick\begin{hscode}\SaveRestoreHook
\column{B}{@{}>{\hspre}l<{\hspost}@{}}%
\column{14}{@{}>{\hspre}c<{\hspost}@{}}%
\column{14E}{@{}l@{}}%
\column{15}{@{}>{\hspre}l<{\hspost}@{}}%
\column{17}{@{}>{\hspre}l<{\hspost}@{}}%
\column{E}{@{}>{\hspre}l<{\hspost}@{}}%
\>[B]{}\textbf{data}{\spacestart}{\nextspace}IntSet{\spacestart}{\nextspace}{\nextspace}{}\<[14]%
\>[14]{}={\spacestart}{\nextspace}{\nextspace}{}\<[14E]%
\>[17]{}Bin{\spacestart}{\nextspace}\mbox{\enskip\{-\# UNPACK  \#-\}\enskip}{\spacestart}{\nextspace}!Prefix{\spacestart}{\nextspace}\mbox{\enskip\{-\# UNPACK  \#-\}\enskip}{\spacestart}{\nextspace}!Mask{\spacestart}{\nextspace}!IntSet{\spacestart}{\nextspace}!IntSet{\spacestart}{\nextspace}{}\<[E]%
\\[-0.3ex]%
\>[B]{}\mbox{\onelinecomment  Invariant: Nil is never found as a child of Bin.}{\spacestart}{\nextspace}{}\<[E]%
\\[-0.3ex]%
\>[B]{}\mbox{\onelinecomment  Invariant: The Mask is a power of 2.  It is the largest bit position at}{\spacestart}{\nextspace}{}\<[E]%
\\[-0.3ex]%
\>[B]{}\mbox{\onelinecomment  \phantom{Invariant:} which two elements of the set differ.}{\spacestart}{\nextspace}{}\<[E]%
\\[-0.3ex]%
\>[B]{}\mbox{\onelinecomment  Invariant: Prefix is the common high-order bits that all elements share to}{\spacestart}{\nextspace}{}\<[E]%
\\[-0.3ex]%
\>[B]{}\mbox{\onelinecomment  \phantom{Invariant:} the left of the Mask bit.}{\spacestart}{\nextspace}{}\<[E]%
\\[-0.3ex]%
\>[B]{}\mbox{\onelinecomment  Invariant: In Bin prefix mask left right, left consists of the elements}{\spacestart}{\nextspace}{}\<[E]%
\\[-0.3ex]%
\>[B]{}\mbox{\onelinecomment  \phantom{Invariant:} that don't have the mask bit set; right is all the elements}{\spacestart}{\nextspace}{}\<[E]%
\\[-0.3ex]%
\>[B]{}\mbox{\onelinecomment  \phantom{Invariant:} that do.}{\spacestart}{\nextspace}{}\<[E]%
\\[-0.3ex]%
\>[B]{}{\spacestart}{\nextspace}{\nextspace}{\nextspace}{\nextspace}{\nextspace}{\nextspace}{\nextspace}{\nextspace}{\nextspace}{\nextspace}{\nextspace}{\nextspace}{\nextspace}{}\<[14]%
\>[14]{}|{\spacestart}{\nextspace}{\nextspace}{}\<[14E]%
\>[17]{}Tip{\spacestart}{\nextspace}\mbox{\enskip\{-\# UNPACK  \#-\}\enskip}{\spacestart}{\nextspace}!Prefix{\spacestart}{\nextspace}\mbox{\enskip\{-\# UNPACK  \#-\}\enskip}{\spacestart}{\nextspace}!BitMap{\spacestart}{\nextspace}{}\<[E]%
\\[-0.3ex]%
\>[B]{}\mbox{\onelinecomment  Invariant: The Prefix is zero for the last 5 (on 32 bit arches) or 6 bits}{\spacestart}{\nextspace}{}\<[E]%
\\[-0.3ex]%
\>[B]{}\mbox{\onelinecomment  \phantom{Invariant:} (on 64 bit arches). The values of the set represented by a tip}{\spacestart}{\nextspace}{}\<[E]%
\\[-0.3ex]%
\>[B]{}\mbox{\onelinecomment  \phantom{Invariant:} are the prefix plus the indices of the set bits in the bit map.}{\spacestart}{\nextspace}{}\<[E]%
\\[-0.3ex]%
\>[B]{}{\spacestart}{\nextspace}{\nextspace}{\nextspace}{\nextspace}{\nextspace}{\nextspace}{\nextspace}{\nextspace}{\nextspace}{\nextspace}{\nextspace}{\nextspace}{\nextspace}{}\<[14]%
\>[14]{}|{\spacestart}{\nextspace}{\nextspace}{}\<[14E]%
\>[17]{}Nil{\spacestart}{\nextspace}{}\<[E]%
\\[\blanklineskip]%
\>[B]{}\mbox{\onelinecomment  A number stored in a set is stored as}{\spacestart}{\nextspace}{}\<[E]%
\\[-0.3ex]%
\>[B]{}\mbox{\onelinecomment  * Prefix (all but last 5-6 bits) and}{\spacestart}{\nextspace}{}\<[E]%
\\[-0.3ex]%
\>[B]{}\mbox{\onelinecomment  * BitMap (last 5-6 bits stored as a bitmask)}{\spacestart}{\nextspace}{}\<[E]%
\\[-0.3ex]%
\>[B]{}\mbox{\onelinecomment  \phantom*  Last 5-6 bits are called a Suffix.}{\spacestart}{\nextspace}{}\<[E]%
\\[-0.3ex]%
\>[B]{}\textbf{type}{\spacestart}{\nextspace}Prefix{\spacestart}{\nextspace}{\nextspace}{\nextspace}{}\<[15]%
\>[15]{}={\spacestart}{\nextspace}Int{\spacestart}{\nextspace}{}\<[E]%
\\[-0.3ex]%
\>[B]{}\textbf{type}{\spacestart}{\nextspace}Mask{\spacestart}{\nextspace}{\nextspace}{\nextspace}{\nextspace}{\nextspace}{}\<[15]%
\>[15]{}={\spacestart}{\nextspace}Int{\spacestart}{\nextspace}{}\<[E]%
\\[-0.3ex]%
\>[B]{}\textbf{type}{\spacestart}{\nextspace}BitMap{\spacestart}{\nextspace}{\nextspace}{\nextspace}{}\<[15]%
\>[15]{}={\spacestart}{\nextspace}Word{\spacestart}{\nextspace}{}\<[E]%
\\[-0.3ex]%
\>[B]{}\textbf{type}{\spacestart}{\nextspace}Key{\spacestart}{\nextspace}{\nextspace}{\nextspace}{\nextspace}{\nextspace}{\nextspace}{}\<[15]%
\>[15]{}={\spacestart}{\nextspace}Int{}\<[E]%
\ColumnHook
\end{hscode}\resethooks\endgroup
\iflong
\begingroup\straighttick\begin{hscode}\SaveRestoreHook
\column{B}{@{}>{\hspre}l<{\hspost}@{}}%
\column{12}{@{}>{\hspre}l<{\hspost}@{}}%
\column{30}{@{}>{\hspre}l<{\hspost}@{}}%
\column{47}{@{}>{\hspre}l<{\hspost}@{}}%
\column{E}{@{}>{\hspre}l<{\hspost}@{}}%
\>[B]{}\mbox{\onelinecomment  | $O(min(n,W))$. Is the value a member of the set?}{\spacestart}{\nextspace}{}\<[E]%
\\[-0.3ex]%
\>[B]{}member{\spacestart}{\nextspace}::{\spacestart}{\nextspace}Key{\spacestart}{\nextspace}->{\spacestart}{\nextspace}IntSet{\spacestart}{\nextspace}->{\spacestart}{\nextspace}Bool{\spacestart}{\nextspace}{}\<[E]%
\\[-0.3ex]%
\>[B]{}member{\spacestart}{\nextspace}!x{\spacestart}{\nextspace}={\spacestart}{\nextspace}go{\spacestart}{\nextspace}{}\<[E]%
\\[-0.3ex]%
\>[B]{}~~{\spacestart}{\nextspace}\textbf{where}{\spacestart}{\nextspace}{\nextspace}{}\<[12]%
\>[12]{}go{\spacestart}{\nextspace}(Bin{\spacestart}{\nextspace}p{\spacestart}{\nextspace}m{\spacestart}{\nextspace}l{\spacestart}{\nextspace}r){\spacestart}{\nextspace}{\nextspace}{}\<[30]%
\>[30]{}|{\spacestart}{\nextspace}nomatch{\spacestart}{\nextspace}x{\spacestart}{\nextspace}p{\spacestart}{\nextspace}m{\spacestart}{\nextspace}{\nextspace}{}\<[47]%
\>[47]{}={\spacestart}{\nextspace}False{\spacestart}{\nextspace}{}\<[E]%
\\[-0.3ex]%
\>[B]{}{\spacestart}{\nextspace}{\nextspace}{\nextspace}{\nextspace}{\nextspace}{\nextspace}{\nextspace}{\nextspace}{\nextspace}{\nextspace}{\nextspace}{\nextspace}{\nextspace}{\nextspace}{\nextspace}{\nextspace}{\nextspace}{\nextspace}{\nextspace}{\nextspace}{\nextspace}{\nextspace}{\nextspace}{\nextspace}{\nextspace}{\nextspace}{\nextspace}{\nextspace}{\nextspace}{}\<[30]%
\>[30]{}|{\spacestart}{\nextspace}zero{\spacestart}{\nextspace}x{\spacestart}{\nextspace}m{\spacestart}{\nextspace}{\nextspace}{\nextspace}{\nextspace}{\nextspace}{\nextspace}{\nextspace}{}\<[47]%
\>[47]{}={\spacestart}{\nextspace}go{\spacestart}{\nextspace}l{\spacestart}{\nextspace}{}\<[E]%
\\[-0.3ex]%
\>[B]{}{\spacestart}{\nextspace}{\nextspace}{\nextspace}{\nextspace}{\nextspace}{\nextspace}{\nextspace}{\nextspace}{\nextspace}{\nextspace}{\nextspace}{\nextspace}{\nextspace}{\nextspace}{\nextspace}{\nextspace}{\nextspace}{\nextspace}{\nextspace}{\nextspace}{\nextspace}{\nextspace}{\nextspace}{\nextspace}{\nextspace}{\nextspace}{\nextspace}{\nextspace}{\nextspace}{}\<[30]%
\>[30]{}|{\spacestart}{\nextspace}otherwise{\spacestart}{\nextspace}{\nextspace}{\nextspace}{\nextspace}{\nextspace}{\nextspace}{}\<[47]%
\>[47]{}={\spacestart}{\nextspace}go{\spacestart}{\nextspace}r{\spacestart}{\nextspace}{}\<[E]%
\\[-0.3ex]%
\>[B]{}{\spacestart}{\nextspace}{\nextspace}{\nextspace}{\nextspace}{\nextspace}{\nextspace}{\nextspace}{\nextspace}{\nextspace}{\nextspace}{\nextspace}{}\<[12]%
\>[12]{}go{\spacestart}{\nextspace}(Tip{\spacestart}{\nextspace}y{\spacestart}{\nextspace}bm){\spacestart}{\nextspace}={\spacestart}{\nextspace}prefixOf{\spacestart}{\nextspace}x{\spacestart}{\nextspace}=={\spacestart}{\nextspace}y{\spacestart}{\nextspace}\&\&{\spacestart}{\nextspace}bitmapOf{\spacestart}{\nextspace}x{\spacestart}{\nextspace}.\&.{\spacestart}{\nextspace}bm{\spacestart}{\nextspace}/={\spacestart}{\nextspace}0{\spacestart}{\nextspace}{}\<[E]%
\\[-0.3ex]%
\>[B]{}{\spacestart}{\nextspace}{\nextspace}{\nextspace}{\nextspace}{\nextspace}{\nextspace}{\nextspace}{\nextspace}{\nextspace}{\nextspace}{\nextspace}{}\<[12]%
\>[12]{}go{\spacestart}{\nextspace}Nil{\spacestart}{\nextspace}={\spacestart}{\nextspace}False{}\<[E]%
\ColumnHook
\end{hscode}\resethooks\endgroup
\fi
\vspace*{-4ex}
\caption[The \ensuretext{\straighttick\ttfamily{}IntSet} data type\iflong{} and its membership function\fi]{%
The \ensuretext{\straighttick\ttfamily{}IntSet} data type\iflong{} and its membership function\fi\footnotemark%
}
\label{fig:comments}
\vspace*{-1ex}
\end{figure} 
\footnotetext{%
From \url{http://hackage.haskell.org/package/containers-0.5.11.0/docs/src/Data.IntSet.Internal.html\#IntSet}}

Instead of storing a single value at the leaf of the tree,
this implementation improves time and space performance
by storing the membership information of consecutive numbers as the bits of a
machine-word-sized bitmap in the \ensuretext{\straighttick\ttfamily{}Tip} constructor.

The \ensuretext{\straighttick\ttfamily{}Nil} constructor is the only way to represent an empty tree, and will
never occur as the child of a \ensuretext{\straighttick\ttfamily{}Bin} constructor.  Every well-formed \ensuretext{\straighttick\ttfamily{}IntSet} is
either made of \ensuretext{\straighttick\ttfamily{}Bin}s and \ensuretext{\straighttick\ttfamily{}Tip}s, or a single \ensuretext{\straighttick\ttfamily{}Nil}.


\subsection{A history of performance tuning}
\label{sec:containers-history}

The history of the \ensuretext{\straighttick\ttfamily{}Data.Set} module can be traced back to 2004, when a number
of competing search tree implementations were debated in the ``Tree Wars''
thread on the Haskell libraries mailing list. Benchmarks
showed that Daan Leijen’s implementation had the best performance, and it was
added to \texttt{containers} in 2005 as \ensuretext{\straighttick\ttfamily{}Data.Set}.\containerCommit{bbbba97c}

In~\citeyear{straka}, Milan Straka thoroughly evaluated the performance of the
\texttt{containers} library and implemented a number of performance
tweaks~\citep{straka}.
\iflong
For example:
\begin{quote}
When balancing a node, the function \ensuretext{\straighttick\ttfamily{}balance} checked the balancing condition and
called one of the four rotating functions, which rebuilt the tree using smart
constructors. This resulted in a repeated pattern matching, which was
unnecessary. We rewrote the \ensuretext{\straighttick\ttfamily{}balance} function to contain all the logic and to
use as few pattern matches as possible. That resulted in significant
performance improvements in all \ensuretext{\straighttick\ttfamily{}Set} methods that modify a given set.
\end{quote}
\fi
This change%
\containerCommit{3535fcbe} replaced
a fairly readable \ensuretext{\straighttick\ttfamily{}balance} and several small and descriptive helper functions
with a single dense block of code. A later change%
\containerCommit{d17d7182} by
\citeauthor{straka} created two copies of this scary-looking \ensuretext{\straighttick\ttfamily{}balance}, each
specialized and optimized for different preconditions.

\Citet{adams-tr} describes two algorithms for \ensuretext{\straighttick\ttfamily{}union}, \ensuretext{\straighttick\ttfamily{}intersection}, and \ensuretext{\straighttick\ttfamily{}difference}: “hedge union” and “divide and conquer”. Originally \texttt{containers}
used the former, but in 2016 its maintainers switched to the
latter,\containerCommit{c3083cfc}
again based on performance measurement.

The module \ensuretext{\straighttick\ttfamily{}Data.IntSet} (and \ensuretext{\straighttick\ttfamily{}Data.IntMap}) has been around even longer.
\citeauthor{okasakigill} mention in their \citeyear{okasakigill}
paper~\citep{okasakigill} that GHC had already made use of \texttt{IntSet} and \texttt{IntMap}
for several years.  In 2011,
the \ensuretext{\straighttick\ttfamily{}Data.IntSet} module was re-written to use machine-words as bit maps in
the leaves of the tree, as discussed at the end of
\cref{sec:tree-structures}.%
\containerPull{3}
This moved the \texttt{containers} library further away
from the literature on Patricia trees and introduced a fair amount of low-level
bit twiddling operations (e.g., \ensuretext{\straighttick\ttfamily{}highestBitMask}, \ensuretext{\straighttick\ttfamily{}lowestBitMask},
and \ensuretext{\straighttick\ttfamily{}revNat}).

\iflong
 \subsection{The test suite of \LITcontainers}
 \label{sec:testsuite}
 \scw{Not sure that this subsection adds that much.}%
 The first tests were added to \texttt{containers} in 2007,%
 \containerCommit{9d6c49b5}
 in the form of a few regressions tests for observed bugs. Three years later, Don Stewart added a comprehensive test suite using QuickCheck~\citep{quickcheck} with 91\% code coverage, and reported “No bugs were found”.%
 \containerCommit{38743e39}
 This test suite helped to maintain a consistently high quality and very few
 bugs crept into released versions of the library. In fact, the only serious bug mentioned in the libraries changelog -- a completely broken implementation of \ensuretext{\straighttick\ttfamily{}Data.IntMap.restrictKeys} -- only occurred because the tests for \ensuretext{\straighttick\ttfamily{}restrictKeys} were accidentially not run as part of the test suite.%
 \containerIssue{392}
\fi

\section{Overview of our verification approach}
\label{sec:verification-overview}

In order to verify \ensuretext{\straighttick\ttfamily{}Set} and \ensuretext{\straighttick\ttfamily{}IntSet}, we use \texttt{hs\char45{}to\char45{}coq} to
translate the unmodified Haskell modules to Gallina and then use Coq to verify
the translated code.
For example, consider the excerpt of the implementation
of \ensuretext{\straighttick\ttfamily{}Set} in \cref{fig:setdef}.
The \texttt{hs\char45{}to\char45{}coq} tool translates this input to the following Coq definitions.%
\footnote{In the file \fileContainers{lib/Data/Set/Internal.v}}
The strictness and unpacking annotations are ignored, as they do not make sense
in Coq, and the type name \ensuretext{\straighttick\ttfamily{}Set} is renamed to \ensuretext{\straighttick\ttfamily{}Set\char95 } to avoid clashing with the
Coq keyword.
\begingroup\straighttick\begin{hscode}\SaveRestoreHook
\column{B}{@{}>{\hspre}l<{\hspost}@{}}%
\column{6}{@{}>{\hspre}l<{\hspost}@{}}%
\column{10}{@{}>{\hspre}l<{\hspost}@{}}%
\column{11}{@{}>{\hspre}c<{\hspost}@{}}%
\column{11E}{@{}l@{}}%
\column{15}{@{}>{\hspre}l<{\hspost}@{}}%
\column{20}{@{}>{\hspre}l<{\hspost}@{}}%
\column{41}{@{}>{\hspre}l<{\hspost}@{}}%
\column{46}{@{}>{\hspre}l<{\hspost}@{}}%
\column{52}{@{}>{\hspre}l<{\hspost}@{}}%
\column{E}{@{}>{\hspre}l<{\hspost}@{}}%
\>[B]{}\textbf{Definition}{\spacestart}{\nextspace}Size{\spacestart}{\nextspace}:={\spacestart}{\nextspace}GHC.Num.Int\%\textbf{type}.{\spacestart}{\nextspace}{}\<[E]%
\\[\blanklineskip]%
\>[B]{}\textbf{Inductive}{\spacestart}{\nextspace}Set\char95 {\spacestart}{\nextspace}a{\spacestart}{\nextspace}:{\spacestart}{\nextspace}\textbf{Type}{\spacestart}{\nextspace}{}\<[E]%
\\[-0.3ex]%
\>[B]{}~~{\spacestart}{\nextspace}{\nextspace}{}\<[6]%
\>[6]{}:={\spacestart}{\nextspace}{\nextspace}{}\<[10]%
\>[10]{}Bin{\spacestart}{\nextspace}:{\spacestart}{\nextspace}Size{\spacestart}{\nextspace}->{\spacestart}{\nextspace}a{\spacestart}{\nextspace}->{\spacestart}{\nextspace}(Set\char95 {\spacestart}{\nextspace}a){\spacestart}{\nextspace}->{\spacestart}{\nextspace}(Set\char95 {\spacestart}{\nextspace}a){\spacestart}{\nextspace}->{\spacestart}{\nextspace}Set\char95 {\spacestart}{\nextspace}a{\spacestart}{\nextspace}{}\<[E]%
\\[-0.3ex]%
\>[B]{}{\spacestart}{\nextspace}{\nextspace}{\nextspace}{\nextspace}{\nextspace}{}\<[6]%
\>[6]{}|{\spacestart}{\nextspace}{\nextspace}{\nextspace}{}\<[10]%
\>[10]{}Tip{\spacestart}{\nextspace}:{\spacestart}{\nextspace}Set\char95 {\spacestart}{\nextspace}a.{\spacestart}{\nextspace}{}\<[E]%
\\[\blanklineskip]%
\>[B]{}\textbf{Definition}{\spacestart}{\nextspace}member{\spacestart}{\nextspace}{\char123}a{\char125}{\spacestart}{\nextspace}{\textasciigrave}{\char123}GHC.Base.Ord{\spacestart}{\nextspace}a{\char125}{\spacestart}{\nextspace}:{\spacestart}{\nextspace}a{\spacestart}{\nextspace}->{\spacestart}{\nextspace}Set\char95 {\spacestart}{\nextspace}a{\spacestart}{\nextspace}->{\spacestart}{\nextspace}bool{\spacestart}{\nextspace}:={\spacestart}{\nextspace}{}\<[E]%
\\[-0.3ex]%
\>[B]{}~~{\spacestart}{\nextspace}{\nextspace}{}\<[6]%
\>[6]{}\textbf{let}{\spacestart}{\nextspace}\textbf{fix}{\spacestart}{\nextspace}go{\spacestart}{\nextspace}arg\char95 0\char95 \char95 {\spacestart}{\nextspace}arg\char95 1\char95 \char95 {\spacestart}{\nextspace}{}\<[E]%
\\[-0.3ex]%
\>[B]{}{\spacestart}{\nextspace}{\nextspace}{\nextspace}{\nextspace}{\nextspace}{}\<[6]%
\>[6]{}~~{\spacestart}{\nextspace}{\nextspace}{}\<[11]%
\>[11]{}:={\spacestart}{\nextspace}{\nextspace}{}\<[11E]%
\>[15]{}\textbf{match}{\spacestart}{\nextspace}arg\char95 0\char95 \char95 ,{\spacestart}{\nextspace}arg\char95 1\char95 \char95 {\spacestart}{\nextspace}\textbf{with}{\spacestart}{\nextspace}{}\<[E]%
\\[-0.3ex]%
\>[B]{}{\spacestart}{\nextspace}{\nextspace}{\nextspace}{\nextspace}{\nextspace}{\nextspace}{\nextspace}{\nextspace}{\nextspace}{\nextspace}{\nextspace}{\nextspace}{\nextspace}{\nextspace}{}\<[15]%
\>[15]{}~~{\spacestart}{\nextspace}{\nextspace}{}\<[20]%
\>[20]{}|{\spacestart}{\nextspace}\char95 ,{\spacestart}{\nextspace}Tip{\spacestart}{\nextspace}=>{\spacestart}{\nextspace}false{\spacestart}{\nextspace}{}\<[E]%
\\[-0.3ex]%
\>[B]{}{\spacestart}{\nextspace}{\nextspace}{\nextspace}{\nextspace}{\nextspace}{\nextspace}{\nextspace}{\nextspace}{\nextspace}{\nextspace}{\nextspace}{\nextspace}{\nextspace}{\nextspace}{\nextspace}{\nextspace}{\nextspace}{\nextspace}{\nextspace}{}\<[20]%
\>[20]{}|{\spacestart}{\nextspace}x,{\spacestart}{\nextspace}Bin{\spacestart}{\nextspace}\char95 {\spacestart}{\nextspace}y{\spacestart}{\nextspace}l{\spacestart}{\nextspace}r{\spacestart}{\nextspace}=>{\spacestart}{\nextspace}{\nextspace}{}\<[41]%
\>[41]{}\textbf{match}{\spacestart}{\nextspace}GHC.Base.compare{\spacestart}{\nextspace}x{\spacestart}{\nextspace}y{\spacestart}{\nextspace}\textbf{with}{\spacestart}{\nextspace}{}\<[E]%
\\[-0.3ex]%
\>[B]{}{\spacestart}{\nextspace}{\nextspace}{\nextspace}{\nextspace}{\nextspace}{\nextspace}{\nextspace}{\nextspace}{\nextspace}{\nextspace}{\nextspace}{\nextspace}{\nextspace}{\nextspace}{\nextspace}{\nextspace}{\nextspace}{\nextspace}{\nextspace}{\nextspace}{\nextspace}{\nextspace}{\nextspace}{\nextspace}{\nextspace}{\nextspace}{\nextspace}{\nextspace}{\nextspace}{\nextspace}{\nextspace}{\nextspace}{\nextspace}{\nextspace}{\nextspace}{\nextspace}{\nextspace}{\nextspace}{\nextspace}{\nextspace}{}\<[41]%
\>[41]{}~~{\spacestart}{\nextspace}{\nextspace}{}\<[46]%
\>[46]{}|{\spacestart}{\nextspace}Lt{\spacestart}{\nextspace}{\nextspace}{}\<[52]%
\>[52]{}=>{\spacestart}{\nextspace}go{\spacestart}{\nextspace}x{\spacestart}{\nextspace}l{\spacestart}{\nextspace}{}\<[E]%
\\[-0.3ex]%
\>[B]{}{\spacestart}{\nextspace}{\nextspace}{\nextspace}{\nextspace}{\nextspace}{\nextspace}{\nextspace}{\nextspace}{\nextspace}{\nextspace}{\nextspace}{\nextspace}{\nextspace}{\nextspace}{\nextspace}{\nextspace}{\nextspace}{\nextspace}{\nextspace}{\nextspace}{\nextspace}{\nextspace}{\nextspace}{\nextspace}{\nextspace}{\nextspace}{\nextspace}{\nextspace}{\nextspace}{\nextspace}{\nextspace}{\nextspace}{\nextspace}{\nextspace}{\nextspace}{\nextspace}{\nextspace}{\nextspace}{\nextspace}{\nextspace}{\nextspace}{\nextspace}{\nextspace}{\nextspace}{\nextspace}{}\<[46]%
\>[46]{}|{\spacestart}{\nextspace}Gt{\spacestart}{\nextspace}{\nextspace}{}\<[52]%
\>[52]{}=>{\spacestart}{\nextspace}go{\spacestart}{\nextspace}x{\spacestart}{\nextspace}r{\spacestart}{\nextspace}{}\<[E]%
\\[-0.3ex]%
\>[B]{}{\spacestart}{\nextspace}{\nextspace}{\nextspace}{\nextspace}{\nextspace}{\nextspace}{\nextspace}{\nextspace}{\nextspace}{\nextspace}{\nextspace}{\nextspace}{\nextspace}{\nextspace}{\nextspace}{\nextspace}{\nextspace}{\nextspace}{\nextspace}{\nextspace}{\nextspace}{\nextspace}{\nextspace}{\nextspace}{\nextspace}{\nextspace}{\nextspace}{\nextspace}{\nextspace}{\nextspace}{\nextspace}{\nextspace}{\nextspace}{\nextspace}{\nextspace}{\nextspace}{\nextspace}{\nextspace}{\nextspace}{\nextspace}{\nextspace}{\nextspace}{\nextspace}{\nextspace}{\nextspace}{}\<[46]%
\>[46]{}|{\spacestart}{\nextspace}Eq{\spacestart}{\nextspace}{\nextspace}{}\<[52]%
\>[52]{}=>{\spacestart}{\nextspace}true{\spacestart}{\nextspace}{}\<[E]%
\\[-0.3ex]%
\>[B]{}{\spacestart}{\nextspace}{\nextspace}{\nextspace}{\nextspace}{\nextspace}{\nextspace}{\nextspace}{\nextspace}{\nextspace}{\nextspace}{\nextspace}{\nextspace}{\nextspace}{\nextspace}{\nextspace}{\nextspace}{\nextspace}{\nextspace}{\nextspace}{\nextspace}{\nextspace}{\nextspace}{\nextspace}{\nextspace}{\nextspace}{\nextspace}{\nextspace}{\nextspace}{\nextspace}{\nextspace}{\nextspace}{\nextspace}{\nextspace}{\nextspace}{\nextspace}{\nextspace}{\nextspace}{\nextspace}{\nextspace}{\nextspace}{}\<[41]%
\>[41]{}\textbf{end}{\spacestart}{\nextspace}{}\<[E]%
\\[-0.3ex]%
\>[B]{}{\spacestart}{\nextspace}{\nextspace}{\nextspace}{\nextspace}{\nextspace}{\nextspace}{\nextspace}{\nextspace}{\nextspace}{\nextspace}{\nextspace}{\nextspace}{\nextspace}{\nextspace}{}\<[15]%
\>[15]{}\textbf{end}{\spacestart}{\nextspace}{}\<[E]%
\\[-0.3ex]%
\>[B]{}{\spacestart}{\nextspace}{\nextspace}{\nextspace}{\nextspace}{\nextspace}{}\<[6]%
\>[6]{}\textbf{in}{\spacestart}{\nextspace}go.{}\<[E]%
\ColumnHook
\end{hscode}\resethooks\endgroup
These definitions depend on \texttt{hs\char45{}to\char45{}coq}'s pre-existing translated version of
GHC's standard library \texttt{base}.  Here, we use the existing translation of Haskell's
\ensuretext{\straighttick\ttfamily{}Int} type, the \ensuretext{\straighttick\ttfamily{}Ord} type class, and \ensuretext{\straighttick\ttfamily{}Ord}'s \ensuretext{\straighttick\ttfamily{}compare} method.

We carry out this translation for the \ensuretext{\straighttick\ttfamily{}Set} and \ensuretext{\straighttick\ttfamily{}IntSet} along with their
attendant functions, and then verify the resulting Gallina code.  In
\cref{sec:specs} we discuss the properties that we prove about the two data
structures.

To further test that the translation from Haskell to Coq,
we also used Coq's extraction mechanism to translate the
generated Gallina code, like that seen above, \emph{back} to Haskell. This process
converts the implicitly-passed type class dictionaries to ordinary
explicitly-passed function arguments, but otherwise preserves the structure of
the code.
%
%
%
By providing an interface that restores the type-class based types, we can run
the original \texttt{containers} test suite against this code. This process helps us
check that \texttt{hs\char45{}to\char45{}coq}
preserves the semantics of the original Haskell program during the translation
process.

\label{sec:scope}

\locfigure

\Cref{fig:locfigure} provides an overview
of the Haskell code that we target, the Gallina code that we translate it into,
and the Coq proofs that we wrote.
The set modules of the \texttt{containers} library contain \funcs{}
functions and \tycls{} type class instances, written in \locHaskell{} lines of
code (excluding comments and blank lines).  Out of these, \funcsUntranslated{}
functions and \tyclsUntranslated{} type class instances
(\locHaskellUntranslated{} loc) were deemed ``out of scope'' and not
translated. (We discuss
untranslated definitions in more detail in
\cref{sec:unwanted-code,sec:untranslated}.)
Our translation produces \locGallina{} lines of Gallina code.


The \ensuretext{\straighttick\ttfamily{}Set} and \ensuretext{\straighttick\ttfamily{}IntSet} data structures come with extensive APIs. We specify and
verify a representative subset of commonly used functions (listed in
\cref{fig:verifiedAPI}), covering \percVerifiedSet{}\% of the \ensuretext{\straighttick\ttfamily{}Set} API and
\percVerifiedIntSet{}\% of the \ensuretext{\straighttick\ttfamily{}IntSet} API.  This verified API is complete
enough to instantiate Coq's specification of finite sets, along with many other
specifications at varying levels of abstraction; for more detail, see
\cref{sec:specs}.


As Coq is not an automated theorem prover, verification of these complex data
structures requires significant effort. In total, we verified
\locHaskellVerified{} lines of Haskell; the verification of \ensuretext{\straighttick\ttfamily{}Set} and \ensuretext{\straighttick\ttfamily{}IntSet}
required \locProofPerHaskell{}~lines of proof per lines of code. This factor is
noticeably higher for \ensuretext{\straighttick\ttfamily{}IntSet} (\locProofPerHaskellIntSet$\times$) than for
\ensuretext{\straighttick\ttfamily{}Set} (\locProofPerHaskellSet$\times$), as the latter is conceptually simpler to
reason about and allowed us to achieve a higher degree of automation using Coq
tactics.

Our proofs also require the formalization of several background theories (not
counted in the proof-to-code ratio above), including: integer arithmetic and
bits (\locArith~loc): lists and sortedness (\locLists~loc); dyadic intervals,
which are used for verifying \ensuretext{\straighttick\ttfamily{}IntSet} (\locDyadic~loc); and support for working
with lawful \ensuretext{\straighttick\ttfamily{}Ord} instances, including a complete decision procedure
(\locOrder~loc).

\verifiedAPIfigure

\section{Specifying \LITSet{} and \LITIntSet}
\label{sec:specs}

The phrase ``we have verified this piece of software'' on its own is
meaningless: the particular specification that a piece of software is
verified against matters.
Good specifications are \emph{rich}, \emph{two-sided}, \emph{formal}, and
\emph{live}~\citep{deepspec}. A specification is \emph{rich} if it ``describ[es]
complex behaviors in detail''. It is \emph{two-sided} if it is ``connected to
both implementations and clients''.  It is \emph{formal} if it is ``written in
a mathematical notation with clear semantics''.  And it is \emph{live} if it
is ``connected via machine-checkable proofs to the implementation''.

All specifications of \ensuretext{\straighttick\ttfamily{}Set} and \ensuretext{\straighttick\ttfamily{}IntSet} that we use
are formal and live by definition.  They are formal because we
express the desired properties using Gallina, the language of the Coq proof
assistant; and they are live because we use \texttt{hs-to-coq} to
automatically convert \texttt{containers} to Coq where we develop
and check our proofs.
But how can we ensure that our specifications of \ensuretext{\straighttick\ttfamily{}Set} and \ensuretext{\straighttick\ttfamily{}IntSet} are
two-sided and rich? How do we know that the specifications are not just
facts that happen to be true, but are useful for the
verification of larger systems? What complex behaviors of the data
structures should we specify?

To ensure that our specifications are two-sided, we use specifications that we
did not invent ourselves. Instead, we draw our specifications from a variety of
diverse sources: from several parts of the \texttt{containers} codebase
(\cref{sec:specs-comments,sec:specs-tests,sec:specs-int,sec:specs-rewrite-rules}),
from Haskell type class laws (\cref{sec:specs-type-classes}), from pre-existing
Coq theories (\cref{sec:specs-coq}), and from a mathematical description of sets
(\cref{sec:specs-math}).  This way, we also ensure that our specifications
are rich because they describe the complex \ensuretext{\straighttick\ttfamily{}Set} and \ensuretext{\straighttick\ttfamily{}IntSet} data structures at
varying levels of abstraction.  Finally, by verifying the code against these
disparate specifications, we not only increase the assurance that we captured
all the important behaviors of \ensuretext{\straighttick\ttfamily{}Set} and \ensuretext{\straighttick\ttfamily{}IntSet}, but we also cross-validate
the specifications against each other.

\subsection{Specifying implementation invariants}
\label{sec:specs-comments}
\label{sec:specs-valid}

\ensuretext{\straighttick\ttfamily{}Set} and \ensuretext{\straighttick\ttfamily{}IntSet} are two examples of abstract types whose
correctness depend on invariants.
Therefore, we define well-formedness predicates
\ensuretext{\straighttick\ttfamily{}WF:{\spacestart}{\nextspace}Set\char95 {\spacestart}{\nextspace}e{\spacestart}{\nextspace}->{\spacestart}{\nextspace}\textbf{Prop}} and \ensuretext{\straighttick\ttfamily{}WF{\spacestart}{\nextspace}:{\spacestart}{\nextspace}IntSet{\spacestart}{\nextspace}->{\spacestart}{\nextspace}\textbf{Prop}}  and show that the
operations preserve these properties.  The definition of
well-formedness differs between the two types,
but specifications of both are already present within \texttt{containers}.

\paragraph{Well-formed weight-balanced trees}

Our definition of \ensuretext{\straighttick\ttfamily{}WF} for \ensuretext{\straighttick\ttfamily{}Set} is derived from the \ensuretext{\straighttick\ttfamily{}valid} function defined
in the \texttt{containers} library.  This function checks whether the input (1) is a
balanced tree, (2) is an ordered tree, and (3) has the correct values in its
size fields.  It is not part of the normal, user-facing API of \texttt{containers}
(since all exported functions preserve well-formedness), but is used
internally by the developers for debugging and testing.  For us, it is
valuable as an executable specification, with less room for ambiguity and
interpretation than comments and documentation.

However, rather than using \ensuretext{\straighttick\ttfamily{}valid} directly, we define well-formedness as an
inductive predicate, because we find it more useful from a proof
engineering perspective. In particular, our definition of \ensuretext{\straighttick\ttfamily{}WF}, as
shown in \cref{fig:set-wf}, relies on the
\ensuretext{\straighttick\ttfamily{}Bounded} inductive family. Its indices express lower and
upper bounds of the elements stored in the tree; \ensuretext{\straighttick\ttfamily{}None} means unbounded.
At the same
time, the property also checks that the sizes of the two subtrees are balanced
in the \ensuretext{\straighttick\ttfamily{}Bin} case.\footnote{In the file \fileContainers{theories/SetProofs.v}}
Nevertheless, we can relate the \ensuretext{\straighttick\ttfamily{}WF} predicate to the \ensuretext{\straighttick\ttfamily{}valid} function found in \texttt{containers}:
\begingroup\straighttick\begin{hscode}\SaveRestoreHook
\column{B}{@{}>{\hspre}l<{\hspost}@{}}%
\column{E}{@{}>{\hspre}l<{\hspost}@{}}%
\>[B]{}\textbf{Lemma}{\spacestart}{\nextspace}Bounded\char95 iff\char95 valid{\spacestart}{\nextspace}:{\spacestart}{\nextspace}\textbf{forall}{\spacestart}{\nextspace}s,{\spacestart}{\nextspace}WF{\spacestart}{\nextspace}s{\spacestart}{\nextspace}<->{\spacestart}{\nextspace}valid{\spacestart}{\nextspace}s{\spacestart}{\nextspace}={\spacestart}{\nextspace}true.{}\<[E]%
\ColumnHook
\end{hscode}\resethooks\endgroup

\begin{figure}[t]
\begingroup\straighttick\begin{hscode}\SaveRestoreHook
\column{B}{@{}>{\hspre}l<{\hspost}@{}}%
\column{3}{@{}>{\hspre}c<{\hspost}@{}}%
\column{3E}{@{}l@{}}%
\column{6}{@{}>{\hspre}l<{\hspost}@{}}%
\column{27}{@{}>{\hspre}l<{\hspost}@{}}%
\column{43}{@{}>{\hspre}l<{\hspost}@{}}%
\column{E}{@{}>{\hspre}l<{\hspost}@{}}%
\>[B]{}\textbf{Inductive}{\spacestart}{\nextspace}Bounded{\spacestart}{\nextspace}:{\spacestart}{\nextspace}Set\char95 {\spacestart}{\nextspace}e{\spacestart}{\nextspace}->{\spacestart}{\nextspace}option{\spacestart}{\nextspace}e{\spacestart}{\nextspace}->{\spacestart}{\nextspace}option{\spacestart}{\nextspace}e{\spacestart}{\nextspace}->{\spacestart}{\nextspace}\textbf{Prop}{\spacestart}{\nextspace}:={\spacestart}{\nextspace}{}\<[E]%
\\[-0.3ex]%
\>[B]{}{\spacestart}{\nextspace}{\nextspace}{}\<[3]%
\>[3]{}|{\spacestart}{\nextspace}{\nextspace}{}\<[3E]%
\>[6]{}BoundedTip{\spacestart}{\nextspace}:{\spacestart}{\nextspace}\textbf{forall}{\spacestart}{\nextspace}lb{\spacestart}{\nextspace}ub,{\spacestart}{\nextspace}{}\<[E]%
\\[-0.3ex]%
\>[B]{}{\spacestart}{\nextspace}{\nextspace}{\nextspace}{\nextspace}{\nextspace}{}\<[6]%
\>[6]{}Bounded{\spacestart}{\nextspace}Tip{\spacestart}{\nextspace}lb{\spacestart}{\nextspace}ub{\spacestart}{\nextspace}{}\<[E]%
\\[-0.3ex]%
\>[B]{}{\spacestart}{\nextspace}{\nextspace}{}\<[3]%
\>[3]{}|{\spacestart}{\nextspace}{\nextspace}{}\<[3E]%
\>[6]{}BoundedBin{\spacestart}{\nextspace}:{\spacestart}{\nextspace}\textbf{forall}{\spacestart}{\nextspace}lb{\spacestart}{\nextspace}ub{\spacestart}{\nextspace}s1{\spacestart}{\nextspace}s2{\spacestart}{\nextspace}x{\spacestart}{\nextspace}sz,{\spacestart}{\nextspace}{}\<[E]%
\\[-0.3ex]%
\>[B]{}{\spacestart}{\nextspace}{\nextspace}{\nextspace}{\nextspace}{\nextspace}{}\<[6]%
\>[6]{}Bounded{\spacestart}{\nextspace}s1{\spacestart}{\nextspace}lb{\spacestart}{\nextspace}(Some{\spacestart}{\nextspace}x){\spacestart}{\nextspace}->{\spacestart}{\nextspace}{}\<[E]%
\\[-0.3ex]%
\>[B]{}{\spacestart}{\nextspace}{\nextspace}{\nextspace}{\nextspace}{\nextspace}{}\<[6]%
\>[6]{}Bounded{\spacestart}{\nextspace}s2{\spacestart}{\nextspace}(Some{\spacestart}{\nextspace}x){\spacestart}{\nextspace}ub{\spacestart}{\nextspace}->{\spacestart}{\nextspace}{}\<[E]%
\\[-0.3ex]%
\>[B]{}{\spacestart}{\nextspace}{\nextspace}{\nextspace}{\nextspace}{\nextspace}{}\<[6]%
\>[6]{}isLB{\spacestart}{\nextspace}lb{\spacestart}{\nextspace}x{\spacestart}{\nextspace}={\spacestart}{\nextspace}true{\spacestart}{\nextspace}->{\spacestart}{\nextspace}{\nextspace}{}\<[27]%
\>[27]{}(*{\spacestart}{\nextspace}If{\spacestart}{\nextspace}lb{\spacestart}{\nextspace}is{\spacestart}{\nextspace}defined,{\spacestart}{\nextspace}it{\spacestart}{\nextspace}is{\spacestart}{\nextspace}less{\spacestart}{\nextspace}than{\spacestart}{\nextspace}x{\spacestart}{\nextspace}*){\spacestart}{\nextspace}{}\<[E]%
\\[-0.3ex]%
\>[B]{}{\spacestart}{\nextspace}{\nextspace}{\nextspace}{\nextspace}{\nextspace}{}\<[6]%
\>[6]{}isUB{\spacestart}{\nextspace}ub{\spacestart}{\nextspace}x{\spacestart}{\nextspace}={\spacestart}{\nextspace}true{\spacestart}{\nextspace}->{\spacestart}{\nextspace}{\nextspace}{}\<[27]%
\>[27]{}(*{\spacestart}{\nextspace}If{\spacestart}{\nextspace}ub{\spacestart}{\nextspace}is{\spacestart}{\nextspace}defined,{\spacestart}{\nextspace}it{\spacestart}{\nextspace}is{\spacestart}{\nextspace}greater{\spacestart}{\nextspace}than{\spacestart}{\nextspace}x{\spacestart}{\nextspace}*){\spacestart}{\nextspace}{}\<[E]%
\\[-0.3ex]%
\>[B]{}{\spacestart}{\nextspace}{\nextspace}{\nextspace}{\nextspace}{\nextspace}{}\<[6]%
\>[6]{}sz{\spacestart}{\nextspace}={\spacestart}{\nextspace}(1{\spacestart}{\nextspace}+{\spacestart}{\nextspace}size{\spacestart}{\nextspace}s1{\spacestart}{\nextspace}+{\spacestart}{\nextspace}size{\spacestart}{\nextspace}s2){\spacestart}{\nextspace}->{\spacestart}{\nextspace}{}\<[E]%
\\[-0.3ex]%
\>[B]{}{\spacestart}{\nextspace}{\nextspace}{\nextspace}{\nextspace}{\nextspace}{}\<[6]%
\>[6]{}balance\char95 prop{\spacestart}{\nextspace}(size{\spacestart}{\nextspace}s1){\spacestart}{\nextspace}(size{\spacestart}{\nextspace}s2){\spacestart}{\nextspace}->{\spacestart}{\nextspace}{\nextspace}{}\<[43]%
\>[43]{}(*{\spacestart}{\nextspace}weights{\spacestart}{\nextspace}\textbf{of}{\spacestart}{\nextspace}tree{\spacestart}{\nextspace}are{\spacestart}{\nextspace}balanced{\spacestart}{\nextspace}*){\spacestart}{\nextspace}{}\<[E]%
\\[-0.3ex]%
\>[B]{}{\spacestart}{\nextspace}{\nextspace}{\nextspace}{\nextspace}{\nextspace}{}\<[6]%
\>[6]{}Bounded{\spacestart}{\nextspace}(Bin{\spacestart}{\nextspace}sz{\spacestart}{\nextspace}x{\spacestart}{\nextspace}s1{\spacestart}{\nextspace}s2){\spacestart}{\nextspace}lb{\spacestart}{\nextspace}ub.{\spacestart}{\nextspace}{}\<[E]%
\\[\blanklineskip]%
\>[B]{}(**{\spacestart}{\nextspace}Any{\spacestart}{\nextspace}set{\spacestart}{\nextspace}that{\spacestart}{\nextspace}has{\spacestart}{\nextspace}bounds{\spacestart}{\nextspace}is{\spacestart}{\nextspace}well-formed{\spacestart}{\nextspace}*){\spacestart}{\nextspace}{}\<[E]%
\\[-0.3ex]%
\>[B]{}\textbf{Definition}{\spacestart}{\nextspace}WF{\spacestart}{\nextspace}(s{\spacestart}{\nextspace}:{\spacestart}{\nextspace}Set\char95 {\spacestart}{\nextspace}e){\spacestart}{\nextspace}:{\spacestart}{\nextspace}\textbf{Prop}{\spacestart}{\nextspace}:={\spacestart}{\nextspace}Bounded{\spacestart}{\nextspace}s{\spacestart}{\nextspace}None{\spacestart}{\nextspace}None.{}\<[E]%
\ColumnHook
\end{hscode}\resethooks\endgroup
\caption{Well-formed weight-balanced sets}
\label{fig:set-wf}
\end{figure}

\paragraph{Well-formed Patricia trees}

Our well-formedness definition for \ensuretext{\straighttick\ttfamily{}IntSet} is derived from the comments
in the \ensuretext{\straighttick\ttfamily{}IntSet} data type, shown in \cref{fig:comments}, where the type declaration of
almost disappears beneath a large swath of comments describing
its invariants.

In this case, the documentation-derived well-formedness predicate is stronger
than the corresponding \ensuretext{\straighttick\ttfamily{}valid} function from the implementation -- the Haskell
function was missing some necessary conditions.  We reported this to the library
authors,\containerIssue{522} who
have since fixed \ensuretext{\straighttick\ttfamily{}valid}.

This fix to \ensuretext{\straighttick\ttfamily{}valid} is an example of how verification allows us to
cross-validate specifications and ensure that the invariants written in the
comments are adequately reflected by the code. On the other hand, sometimes we
discover that it is the comments in the code that are incomplete.  For example,
the comment describing the precondition for \ensuretext{\straighttick\ttfamily{}balanceL} in the \ensuretext{\straighttick\ttfamily{}Data.Set} module
was misleading and too vague; for more detail, see \cref{sec:balanceL}.

\subsection{Property-based testing}\label{sec:specs-tests}

At the next level of verification, we would like to show that the
implementations of \ensuretext{\straighttick\ttfamily{}Set} and \ensuretext{\straighttick\ttfamily{}IntSet} are correct according to the
implementors of the module. We specify correctness by
deriving a definition directly from the test suite that is distributed with
the \texttt{containers} library.

Thanks to the popularity of property-based testing within the Haskell community,
this test suite contains a wealth of precisely specified general
properties expressed using QuickCheck~\citep{quickcheck}.
For example, one such property states that the \ensuretext{\straighttick\ttfamily{}union} operation is associative:
\begingroup\straighttick\begin{hscode}\SaveRestoreHook
\column{B}{@{}>{\hspre}l<{\hspost}@{}}%
\column{E}{@{}>{\hspre}l<{\hspost}@{}}%
\>[B]{}prop\char95 UnionAssoc{\spacestart}{\nextspace}::{\spacestart}{\nextspace}IntSet{\spacestart}{\nextspace}->{\spacestart}{\nextspace}IntSet{\spacestart}{\nextspace}->{\spacestart}{\nextspace}IntSet{\spacestart}{\nextspace}->{\spacestart}{\nextspace}Bool{\spacestart}{\nextspace}{}\<[E]%
\\[-0.3ex]%
\>[B]{}prop\char95 UnionAssoc{\spacestart}{\nextspace}t1{\spacestart}{\nextspace}t2{\spacestart}{\nextspace}t3{\spacestart}{\nextspace}={\spacestart}{\nextspace}union{\spacestart}{\nextspace}t1{\spacestart}{\nextspace}(union{\spacestart}{\nextspace}t2{\spacestart}{\nextspace}t3){\spacestart}{\nextspace}=={\spacestart}{\nextspace}union{\spacestart}{\nextspace}(union{\spacestart}{\nextspace}t1{\spacestart}{\nextspace}t2){\spacestart}{\nextspace}t3{}\<[E]%
\ColumnHook
\end{hscode}\resethooks\endgroup
which we can interpret as a theorem about \ensuretext{\straighttick\ttfamily{}union}:%
\footnote{In the file \fileContainers{theories/IntSetPropertyProofs.v}}
\begingroup\straighttick\begin{hscode}\SaveRestoreHook
\column{B}{@{}>{\hspre}l<{\hspost}@{}}%
\column{6}{@{}>{\hspre}l<{\hspost}@{}}%
\column{E}{@{}>{\hspre}l<{\hspost}@{}}%
\>[B]{}\textbf{Theorem}{\spacestart}{\nextspace}thm\char95 UnionAssoc:{\spacestart}{\nextspace}{}\<[E]%
\\[-0.3ex]%
\>[B]{}~~{\spacestart}{\nextspace}{\nextspace}{}\<[6]%
\>[6]{}\textbf{forall}{\spacestart}{\nextspace}t1,{\spacestart}{\nextspace}WF{\spacestart}{\nextspace}t1{\spacestart}{\nextspace}->{\spacestart}{\nextspace}\textbf{forall}{\spacestart}{\nextspace}t2,{\spacestart}{\nextspace}WF{\spacestart}{\nextspace}t2{\spacestart}{\nextspace}->{\spacestart}{\nextspace}\textbf{forall}{\spacestart}{\nextspace}t3,{\spacestart}{\nextspace}WF{\spacestart}{\nextspace}t3{\spacestart}{\nextspace}->{\spacestart}{\nextspace}{}\<[E]%
\\[-0.3ex]%
\>[B]{}{\spacestart}{\nextspace}{\nextspace}{\nextspace}{\nextspace}{\nextspace}{}\<[6]%
\>[6]{}union{\spacestart}{\nextspace}t1{\spacestart}{\nextspace}(union{\spacestart}{\nextspace}t2{\spacestart}{\nextspace}t3){\spacestart}{\nextspace}=={\spacestart}{\nextspace}union{\spacestart}{\nextspace}(union{\spacestart}{\nextspace}t1{\spacestart}{\nextspace}t2){\spacestart}{\nextspace}t3{\spacestart}{\nextspace}={\spacestart}{\nextspace}true.{}\<[E]%
\ColumnHook
\end{hscode}\resethooks\endgroup
We do not have to write these theorems by hand:
as we describe in \cref{sec:translating-quickcheck}, we use \texttt{hs\char45{}to\char45{}coq} in a
nonstandard way to automatically turn these executable tests into Gallina
propositions (i.e.\ types). We have have translated \ensuretext{\straighttick\ttfamily{}IntSet}'s test suite in this manner and have proven that all QuickCheck properties about verified \ensuretext{\straighttick\ttfamily{}IntSet} functions are theorems
(with one exception due to our choice of integer representation -- see
\cref{sec:number-types}).

\subsection{Numeric overflow in \ensuretext{\straighttick\ttfamily{}Set}}\label{sec:specs-int}

There is one way in which we have diverged from the specification of
correctness given by the comments of the \texttt{containers} library.  The \ensuretext{\straighttick\ttfamily{}Data.Set}
module states:%
\footnote{\url{http://hackage.haskell.org/package/containers-0.5.11.0/docs/Data-Set.html}}
\begin{quote}
  Warning: The size of the set must not exceed \ensuretext{\straighttick\ttfamily{}maxBound::Int}. Violation of
  this condition is not detected and if the size limit is exceeded, its
  behavior is undefined.
\end{quote}
In practice, it makes no difference whether \ensuretext{\straighttick\ttfamily{}Int} is bounded or not, as a set
with $(2^{63}-1)$ elements would require at least 368 exabytes of storage.
%
What does this imply for our specification of \ensuretext{\straighttick\ttfamily{}Set}? Should we use fixed-width
integers to represent \ensuretext{\straighttick\ttfamily{}Int}? This choice would closely match the implementation, but
we would have to carefully add preconditions to all our lemmas to avoid
integer overflow, greatly complicating the proofs, with little verification
insight to be gained. Furthermore, such a specification would be difficult to
use by clients, who themselves would need to prove that they satisfy such
preconditions.

Instead, we translate Haskell's \ensuretext{\straighttick\ttfamily{}Int} type to Coq's type of unbounded integers
(called \ensuretext{\straighttick\ttfamily{}Z}). This mapping avoids the problem of integer overflow altogether
and is arguably consistent with the comment, as this choice replaces undefined
behavior with concrete behavior.  (The situation is slightly different for
\ensuretext{\straighttick\ttfamily{}IntSet}; see \cref{sec:number-types}.)

\subsection{Rewrite rules}\label{sec:specs-rewrite-rules}

So far, we have only been concerned with specifying the correctness of the
data structures using the definition of correctness that is present in the
original source code; the comments, the \ensuretext{\straighttick\ttfamily{}valid} functions and the QuickCheck
properties.  However, to be able to claim that our specifications are
two-sided, we need to show that the properties that we prove are
useful to clients of the module.

One source of such properties is \emph{rewrite rules}~\citep{rewriterules}.
The \texttt{containers} library includes a small number of such rules.  These
annotations instruct the compiler
to replace any occurrence of the pattern on the left-hand side in the rule by
the expression on the right hand side. The standard example is
\begingroup\straighttick\begin{hscode}\SaveRestoreHook
\column{B}{@{}>{\hspre}l<{\hspost}@{}}%
\column{E}{@{}>{\hspre}l<{\hspost}@{}}%
\>[B]{}\{-\# RULES \char34\textit{\!map/map}{\char34}{\spacestart}{\nextspace}\textbf{forall}{\spacestart}{\nextspace}f{\spacestart}{\nextspace}g{\spacestart}{\nextspace}xs.{\spacestart}{\nextspace}map{\spacestart}{\nextspace}f{\spacestart}{\nextspace}(map{\spacestart}{\nextspace}g{\spacestart}{\nextspace}xs){\spacestart}{\nextspace}={\spacestart}{\nextspace}map{\spacestart}{\nextspace}(f{\spacestart}{\nextspace}.{\spacestart}{\nextspace}g){\spacestart}{\nextspace}xs \#-\}{}\<[E]%
\ColumnHook
\end{hscode}\resethooks\endgroup
which fuses two adjacent calls to \ensuretext{\straighttick\ttfamily{}map} into one, eliminating the intermediate
list. 

The program transformation \emph{list fusion} \citep{rewriterules} is implemented
completely in terms of rewrite rules, and the rules in the \texttt{containers} library setup its functions for fusion; an example is
\begingroup\straighttick\begin{hscode}\SaveRestoreHook
\column{B}{@{}>{\hspre}l<{\hspost}@{}}%
\column{E}{@{}>{\hspre}l<{\hspost}@{}}%
\>[B]{}\{-\# RULES \char34\textit{\!Set.toAscList}{\char34}{\spacestart}{\nextspace}\textbf{forall}{\spacestart}{\nextspace}s{\spacestart}{\nextspace}.{\spacestart}{\nextspace}toAscList{\spacestart}{\nextspace}s{\spacestart}{\nextspace}={\spacestart}{\nextspace}build{\spacestart}{\nextspace}(\char92 c{\spacestart}{\nextspace}n{\spacestart}{\nextspace}->{\spacestart}{\nextspace}foldrFB{\spacestart}{\nextspace}c{\spacestart}{\nextspace}n{\spacestart}{\nextspace}s) \#-\}{}\<[E]%
\ColumnHook
\end{hscode}\resethooks\endgroup
which transforms the \ensuretext{\straighttick\ttfamily{}toAscList} function into an equivalent representation in
terms of \ensuretext{\straighttick\ttfamily{}build}.

We can view these rewrite rules as a direct specification of properties that
the compiler assumes are true during compilation. Rewrite rules are used by GHC
during optimization; if any of these properties are
actually false, GHC will silently produce incorrect code.
Therefore, any proof about the correctness of GHC's compilation of these files
depends on a proof of these properties. We have manually translated all the
rules into Coq -- there are only few of them, so manual translation is viable
-- and have proved that the translated operations satisfy this
specification.%
\footnote{In the files \fileContainers{theories/SetProofs.v} and \fileContainers{theories/IntSetProofs.v}}

\subsection{Type classes with laws}\label{sec:specs-type-classes}

\begin{figure}
\abovedisplayskip=0pt
\belowdisplayskip=0pt
\raggedright
\begingroup\straighttick\begin{hscode}\SaveRestoreHook
\column{B}{@{}>{\hspre}l<{\hspost}@{}}%
\column{6}{@{}>{\hspre}l<{\hspost}@{}}%
\column{22}{@{}>{\hspre}l<{\hspost}@{}}%
\column{39}{@{}>{\hspre}l<{\hspost}@{}}%
\column{43}{@{}>{\hspre}c<{\hspost}@{}}%
\column{43E}{@{}l@{}}%
\column{47}{@{}>{\hspre}l<{\hspost}@{}}%
\column{57}{@{}>{\hspre}l<{\hspost}@{}}%
\column{E}{@{}>{\hspre}l<{\hspost}@{}}%
\>[B]{}\textbf{Class}{\spacestart}{\nextspace}OrdLaws{\spacestart}{\nextspace}(t{\spacestart}{\nextspace}:{\spacestart}{\nextspace}\textbf{Type}){\spacestart}{\nextspace}{\char123}HEq{\spacestart}{\nextspace}:{\spacestart}{\nextspace}Eq\char95 {\spacestart}{\nextspace}t{\char125}{\spacestart}{\nextspace}{\char123}HOrd{\spacestart}{\nextspace}:{\spacestart}{\nextspace}Ord{\spacestart}{\nextspace}t{\char125}{\spacestart}{\nextspace}{\char123}HEqLaw{\spacestart}{\nextspace}:{\spacestart}{\nextspace}EqLaws{\spacestart}{\nextspace}t{\char125}{\spacestart}{\nextspace}:={\spacestart}{\nextspace}{\char123}{\spacestart}{\nextspace}{}\<[E]%
\\[-0.3ex]%
\>[B]{}~~{\spacestart}{\nextspace}{\nextspace}{}\<[6]%
\>[6]{}(*{\spacestart}{\nextspace}The{\spacestart}{\nextspace}axioms{\spacestart}{\nextspace}*){\spacestart}{\nextspace}{}\<[E]%
\\[-0.3ex]%
\>[B]{}{\spacestart}{\nextspace}{\nextspace}{\nextspace}{\nextspace}{\nextspace}{}\<[6]%
\>[6]{}Ord\char95 antisym{\spacestart}{\nextspace}{\nextspace}{\nextspace}{\nextspace}{\nextspace}{}\<[22]%
\>[22]{}:{\spacestart}{\nextspace}\textbf{forall}{\spacestart}{\nextspace}a{\spacestart}{\nextspace}b,{\spacestart}{\nextspace}{\nextspace}{\nextspace}{\nextspace}{}\<[39]%
\>[39]{}a{\spacestart}{\nextspace}<={\spacestart}{\nextspace}b{\spacestart}{\nextspace}={\spacestart}{\nextspace}true{\spacestart}{\nextspace}->{\spacestart}{\nextspace}b{\spacestart}{\nextspace}<={\spacestart}{\nextspace}a{\spacestart}{\nextspace}={\spacestart}{\nextspace}true{\spacestart}{\nextspace}->{\spacestart}{\nextspace}a{\spacestart}{\nextspace}=={\spacestart}{\nextspace}b{\spacestart}{\nextspace}={\spacestart}{\nextspace}true;{\spacestart}{\nextspace}{}\<[E]%
\\[-0.3ex]%
\>[B]{}{\spacestart}{\nextspace}{\nextspace}{\nextspace}{\nextspace}{\nextspace}{}\<[6]%
\>[6]{}Ord\char95 trans\char95 le{\spacestart}{\nextspace}{\nextspace}{\nextspace}{\nextspace}{}\<[22]%
\>[22]{}:{\spacestart}{\nextspace}\textbf{forall}{\spacestart}{\nextspace}a{\spacestart}{\nextspace}b{\spacestart}{\nextspace}c,{\spacestart}{\nextspace}{\nextspace}{}\<[39]%
\>[39]{}a{\spacestart}{\nextspace}<={\spacestart}{\nextspace}b{\spacestart}{\nextspace}={\spacestart}{\nextspace}true{\spacestart}{\nextspace}->{\spacestart}{\nextspace}b{\spacestart}{\nextspace}<={\spacestart}{\nextspace}c{\spacestart}{\nextspace}={\spacestart}{\nextspace}true{\spacestart}{\nextspace}->{\spacestart}{\nextspace}a{\spacestart}{\nextspace}<={\spacestart}{\nextspace}c{\spacestart}{\nextspace}={\spacestart}{\nextspace}true;{\spacestart}{\nextspace}{}\<[E]%
\\[-0.3ex]%
\>[B]{}{\spacestart}{\nextspace}{\nextspace}{\nextspace}{\nextspace}{\nextspace}{}\<[6]%
\>[6]{}Ord\char95 total{\spacestart}{\nextspace}{\nextspace}{\nextspace}{\nextspace}{\nextspace}{\nextspace}{\nextspace}{}\<[22]%
\>[22]{}:{\spacestart}{\nextspace}\textbf{forall}{\spacestart}{\nextspace}a{\spacestart}{\nextspace}b,{\spacestart}{\nextspace}{\nextspace}{\nextspace}{\nextspace}{}\<[39]%
\>[39]{}a{\spacestart}{\nextspace}<={\spacestart}{\nextspace}b{\spacestart}{\nextspace}={\spacestart}{\nextspace}true{\spacestart}{\nextspace}\char92 /{\spacestart}{\nextspace}b{\spacestart}{\nextspace}<={\spacestart}{\nextspace}a{\spacestart}{\nextspace}={\spacestart}{\nextspace}true;{\spacestart}{\nextspace}{}\<[E]%
\\[-0.3ex]%
\>[B]{}{\spacestart}{\nextspace}{\nextspace}{\nextspace}{\nextspace}{\nextspace}{}\<[6]%
\>[6]{}(*{\spacestart}{\nextspace}The{\spacestart}{\nextspace}other{\spacestart}{\nextspace}operations,{\spacestart}{\nextspace}\textbf{in}{\spacestart}{\nextspace}terms{\spacestart}{\nextspace}\textbf{of}{\spacestart}{\nextspace}<={\spacestart}{\nextspace}or{\spacestart}{\nextspace}=={\spacestart}{\nextspace}*){\spacestart}{\nextspace}{}\<[E]%
\\[-0.3ex]%
\>[B]{}{\spacestart}{\nextspace}{\nextspace}{\nextspace}{\nextspace}{\nextspace}{}\<[6]%
\>[6]{}Ord\char95 compare\char95 Lt{\spacestart}{\nextspace}{\nextspace}{}\<[22]%
\>[22]{}:{\spacestart}{\nextspace}\textbf{forall}{\spacestart}{\nextspace}a{\spacestart}{\nextspace}b,{\spacestart}{\nextspace}{\nextspace}{\nextspace}{\nextspace}{}\<[39]%
\>[39]{}compare{\spacestart}{\nextspace}a{\spacestart}{\nextspace}b{\spacestart}{\nextspace}={\spacestart}{\nextspace}Lt{\spacestart}{\nextspace}<->{\spacestart}{\nextspace}b{\spacestart}{\nextspace}<={\spacestart}{\nextspace}a{\spacestart}{\nextspace}={\spacestart}{\nextspace}false;{\spacestart}{\nextspace}{}\<[E]%
\\[-0.3ex]%
\>[B]{}{\spacestart}{\nextspace}{\nextspace}{\nextspace}{\nextspace}{\nextspace}{}\<[6]%
\>[6]{}Ord\char95 compare\char95 Eq{\spacestart}{\nextspace}{\nextspace}{}\<[22]%
\>[22]{}:{\spacestart}{\nextspace}\textbf{forall}{\spacestart}{\nextspace}a{\spacestart}{\nextspace}b,{\spacestart}{\nextspace}{\nextspace}{\nextspace}{\nextspace}{}\<[39]%
\>[39]{}compare{\spacestart}{\nextspace}a{\spacestart}{\nextspace}b{\spacestart}{\nextspace}={\spacestart}{\nextspace}Eq{\spacestart}{\nextspace}<->{\spacestart}{\nextspace}a{\spacestart}{\nextspace}=={\spacestart}{\nextspace}b{\spacestart}{\nextspace}={\spacestart}{\nextspace}true;{\spacestart}{\nextspace}{}\<[E]%
\\[-0.3ex]%
\>[B]{}{\spacestart}{\nextspace}{\nextspace}{\nextspace}{\nextspace}{\nextspace}{}\<[6]%
\>[6]{}Ord\char95 compare\char95 Gt{\spacestart}{\nextspace}{\nextspace}{}\<[22]%
\>[22]{}:{\spacestart}{\nextspace}\textbf{forall}{\spacestart}{\nextspace}a{\spacestart}{\nextspace}b,{\spacestart}{\nextspace}{\nextspace}{\nextspace}{\nextspace}{}\<[39]%
\>[39]{}compare{\spacestart}{\nextspace}a{\spacestart}{\nextspace}b{\spacestart}{\nextspace}={\spacestart}{\nextspace}Gt{\spacestart}{\nextspace}<->{\spacestart}{\nextspace}a{\spacestart}{\nextspace}<={\spacestart}{\nextspace}b{\spacestart}{\nextspace}={\spacestart}{\nextspace}false;{\spacestart}{\nextspace}{}\<[E]%
\\[-0.3ex]%
\>[B]{}{\spacestart}{\nextspace}{\nextspace}{\nextspace}{\nextspace}{\nextspace}{}\<[6]%
\>[6]{}Ord\char95 lt\char95 le{\spacestart}{\nextspace}{\nextspace}{\nextspace}{\nextspace}{\nextspace}{\nextspace}{\nextspace}{}\<[22]%
\>[22]{}:{\spacestart}{\nextspace}\textbf{forall}{\spacestart}{\nextspace}a{\spacestart}{\nextspace}b,{\spacestart}{\nextspace}{\nextspace}{\nextspace}{\nextspace}{}\<[39]%
\>[39]{}a{\spacestart}{\nextspace}{\nextspace}{\nextspace}{}\<[43]%
\>[43]{}<{\spacestart}{\nextspace}{\nextspace}{\nextspace}{}\<[43E]%
\>[47]{}b{\spacestart}{\nextspace}={\spacestart}{\nextspace}negb{\spacestart}{\nextspace}{\nextspace}{}\<[57]%
\>[57]{}(b{\spacestart}{\nextspace}<={\spacestart}{\nextspace}a);{\spacestart}{\nextspace}{}\<[E]%
\\[-0.3ex]%
\>[B]{}{\spacestart}{\nextspace}{\nextspace}{\nextspace}{\nextspace}{\nextspace}{}\<[6]%
\>[6]{}Ord\char95 ge\char95 le{\spacestart}{\nextspace}{\nextspace}{\nextspace}{\nextspace}{\nextspace}{\nextspace}{\nextspace}{}\<[22]%
\>[22]{}:{\spacestart}{\nextspace}\textbf{forall}{\spacestart}{\nextspace}a{\spacestart}{\nextspace}b,{\spacestart}{\nextspace}{\nextspace}{\nextspace}{\nextspace}{}\<[39]%
\>[39]{}a{\spacestart}{\nextspace}{\nextspace}{\nextspace}{}\<[43]%
\>[43]{}>={\spacestart}{\nextspace}{\nextspace}{}\<[43E]%
\>[47]{}b{\spacestart}{\nextspace}={\spacestart}{\nextspace}{\nextspace}{\nextspace}{\nextspace}{\nextspace}{\nextspace}{\nextspace}{}\<[57]%
\>[57]{}(b{\spacestart}{\nextspace}<={\spacestart}{\nextspace}a);{\spacestart}{\nextspace}{}\<[E]%
\\[-0.3ex]%
\>[B]{}{\spacestart}{\nextspace}{\nextspace}{\nextspace}{\nextspace}{\nextspace}{}\<[6]%
\>[6]{}Ord\char95 gt\char95 le{\spacestart}{\nextspace}{\nextspace}{\nextspace}{\nextspace}{\nextspace}{\nextspace}{\nextspace}{}\<[22]%
\>[22]{}:{\spacestart}{\nextspace}\textbf{forall}{\spacestart}{\nextspace}a{\spacestart}{\nextspace}b,{\spacestart}{\nextspace}{\nextspace}{\nextspace}{\nextspace}{}\<[39]%
\>[39]{}a{\spacestart}{\nextspace}{\nextspace}{\nextspace}{}\<[43]%
\>[43]{}>{\spacestart}{\nextspace}{\nextspace}{\nextspace}{}\<[43E]%
\>[47]{}b{\spacestart}{\nextspace}={\spacestart}{\nextspace}negb{\spacestart}{\nextspace}{\nextspace}{}\<[57]%
\>[57]{}(a{\spacestart}{\nextspace}<={\spacestart}{\nextspace}b){\char125}.{}\<[E]%
\ColumnHook
\end{hscode}\resethooks\endgroup
\caption{Our codification of the \ensuretext{\straighttick\ttfamily{}Ord} type class laws}
\label{fig:OrdLaws}
\end{figure}

Many Haskell type classes come with \emph{laws} that all
instances of the type class should satisfy, which provides another source
of external specification that we can use.
For example, an instance of
\ensuretext{\straighttick\ttfamily{}Eq} is expected to implement an equivalence relation, an instance of
\ensuretext{\straighttick\ttfamily{}Ord} should describe a linear order, and an instance of \ensuretext{\straighttick\ttfamily{}Monoid} should be, well, a monoid.

We reflect these laws using type classes whose members are the required
properties. For example, we have defined \ensuretext{\straighttick\ttfamily{}EqLaws}, \ensuretext{\straighttick\ttfamily{}OrdLaws} (shown
in \cref{fig:OrdLaws}), and \ensuretext{\straighttick\ttfamily{}MonoidLaws}. These classes can only be
instantiated if the corresponding instance is lawful.

Even though we have defined these laws ourselves, using our understanding of
what they mean for the Haskell standard library, we argue that they form a
two-sided specification. In particular, we have been clients to the \ensuretext{\straighttick\ttfamily{}Ord} laws
in our verification of the \ensuretext{\straighttick\ttfamily{}Set} data structure. Almost all theorems about \ensuretext{\straighttick\ttfamily{}Set}
must constrain the element type to one that is an instance of
\ensuretext{\straighttick\ttfamily{}OrdLaws}. Therefore, we know our specification of these laws is sufficiently
strong to verify this library.

At the same time, we have also shown that multiple type class instances
satisfy these laws including both set modules%
\footnote{In the files \fileContainers{theories/SetProofs.v} and \fileContainers{theories/IntSetProofs.v}}
and other types such as \ensuretext{\straighttick\ttfamily{}Z},
\ensuretext{\straighttick\ttfamily{}unit}, tuples, \ensuretext{\straighttick\ttfamily{}option}, and \ensuretext{\straighttick\ttfamily{}list}.%
\footnote{In the file \fileHsToCoq{base-thy/GHC/Base.v}}
Because we successfully instantiated these type classes for many types, we
also know that they are not too strong.

\label{sec:specs-type-classes-WF}

That said, nailing down the precise form of the type class laws can be
tricky. Consider the case of a \ensuretext{\straighttick\ttfamily{}Monoid} instance for a type \ensuretext{\straighttick\ttfamily{}T}. The
associativity law can be stated as “for all elements \ensuretext{\straighttick\ttfamily{}x}, \ensuretext{\straighttick\ttfamily{}y} and \ensuretext{\straighttick\ttfamily{}z} of \ensuretext{\straighttick\ttfamily{}T},
we have that %
\ensuretext{\straighttick\ttfamily{}x{\spacestart}{\nextspace}<>{\spacestart}{\nextspace}(y{\spacestart}{\nextspace}<>{\spacestart}{\nextspace}z)} is equal to \ensuretext{\straighttick\ttfamily{}(x{\spacestart}{\nextspace}<>{\spacestart}{\nextspace}y){\spacestart}{\nextspace}<>{\spacestart}{\nextspace}z}.”  But in order to write this down
as part of \ensuretext{\straighttick\ttfamily{}MonoidLaws} in Coq, we need to make two decisions:
\begin{enumerate}
\item What do we mean by “equal”? The first option is to use Coq’s propositional
  equality and require that \ensuretext{\straighttick\ttfamily{}x{\spacestart}{\nextspace}<>{\spacestart}{\nextspace}(y{\spacestart}{\nextspace}<>{\spacestart}{\nextspace}z){\spacestart}{\nextspace}={\spacestart}{\nextspace}(x{\spacestart}{\nextspace}<>{\spacestart}{\nextspace}y){\spacestart}{\nextspace}<>{\spacestart}{\nextspace}z}. This would,
  however, prevent us from making \ensuretext{\straighttick\ttfamily{}Set\char95 }, with \ensuretext{\straighttick\ttfamily{}(<>){\spacestart}{\nextspace}={\spacestart}{\nextspace}union}, a member of
  \ensuretext{\straighttick\ttfamily{}MonoidLaws}: two extensionally-equal sets may be represented by differently
  structured trees. Therefore, we instead require that the two expressions are
  equal according to their \ensuretext{\straighttick\ttfamily{}Eq} instance:%
  \ensuretext{\straighttick\ttfamily{}(x{\spacestart}{\nextspace}<>{\spacestart}{\nextspace}(y{\spacestart}{\nextspace}<>{\spacestart}{\nextspace}z){\spacestart}{\nextspace}=={\spacestart}{\nextspace}(x{\spacestart}{\nextspace}<>{\spacestart}{\nextspace}y){\spacestart}{\nextspace}<>{\spacestart}{\nextspace}z){\spacestart}{\nextspace}={\spacestart}{\nextspace}true}.
  The tradeoff with this approach is that it precludes instances like
  \ensuretext{\straighttick\ttfamily{}MonoidLaws{\spacestart}{\nextspace}b{\spacestart}{\nextspace}->{\spacestart}{\nextspace}MonoidLaws{\spacestart}{\nextspace}(a{\spacestart}{\nextspace}->{\spacestart}{\nextspace}b)}, since functions have no instance of
  \ensuretext{\straighttick\ttfamily{}Eq} (and indeed, cannot have decidable equality).

  For many types, however, this distinction is moot, since Haskell's
  equality coincides with structural equality; for example, this is the case for
  \ensuretext{\straighttick\ttfamily{}Bool}, for \ensuretext{\straighttick\ttfamily{}Integer}, and for \ensuretext{\straighttick\ttfamily{}IntSet} itself (although not for \ensuretext{\straighttick\ttfamily{}Set}, as
  mentioned above).  To facilitate reasoning about such types, we provide the
  type class \ensuretext{\straighttick\ttfamily{}EqExact}, which states that%
\begingroup\straighttick\begin{hscode}\SaveRestoreHook
\column{B}{@{}>{\hspre}l<{\hspost}@{}}%
\column{E}{@{}>{\hspre}l<{\hspost}@{}}%
\>[B]{}\textbf{forall}{\spacestart}{\nextspace}{\char123}a{\char125}{\spacestart}{\nextspace}{\textasciigrave}{\char123}EqExact{\spacestart}{\nextspace}a{\char125},{\spacestart}{\nextspace}x{\spacestart}{\nextspace}=={\spacestart}{\nextspace}y{\spacestart}{\nextspace}={\spacestart}{\nextspace}true{\spacestart}{\nextspace}<->{\spacestart}{\nextspace}x{\spacestart}{\nextspace}={\spacestart}{\nextspace}y{}\<[E]%
\ColumnHook
\end{hscode}\resethooks\endgroup

\item What do we mean by “For all elements of \ensuretext{\straighttick\ttfamily{}T}”? The obvious choice is
  universal quantification over all elements of \ensuretext{\straighttick\ttfamily{}T}:
\begingroup\straighttick\begin{hscode}\SaveRestoreHook
\column{B}{@{}>{\hspre}l<{\hspost}@{}}%
\column{E}{@{}>{\hspre}l<{\hspost}@{}}%
\>[B]{}\textbf{forall}{\spacestart}{\nextspace}(x{\spacestart}{\nextspace}y{\spacestart}{\nextspace}z{\spacestart}{\nextspace}:{\spacestart}{\nextspace}T),{\spacestart}{\nextspace}x{\spacestart}{\nextspace}<>{\spacestart}{\nextspace}(y{\spacestart}{\nextspace}<>{\spacestart}{\nextspace}z){\spacestart}{\nextspace}=={\spacestart}{\nextspace}(x{\spacestart}{\nextspace}<>{\spacestart}{\nextspace}y){\spacestart}{\nextspace}<>{\spacestart}{\nextspace}z{\spacestart}{\nextspace}={\spacestart}{\nextspace}true{}\<[E]%
\ColumnHook
\end{hscode}\resethooks\endgroup
But again, this collides with common practice in Haskell. Once again, consider
\ensuretext{\straighttick\ttfamily{}Set}: \ensuretext{\straighttick\ttfamily{}union} only works correctly on
well-formed sets. 
Therefore, our approach is to define an instance of this and other classes not
for the type \ensuretext{\straighttick\ttfamily{}Set\char95 {\spacestart}{\nextspace}e}, but for the type of well-formed sets, \ensuretext{\straighttick\ttfamily{}{\char123}s{\spacestart}{\nextspace}:{\spacestart}{\nextspace}Set\char95 {\spacestart}{\nextspace}e{\spacestart}{\nextspace}|{\spacestart}{\nextspace}WF{\spacestart}{\nextspace}{\nextspace}{\nextspace}s{\char125}}, where type class laws hold universally.  This instance reflects the
``external view'' of the data structure -- clients should only have access to
well-formed sets.

An alternative could be to instead constrain \ensuretext{\straighttick\ttfamily{}MonoidLaws}'s
theorems to hold only on members of \ensuretext{\straighttick\ttfamily{}T} that are well-formed in some general way
(e.g., according to an \ensuretext{\straighttick\ttfamily{}IsWF} type class that could be instantiated at different
types). In this way, we could instantiate \ensuretext{\straighttick\ttfamily{}MonoidLaws} directly with types that
require well-formedness, without the need for subset types.
\end{enumerate}


\subsection{Specifications from the Coq standard library}\label{sec:specs-coq}

Because we are working in Coq, we have access to a standard library of
specifications for finite sets, which we know are two-sided because they have
already appeared in larger Coq developments.
The \ensuretext{\straighttick\ttfamily{}Coq.FSets.FSetInterface}
module\footnote{\url{https://coq.inria.fr/library/Coq.FSets.FSetInterface.html}}
provides module types that cover many common operations and their properties.
The module types come in two varieties: one that specifies sets of elements that
merely have decidable equality (\ensuretext{\straighttick\ttfamily{}WSfun}, \ensuretext{\straighttick\ttfamily{}WS}), and one that specifies sets of
elements that can be linearly ordered (\ensuretext{\straighttick\ttfamily{}Sfun}, \ensuretext{\straighttick\ttfamily{}S}).  The \ensuretext{\straighttick\ttfamily{}WSfun} and \ensuretext{\straighttick\ttfamily{}Sfun}
modules are presented as module functors that take an \ensuretext{\straighttick\ttfamily{}OrderedType} module,
containing the linearly-ordered element type, as an input; the \ensuretext{\straighttick\ttfamily{}WS} and \ensuretext{\straighttick\ttfamily{}S}
modules are the same, but they inline this information.

For example, the parameterized module type \ensuretext{\straighttick\ttfamily{}WSfun} provides one specification of
a finite set type, called \ensuretext{\straighttick\ttfamily{}t} in the excerpt from this interface below.  The
element type of this set, \ensuretext{\straighttick\ttfamily{}E.t}, is required to have decidable
equality.

\begingroup\straighttick\begin{hscode}\SaveRestoreHook
\column{B}{@{}>{\hspre}l<{\hspost}@{}}%
\column{6}{@{}>{\hspre}l<{\hspost}@{}}%
\column{41}{@{}>{\hspre}l<{\hspost}@{}}%
\column{E}{@{}>{\hspre}l<{\hspost}@{}}%
\>[B]{}\textbf{Module}{\spacestart}{\nextspace}\textbf{Type}{\spacestart}{\nextspace}WSfun{\spacestart}{\nextspace}(E{\spacestart}{\nextspace}:{\spacestart}{\nextspace}DecidableType).{\spacestart}{\nextspace}{}\<[E]%
\\[\blanklineskip]%
\>[B]{}~~{\spacestart}{\nextspace}{\nextspace}{}\<[6]%
\>[6]{}\textbf{Definition}{\spacestart}{\nextspace}elt{\spacestart}{\nextspace}:={\spacestart}{\nextspace}E.t.{\spacestart}{\nextspace}{}\<[E]%
\\[\blanklineskip]%
\>[B]{}{\spacestart}{\nextspace}{\nextspace}{\nextspace}{\nextspace}{\nextspace}{}\<[6]%
\>[6]{}\textbf{Parameter}{\spacestart}{\nextspace}t{\spacestart}{\nextspace}:{\spacestart}{\nextspace}\textbf{Type}.{\spacestart}{\nextspace}{\nextspace}{\nextspace}{\nextspace}{\nextspace}{\nextspace}{\nextspace}{\nextspace}{\nextspace}{\nextspace}{\nextspace}{\nextspace}{\nextspace}{\nextspace}{\nextspace}{\nextspace}{}\<[41]%
\>[41]{}(*{\spacestart}{\nextspace}Set{\spacestart}{\nextspace}type{\spacestart}{\nextspace}*){\spacestart}{\nextspace}{}\<[E]%
\\[-0.3ex]%
\>[B]{}{\spacestart}{\nextspace}{\nextspace}{\nextspace}{\nextspace}{\nextspace}{}\<[6]%
\>[6]{}\textbf{Parameter}{\spacestart}{\nextspace}In{\spacestart}{\nextspace}:{\spacestart}{\nextspace}elt{\spacestart}{\nextspace}->{\spacestart}{\nextspace}t{\spacestart}{\nextspace}->{\spacestart}{\nextspace}\textbf{Prop}.{\spacestart}{\nextspace}{\nextspace}{\nextspace}{}\<[41]%
\>[41]{}(*{\spacestart}{\nextspace}Characteristic{\spacestart}{\nextspace}function{\spacestart}{\nextspace}for{\spacestart}{\nextspace}the{\spacestart}{\nextspace}Set{\spacestart}{\nextspace}*){\spacestart}{\nextspace}{}\<[E]%
\\[\blanklineskip]%
\>[B]{}{\spacestart}{\nextspace}{\nextspace}{\nextspace}{\nextspace}{\nextspace}{}\<[6]%
\>[6]{}\textbf{Parameter}{\spacestart}{\nextspace}mem{\spacestart}{\nextspace}:{\spacestart}{\nextspace}elt{\spacestart}{\nextspace}->{\spacestart}{\nextspace}t{\spacestart}{\nextspace}->{\spacestart}{\nextspace}bool.{\spacestart}{\nextspace}{\nextspace}{}\<[41]%
\>[41]{}(*{\spacestart}{\nextspace}Membership{\spacestart}{\nextspace}function{\spacestart}{\nextspace}*){\spacestart}{\nextspace}{}\<[E]%
\\[\blanklineskip]%
\>[B]{}{\spacestart}{\nextspace}{\nextspace}{\nextspace}{\nextspace}{\nextspace}{}\<[6]%
\>[6]{}(*{\spacestart}{\nextspace}Specification{\spacestart}{\nextspace}of{\spacestart}{\nextspace}mem{\spacestart}{\nextspace}*){\spacestart}{\nextspace}{}\<[E]%
\\[-0.3ex]%
\>[B]{}{\spacestart}{\nextspace}{\nextspace}{\nextspace}{\nextspace}{\nextspace}{}\<[6]%
\>[6]{}\textbf{Parameter}{\spacestart}{\nextspace}mem\char95 1{\spacestart}{\nextspace}:{\spacestart}{\nextspace}\textbf{forall}{\spacestart}{\nextspace}x{\spacestart}{\nextspace}s,{\spacestart}{\nextspace}In{\spacestart}{\nextspace}x{\spacestart}{\nextspace}s{\spacestart}{\nextspace}->{\spacestart}{\nextspace}mem{\spacestart}{\nextspace}x{\spacestart}{\nextspace}s{\spacestart}{\nextspace}={\spacestart}{\nextspace}true.{\spacestart}{\nextspace}{}\<[E]%
\\[-0.3ex]%
\>[B]{}{\spacestart}{\nextspace}{\nextspace}{\nextspace}{\nextspace}{\nextspace}{}\<[6]%
\>[6]{}\textbf{Parameter}{\spacestart}{\nextspace}mem\char95 2{\spacestart}{\nextspace}:{\spacestart}{\nextspace}\textbf{forall}{\spacestart}{\nextspace}x{\spacestart}{\nextspace}s,{\spacestart}{\nextspace}mem{\spacestart}{\nextspace}x{\spacestart}{\nextspace}s{\spacestart}{\nextspace}={\spacestart}{\nextspace}true{\spacestart}{\nextspace}->{\spacestart}{\nextspace}In{\spacestart}{\nextspace}x{\spacestart}{\nextspace}s.{\spacestart}{\nextspace}{}\<[E]%
\\[\blanklineskip]%
\>[B]{}{\spacestart}{\nextspace}{\nextspace}{\nextspace}{\nextspace}{\nextspace}{}\<[6]%
\>[6]{}\textrm{\ldots}{\spacestart}{\nextspace}{}\<[E]%
\\[-0.3ex]%
\>[B]{}\textbf{End}{\spacestart}{\nextspace}WSFun.{}\<[E]%
\ColumnHook
\end{hscode}\resethooks\endgroup
Every operation in this interface, such as \ensuretext{\straighttick\ttfamily{}mem} above, is accompanied
by a small number of properties that specify the behavior of the
operation.

We instantiate all four interfaces for \ensuretext{\straighttick\ttfamily{}Set}
and \ensuretext{\straighttick\ttfamily{}IntSet}.\footnote{In the files \fileContainers{theories/SetProofs.v} and \fileContainers{theories/IntSetProofs.v}}
For example, the instance for \ensuretext{\straighttick\ttfamily{}Set} starts out:
\begingroup\straighttick\begin{hscode}\SaveRestoreHook
\column{B}{@{}>{\hspre}l<{\hspost}@{}}%
\column{6}{@{}>{\hspre}l<{\hspost}@{}}%
\column{E}{@{}>{\hspre}l<{\hspost}@{}}%
\>[B]{}\textbf{Module}{\spacestart}{\nextspace}SetFSet{\spacestart}{\nextspace}(E{\spacestart}{\nextspace}:{\spacestart}{\nextspace}OrderedType){\spacestart}{\nextspace}<:{\spacestart}{\nextspace}WSfun(E){\spacestart}{\nextspace}<:{\spacestart}{\nextspace}WS{\spacestart}{\nextspace}<:{\spacestart}{\nextspace}Sfun(E){\spacestart}{\nextspace}<:{\spacestart}{\nextspace}S.{\spacestart}{\nextspace}{}\<[E]%
\\[\blanklineskip]%
\>[B]{}~~{\spacestart}{\nextspace}{\nextspace}{}\<[6]%
\>[6]{}\textbf{Definition}{\spacestart}{\nextspace}t{\spacestart}{\nextspace}:={\spacestart}{\nextspace}{\char123}s{\spacestart}{\nextspace}:{\spacestart}{\nextspace}Set\char95 {\spacestart}{\nextspace}elt{\spacestart}{\nextspace}|{\spacestart}{\nextspace}WF{\spacestart}{\nextspace}s{\char125}.{\spacestart}{\nextspace}{}\<[E]%
\\[\blanklineskip]%
\>[B]{}{\spacestart}{\nextspace}{\nextspace}{\nextspace}{\nextspace}{\nextspace}{}\<[6]%
\>[6]{}\textbf{Program}{\spacestart}{\nextspace}\textbf{Definition}{\spacestart}{\nextspace}In{\spacestart}{\nextspace}(x{\spacestart}{\nextspace}:elt){\spacestart}{\nextspace}(s{\spacestart}{\nextspace}:{\spacestart}{\nextspace}t){\spacestart}{\nextspace}:{\spacestart}{\nextspace}\textbf{Prop}{\spacestart}{\nextspace}:={\spacestart}{\nextspace}\textrm{\ldots}.{\spacestart}{\nextspace}{}\<[E]%
\\[\blanklineskip]%
\>[B]{}{\spacestart}{\nextspace}{\nextspace}{\nextspace}{\nextspace}{\nextspace}{}\<[6]%
\>[6]{}\textbf{Program}{\spacestart}{\nextspace}\textbf{Definition}{\spacestart}{\nextspace}mem{\spacestart}{\nextspace}:{\spacestart}{\nextspace}elt{\spacestart}{\nextspace}->{\spacestart}{\nextspace}t{\spacestart}{\nextspace}->{\spacestart}{\nextspace}bool{\spacestart}{\nextspace}:={\spacestart}{\nextspace}member.{\spacestart}{\nextspace}{}\<[E]%
\\[\blanklineskip]%
\>[B]{}{\spacestart}{\nextspace}{\nextspace}{\nextspace}{\nextspace}{\nextspace}{}\<[6]%
\>[6]{}\textbf{Lemma}{\spacestart}{\nextspace}mem\char95 1{\spacestart}{\nextspace}:{\spacestart}{\nextspace}\textbf{forall}{\spacestart}{\nextspace}(s{\spacestart}{\nextspace}:{\spacestart}{\nextspace}t){\spacestart}{\nextspace}(x{\spacestart}{\nextspace}:{\spacestart}{\nextspace}elt),{\spacestart}{\nextspace}In{\spacestart}{\nextspace}x{\spacestart}{\nextspace}s{\spacestart}{\nextspace}->{\spacestart}{\nextspace}mem{\spacestart}{\nextspace}x{\spacestart}{\nextspace}s{\spacestart}{\nextspace}={\spacestart}{\nextspace}true.{\spacestart}{\nextspace}{}\<[E]%
\\[-0.3ex]%
\>[B]{}{\spacestart}{\nextspace}{\nextspace}{\nextspace}{\nextspace}{\nextspace}{}\<[6]%
\>[6]{}\textbf{Proof}.{\spacestart}{\nextspace}\textrm{\ldots}{\spacestart}{\nextspace}(*{\spacestart}{\nextspace}Proof{\spacestart}{\nextspace}may{\spacestart}{\nextspace}assume{\spacestart}{\nextspace}that{\spacestart}{\nextspace}s{\spacestart}{\nextspace}is{\spacestart}{\nextspace}well-formed{\spacestart}{\nextspace}*){\spacestart}{\nextspace}\textrm{\ldots}{\spacestart}{\nextspace}\textbf{Qed}.{\spacestart}{\nextspace}{}\<[E]%
\\[\blanklineskip]%
\>[B]{}{\spacestart}{\nextspace}{\nextspace}{\nextspace}{\nextspace}{\nextspace}{}\<[6]%
\>[6]{}\textrm{\ldots}{\spacestart}{\nextspace}{}\<[E]%
\\[-0.3ex]%
\>[B]{}\textbf{End}{\spacestart}{\nextspace}SetFSet.{}\<[E]%
\ColumnHook
\end{hscode}\resethooks\endgroup

Instantiating these interfaces runs into two small hiccups.  The first is that
they talk about \emph{all} sets, not simply all \emph{well-formed} sets.
Therefore, as in the previous section, we instantiate these interfaces with the
subset type \ensuretext{\straighttick\ttfamily{}{\char123}s{\spacestart}{\nextspace}:{\spacestart}{\nextspace}Set\char95 {\spacestart}{\nextspace}e{\spacestart}{\nextspace}|{\spacestart}{\nextspace}WF{\spacestart}{\nextspace}s{\char125}}.  The second is that Coq's module system does
not interact with type classes, and \ensuretext{\straighttick\ttfamily{}Set\char95 } is defined such that its element type
must be an instance of the \ensuretext{\straighttick\ttfamily{}Ord} type class.  This impedance mismatch requires
us to write a module which can generate an \ensuretext{\straighttick\ttfamily{}Ord} instance from a Coq
\ensuretext{\straighttick\ttfamily{}OrderedType} module.


By successfully instantiating this module interface, we obtain two benefits.
First, we must prove theorems that cover many of the main functions provided
by \texttt{containers}; these theorems are particularly valuable, since as the
interface itself is heavily used by Coq users. Second, by instantiating this
interface, we connect our injected Haskell code to the Coq ecosystem, enabling
Coq users to easily use the \texttt{containers}-derived data structures in their
developments, should they so desire.

\subsection{Abstract models as specifications}\label{sec:specs-math}

Tests, type classes and the other sources of specifications do not fully
describe the intended behavior of all functions. We therefore also have to also come up
with specifications on our own. We do this by relating a concrete search tree to
the abstract set that it represents; that is, we provide a denotational
semantics.  We denote a set with elements of type \ensuretext{\straighttick\ttfamily{}e} as its indicator function
of type \ensuretext{\straighttick\ttfamily{}e{\spacestart}{\nextspace}->{\spacestart}{\nextspace}bool}; for \ensuretext{\straighttick\ttfamily{}Set{\spacestart}{\nextspace}e}, we provide a denotation function %
\ensuretext{\straighttick\ttfamily{}sem{\spacestart}{\nextspace}:{\spacestart}{\nextspace}\textbf{forall}{\spacestart}{\nextspace}{\char123}e{\char125}{\spacestart}{\nextspace}{\textasciigrave}{\char123}Eq\char95 {\spacestart}{\nextspace}e{\char125},{\spacestart}{\nextspace}Set\char95 {\spacestart}{\nextspace}e{\spacestart}{\nextspace}->{\spacestart}{\nextspace}(e{\spacestart}{\nextspace}->{\spacestart}{\nextspace}bool)}, and for \ensuretext{\straighttick\ttfamily{}IntSet}, we provide
a denotation relation \ensuretext{\straighttick\ttfamily{}Sem{\spacestart}{\nextspace}:{\spacestart}{\nextspace}IntSet{\spacestart}{\nextspace}->{\spacestart}{\nextspace}(N{\spacestart}{\nextspace}->{\spacestart}{\nextspace}bool){\spacestart}{\nextspace}->{\spacestart}{\nextspace}\textbf{Prop}}.

This approach allows us to abstractly describe the meaning of operations like
\ensuretext{\straighttick\ttfamily{}insert}.  For \ensuretext{\straighttick\ttfamily{}Set\char95 }, we do this by providing a theorem like
\begingroup\straighttick\begin{hscode}\SaveRestoreHook
\column{B}{@{}>{\hspre}l<{\hspost}@{}}%
\column{6}{@{}>{\hspre}l<{\hspost}@{}}%
\column{E}{@{}>{\hspre}l<{\hspost}@{}}%
\>[B]{}\textbf{Theorem}{\spacestart}{\nextspace}insert\char95 sem:{\spacestart}{\nextspace}{}\<[E]%
\\[-0.3ex]%
\>[B]{}~~{\spacestart}{\nextspace}{\nextspace}{}\<[6]%
\>[6]{}\textbf{forall}{\spacestart}{\nextspace}{\char123}a{\char125}{\spacestart}{\nextspace}{\textasciigrave}{\char123}OrdLaws{\spacestart}{\nextspace}a{\char125}{\spacestart}{\nextspace}(s{\spacestart}{\nextspace}:{\spacestart}{\nextspace}Set\char95 {\spacestart}{\nextspace}a){\spacestart}{\nextspace}(x{\spacestart}{\nextspace}:{\spacestart}{\nextspace}a),{\spacestart}{\nextspace}WF{\spacestart}{\nextspace}s{\spacestart}{\nextspace}->{\spacestart}{\nextspace}{}\<[E]%
\\[-0.3ex]%
\>[B]{}{\spacestart}{\nextspace}{\nextspace}{\nextspace}{\nextspace}{\nextspace}{}\<[6]%
\>[6]{}\textbf{forall}{\spacestart}{\nextspace}(i{\spacestart}{\nextspace}:{\spacestart}{\nextspace}a),{\spacestart}{\nextspace}sem{\spacestart}{\nextspace}(insert{\spacestart}{\nextspace}x{\spacestart}{\nextspace}s){\spacestart}{\nextspace}i{\spacestart}{\nextspace}={\spacestart}{\nextspace}(i{\spacestart}{\nextspace}=={\spacestart}{\nextspace}x){\spacestart}{\nextspace}||{\spacestart}{\nextspace}sem{\spacestart}{\nextspace}s{\spacestart}{\nextspace}i.{}\<[E]%
\ColumnHook
\end{hscode}\resethooks\endgroup
(For \ensuretext{\straighttick\ttfamily{}IntSet}, some techincal details differ.)
However, there is more we need to know about \ensuretext{\straighttick\ttfamily{}insert} than just its
denotation.  We also need to know that it preserves well-formedness and bounds,
and -- to reason about balancing -- its size. To avoid having to prove these
properties independently, we define a relation \ensuretext{\straighttick\ttfamily{}Desc} that completely describes
a set, by asserting that it is well-founded and relating it to bounds, its size
and its denotation:
\begingroup\straighttick\begin{hscode}\SaveRestoreHook
\column{B}{@{}>{\hspre}l<{\hspost}@{}}%
\column{6}{@{}>{\hspre}l<{\hspost}@{}}%
\column{E}{@{}>{\hspre}l<{\hspost}@{}}%
\>[B]{}\textbf{Definition}{\spacestart}{\nextspace}Desc{\spacestart}{\nextspace}(s{\spacestart}{\nextspace}:{\spacestart}{\nextspace}Set{\spacestart}{\nextspace}\char95 e){\spacestart}{\nextspace}(lb{\spacestart}{\nextspace}ub{\spacestart}{\nextspace}:{\spacestart}{\nextspace}option{\spacestart}{\nextspace}e){\spacestart}{\nextspace}(sz{\spacestart}{\nextspace}:{\spacestart}{\nextspace}Z){\spacestart}{\nextspace}(f{\spacestart}{\nextspace}:{\spacestart}{\nextspace}e{\spacestart}{\nextspace}->{\spacestart}{\nextspace}bool){\spacestart}{\nextspace}:={\spacestart}{\nextspace}{}\<[E]%
\\[-0.3ex]%
\>[B]{}~~{\spacestart}{\nextspace}{\nextspace}{}\<[6]%
\>[6]{}Bounded{\spacestart}{\nextspace}s{\spacestart}{\nextspace}lb{\spacestart}{\nextspace}ub{\spacestart}{\nextspace}/\char92 {\spacestart}{\nextspace}size{\spacestart}{\nextspace}s{\spacestart}{\nextspace}={\spacestart}{\nextspace}sz{\spacestart}{\nextspace}/\char92 {\spacestart}{\nextspace}(\textbf{forall}{\spacestart}{\nextspace}i,{\spacestart}{\nextspace}sem{\spacestart}{\nextspace}s{\spacestart}{\nextspace}i{\spacestart}{\nextspace}={\spacestart}{\nextspace}f{\spacestart}{\nextspace}i).{}\<[E]%
\ColumnHook
\end{hscode}\resethooks\endgroup
This allows us to state a single theorem about \ensuretext{\straighttick\ttfamily{}insert}, namely
\begingroup\straighttick\begin{hscode}\SaveRestoreHook
\column{B}{@{}>{\hspre}l<{\hspost}@{}}%
\column{6}{@{}>{\hspre}l<{\hspost}@{}}%
\column{11}{@{}>{\hspre}l<{\hspost}@{}}%
\column{27}{@{}>{\hspre}l<{\hspost}@{}}%
\column{E}{@{}>{\hspre}l<{\hspost}@{}}%
\>[B]{}\textbf{Lemma}{\spacestart}{\nextspace}insert\char95 Desc:{\spacestart}{\nextspace}\textbf{forall}{\spacestart}{\nextspace}x{\spacestart}{\nextspace}s{\spacestart}{\nextspace}lb{\spacestart}{\nextspace}ub,{\spacestart}{\nextspace}{}\<[E]%
\\[-0.3ex]%
\>[B]{}~~{\spacestart}{\nextspace}{\nextspace}{}\<[6]%
\>[6]{}Bounded{\spacestart}{\nextspace}s{\spacestart}{\nextspace}lb{\spacestart}{\nextspace}ub{\spacestart}{\nextspace}->{\spacestart}{\nextspace}{}\<[E]%
\\[-0.3ex]%
\>[B]{}{\spacestart}{\nextspace}{\nextspace}{\nextspace}{\nextspace}{\nextspace}{}\<[6]%
\>[6]{}isLB{\spacestart}{\nextspace}lb{\spacestart}{\nextspace}x{\spacestart}{\nextspace}={\spacestart}{\nextspace}true{\spacestart}{\nextspace}->{\spacestart}{\nextspace}{\nextspace}{}\<[27]%
\>[27]{}(*{\spacestart}{\nextspace}If{\spacestart}{\nextspace}lb{\spacestart}{\nextspace}is{\spacestart}{\nextspace}defined,{\spacestart}{\nextspace}it{\spacestart}{\nextspace}is{\spacestart}{\nextspace}less{\spacestart}{\nextspace}than{\spacestart}{\nextspace}x{\spacestart}{\nextspace}*){\spacestart}{\nextspace}{}\<[E]%
\\[-0.3ex]%
\>[B]{}{\spacestart}{\nextspace}{\nextspace}{\nextspace}{\nextspace}{\nextspace}{}\<[6]%
\>[6]{}isUB{\spacestart}{\nextspace}ub{\spacestart}{\nextspace}x{\spacestart}{\nextspace}={\spacestart}{\nextspace}true{\spacestart}{\nextspace}->{\spacestart}{\nextspace}{\nextspace}{}\<[27]%
\>[27]{}(*{\spacestart}{\nextspace}If{\spacestart}{\nextspace}ub{\spacestart}{\nextspace}is{\spacestart}{\nextspace}defined,{\spacestart}{\nextspace}it{\spacestart}{\nextspace}is{\spacestart}{\nextspace}greater{\spacestart}{\nextspace}than{\spacestart}{\nextspace}x{\spacestart}{\nextspace}*){\spacestart}{\nextspace}{}\<[E]%
\\[-0.3ex]%
\>[B]{}{\spacestart}{\nextspace}{\nextspace}{\nextspace}{\nextspace}{\nextspace}{}\<[6]%
\>[6]{}Desc{\spacestart}{\nextspace}(insert{\spacestart}{\nextspace}x{\spacestart}{\nextspace}s){\spacestart}{\nextspace}lb{\spacestart}{\nextspace}ub{\spacestart}{\nextspace}{}\<[E]%
\\[-0.3ex]%
\>[B]{}{\spacestart}{\nextspace}{\nextspace}{\nextspace}{\nextspace}{\nextspace}{}\<[6]%
\>[6]{}~~{\spacestart}{\nextspace}{\nextspace}{}\<[11]%
\>[11]{}(\textbf{if}{\spacestart}{\nextspace}sem{\spacestart}{\nextspace}s{\spacestart}{\nextspace}y{\spacestart}{\nextspace}\textbf{then}{\spacestart}{\nextspace}size{\spacestart}{\nextspace}s{\spacestart}{\nextspace}\textbf{else}{\spacestart}{\nextspace}(1{\spacestart}{\nextspace}+{\spacestart}{\nextspace}size{\spacestart}{\nextspace}s)){\spacestart}{\nextspace}(\textbf{fun}{\spacestart}{\nextspace}i{\spacestart}{\nextspace}=>{\spacestart}{\nextspace}(i{\spacestart}{\nextspace}=={\spacestart}{\nextspace}x){\spacestart}{\nextspace}||{\spacestart}{\nextspace}sem{\spacestart}{\nextspace}s{\spacestart}{\nextspace}i).{}\<[E]%
\ColumnHook
\end{hscode}\resethooks\endgroup
and prove everything we need to know about \ensuretext{\straighttick\ttfamily{}insert} in one single inductive
proof.%
\footnote{In the file \fileContainers{theories/SetProofs.v}}

Since \ensuretext{\straighttick\ttfamily{}Desc} describes \ensuretext{\straighttick\ttfamily{}insert} completely, it introduces a layer of
abstraction that we can build upon. In fact, we specify \emph{all} functions this way, and use these specifications, rather than the concrete implementation, to prove the other specifications.
(We have an analogous \ensuretext{\straighttick\ttfamily{}Desc} relation for \ensuretext{\straighttick\ttfamily{}IntSet} that describes the properties of Patricia trees.)

An alternative abstract model for finite sets is
the sorted list of their elements, i.e.\ the result of \ensuretext{\straighttick\ttfamily{}toAscList}. The meaning
of certain operations, like \ensuretext{\straighttick\ttfamily{}foldr}, \ensuretext{\straighttick\ttfamily{}take} or \ensuretext{\straighttick\ttfamily{}size}, can naturally be
expressed in terms of \ensuretext{\straighttick\ttfamily{}toAscList}, but would be very convoluted to state in
terms of the indicator function, and we use this denotation -- or both -- where
appropriate.

\section{Producing verifiable code with \LIThstocoq}
\label{sec:translating}

Identifying what to prove about the code is only half of the challenge -- we
also need to get the Haskell code into Coq.
Ideally, the translation of Haskell code into Gallina using
\texttt{hs\char45{}to\char45{}coq} would be completely automatic and produce code that can be
verified as easily as code written directly in Coq -- and for textbook-level
examples, that is the case~\citep{hs-to-coq-cpp}. However, working with
real-world code requires adjustments to the translation process to make sure
that the output is both accepted by Coq and amenable to verification.

A core principle of our approach is that the Haskell source code \emph{does not
  need to be modified} in order to be verified. This principle ensures that we
verify “the” \texttt{containers} library (not a “verified fork”) and that the
verification can be ported to a newer version of the library.

The crucial feature of \texttt{hs\char45{}to\char45{}coq} that enables this approach is the support for
\emph{edits}: instructions to treat some code differently during translation.
Edits are specified in plain text files, which also serve as a concise summary
of our interventions.  The \texttt{hs\char45{}to\char45{}coq} tool already supported many forms of
edits; for example, specifying when names need to be changed, when parts of the module
should be ignored or replaced by some other term, when we want to map Haskell
types to existing Coq types, or when a recursive function definition needs an
explicit termination proof. In the course of this work, we added new features to
\texttt{hs\char45{}to\char45{}coq} -- such as the ability to apply rewrite rules, to handle partiality
and to defer termination proofs to the verification stage -- and extended the
provided \texttt{base} library.

In this section we demonstrate some of the challenges posed by translating
real-world code, and show how \texttt{hs\char45{}to\char45{}coq}’s flexibility allowed us to not only to
overcome them, but also to facilitate subsequently proving the input correct.

%

\subsection{Unsafe pointer equality}\label{sec:ptrEq}

An example of a Haskell feature that we cannot expect to translate
without intervention is \emph{unsafe pointer equality}. GHC’s runtime provides the
scarily named function \ensuretext{\straighttick\ttfamily{}reallyUnsafePtrEqualty\#}, which the \texttt{containers} library
wraps as \ensuretext{\straighttick\ttfamily{}ptrEq{\spacestart}{\nextspace}::{\spacestart}{\nextspace}a{\spacestart}{\nextspace}->{\spacestart}{\nextspace}a{\spacestart}{\nextspace}->{\spacestart}{\nextspace}Bool}.
If this function returns \ensuretext{\straighttick\ttfamily{}True}, then both arguments are represented in memory
by the same pointer.  If this function returns \ensuretext{\straighttick\ttfamily{}False}, we know nothing -- this
function is underspecified and may return \ensuretext{\straighttick\ttfamily{}False} even if the two pointers are
equal.
This operation is used, for example, in \ensuretext{\straighttick\ttfamily{}Set.insert{\spacestart}{\nextspace}x{\spacestart}{\nextspace}s} when \ensuretext{\straighttick\ttfamily{}x} is already a member of \ensuretext{\straighttick\ttfamily{}s}:
If \ensuretext{\straighttick\ttfamily{}ptrEq} indicates that the subtree is unchanged, the function skips the
redundant re-balancing step -- which enhances performance -- and returns the
original set
rather than constructing a semantically equivalent copy -- which increases sharing.

Coq does not provide any way of reasoning about memory,
so when we use \texttt{hs\char45{}to\char45{}coq}, we must replace \ensuretext{\straighttick\ttfamily{}ptrEq} with something else.
But what?

One option is to replace the definition of \ensuretext{\straighttick\ttfamily{}ptrEq} with a definition that does
not do any computation and simply always returns \ensuretext{\straighttick\ttfamily{}False}. This can be done using
the following edit:
\begingroup\straighttick\begin{hscode}\SaveRestoreHook
\column{B}{@{}>{\hspre}l<{\hspost}@{}}%
\column{E}{@{}>{\hspre}l<{\hspost}@{}}%
\>[B]{}\textbf{replace}{\spacestart}{\nextspace}\textbf{Definition}{\spacestart}{\nextspace}ptrEq{\spacestart}{\nextspace}:{\spacestart}{\nextspace}\textbf{forall}{\spacestart}{\nextspace}{\char123}a{\char125},{\spacestart}{\nextspace}a{\spacestart}{\nextspace}->{\spacestart}{\nextspace}a{\spacestart}{\nextspace}->{\spacestart}{\nextspace}bool{\spacestart}{\nextspace}:={\spacestart}{\nextspace}\textbf{fun}{\spacestart}{\nextspace}\char95 {\spacestart}{\nextspace}\char95 {\spacestart}{\nextspace}\char95 {\spacestart}{\nextspace}=>{\spacestart}{\nextspace}false.{}\<[E]%
\ColumnHook
\end{hscode}\resethooks\endgroup

This replacement would behave in a way that is consistent with
\ensuretext{\straighttick\ttfamily{}reallyUnsafePtrEqualty\#} and allows us to proceed with translation and
verification.
However, the code in the \ensuretext{\straighttick\ttfamily{}True} branch of an unsafe pointer equality test would
be dead code in Coq, and our verification would miss bugs possibly lurking
there.

Consequently, we choose a different encoding that captures the semantics
of \ensuretext{\straighttick\ttfamily{}ptrEq} more precisely.
We make the definition of \texttt{ptrEq} \emph{opaque} and partially
specify its behavior.\footnote{In the file \fileContainers{lib/Utils/Containers/Internal/PtrEquality.v}}
\begingroup\straighttick\begin{hscode}\SaveRestoreHook
\column{B}{@{}>{\hspre}l<{\hspost}@{}}%
\column{6}{@{}>{\hspre}l<{\hspost}@{}}%
\column{8}{@{}>{\hspre}l<{\hspost}@{}}%
\column{9}{@{}>{\hspre}l<{\hspost}@{}}%
\column{E}{@{}>{\hspre}l<{\hspost}@{}}%
\>[B]{}\textbf{Definition}{\spacestart}{\nextspace}ptrEq\char95 spec{\spacestart}{\nextspace}:{\spacestart}{\nextspace}{}\<[E]%
\\[-0.3ex]%
\>[B]{}~~{\spacestart}{\nextspace}{\nextspace}{}\<[6]%
\>[6]{}{\char123}ptrEq{\spacestart}{\nextspace}:{\spacestart}{\nextspace}\textbf{forall}{\spacestart}{\nextspace}a,{\spacestart}{\nextspace}a{\spacestart}{\nextspace}->{\spacestart}{\nextspace}a{\spacestart}{\nextspace}->{\spacestart}{\nextspace}bool{\spacestart}{\nextspace}|{\spacestart}{\nextspace}\textbf{forall}{\spacestart}{\nextspace}a{\spacestart}{\nextspace}(x{\spacestart}{\nextspace}y{\spacestart}{\nextspace}:{\spacestart}{\nextspace}a),{\spacestart}{\nextspace}ptrEq{\spacestart}{\nextspace}\char95 {\spacestart}{\nextspace}x{\spacestart}{\nextspace}y{\spacestart}{\nextspace}={\spacestart}{\nextspace}true{\spacestart}{\nextspace}->{\spacestart}{\nextspace}x{\spacestart}{\nextspace}={\spacestart}{\nextspace}y{\char125}.{\spacestart}{\nextspace}{}\<[E]%
\\[-0.3ex]%
\>[B]{}\textbf{Proof}.{\spacestart}{\nextspace}{\nextspace}{}\<[9]%
\>[9]{}apply{\spacestart}{\nextspace}(exist{\spacestart}{\nextspace}\char95 {\spacestart}{\nextspace}(\textbf{fun}{\spacestart}{\nextspace}\char95 {\spacestart}{\nextspace}\char95 {\spacestart}{\nextspace}\char95 {\spacestart}{\nextspace}=>{\spacestart}{\nextspace}false)).{\spacestart}{\nextspace}intros;{\spacestart}{\nextspace}congruence.{\spacestart}{\nextspace}\textbf{Qed}.{\spacestart}{\nextspace}{}\<[E]%
\\[\blanklineskip]%
\>[B]{}\textbf{Definition}{\spacestart}{\nextspace}ptrEq{\spacestart}{\nextspace}:{\spacestart}{\nextspace}\textbf{forall}{\spacestart}{\nextspace}{\char123}a{\char125},{\spacestart}{\nextspace}a{\spacestart}{\nextspace}->{\spacestart}{\nextspace}a{\spacestart}{\nextspace}->{\spacestart}{\nextspace}bool{\spacestart}{\nextspace}:={\spacestart}{\nextspace}proj1\char95 sig{\spacestart}{\nextspace}ptrEq\char95 spec.{\spacestart}{\nextspace}{}\<[E]%
\\[-0.3ex]%
\>[B]{}\textbf{Lemma}{\spacestart}{\nextspace}{\nextspace}{}\<[8]%
\>[8]{}ptrEq\char95 eq{\spacestart}{\nextspace}:{\spacestart}{\nextspace}\textbf{forall}{\spacestart}{\nextspace}{\char123}a{\char125}{\spacestart}{\nextspace}(x:a)(y:a),{\spacestart}{\nextspace}ptrEq{\spacestart}{\nextspace}x{\spacestart}{\nextspace}y{\spacestart}{\nextspace}={\spacestart}{\nextspace}true{\spacestart}{\nextspace}->{\spacestart}{\nextspace}x{\spacestart}{\nextspace}={\spacestart}{\nextspace}y.{\spacestart}{\nextspace}{}\<[E]%
\\[-0.3ex]%
\>[B]{}\textbf{Proof}.{\spacestart}{\nextspace}exact{\spacestart}{\nextspace}(proj2\char95 sig{\spacestart}{\nextspace}ptrEq\char95 spec).{\spacestart}{\nextspace}\textbf{Qed}.{}\<[E]%
\ColumnHook
\end{hscode}\resethooks\endgroup

Here, we define the \ensuretext{\straighttick\ttfamily{}ptrEq} function together with a specification as
\ensuretext{\straighttick\ttfamily{}ptrEq\char95 spec}.  Although the function in \ensuretext{\straighttick\ttfamily{}ptrEq\char95 spec} also always returns
\ensuretext{\straighttick\ttfamily{}false}, the \ensuretext{\straighttick\ttfamily{}\textbf{Qed}} at the end of its definition completely hides this
implementation of \ensuretext{\straighttick\ttfamily{}ptrEq}.  While the specification \ensuretext{\straighttick\ttfamily{}ptrEq\char95 eq} is vacuously
true, making this definition opaque forces verification to proceed down
both paths. While could achieve the same using \ensuretext{\straighttick\ttfamily{}\textbf{Axiom}}, our variant protects us from accidentially introducing inconsistencies to Coq.

We do have to trust that GHC's definition of pointer equality has
this specification, but given this assumption, we can soundly verify that
pointer equality is used correctly.

\subsection{Evaluation order}
\label{sec:strict-types-and-functions}

A shallow embedding of Haskell into Coq makes the difference between strict and
lazy code vanish, because Gallina is a total language and does not care about
evaluation order. \Citet{fast-and-loose} show that such “fast and loose”
reasoning does not invalidate our theorems.

Haskell has “magic” functions like \ensuretext{\straighttick\ttfamily{}seq} that allow the programmer to explicitly
control strictness, and the \texttt{containers} library uses it. Its effect is
irrelevant in Coq, and we instruct \texttt{hs\char45{}to\char45{}coq} to use this simple, magic-free
implementation for it:
\begingroup\straighttick\begin{hscode}\SaveRestoreHook
\column{B}{@{}>{\hspre}l<{\hspost}@{}}%
\column{E}{@{}>{\hspre}l<{\hspost}@{}}%
\>[B]{}\textbf{Definition}{\spacestart}{\nextspace}seq{\spacestart}{\nextspace}{\char123}a{\char125}{\spacestart}{\nextspace}{\char123}b{\char125}{\spacestart}{\nextspace}(x{\spacestart}{\nextspace}:{\spacestart}{\nextspace}a){\spacestart}{\nextspace}(y{\spacestart}{\nextspace}:{\spacestart}{\nextspace}b){\spacestart}{\nextspace}:={\spacestart}{\nextspace}y.{}\<[E]%
\ColumnHook
\end{hscode}\resethooks\endgroup

\subsection{Eliminating unwanted parts of the code}
\label{sec:unwanted-code}
\untranslatedAPIfigure

\Cref{fig:untranslatedAPI} lists the untranslated portions of the \ensuretext{\straighttick\ttfamily{}Set} and
\ensuretext{\straighttick\ttfamily{}IntSet} modules.
Many of these operations are functions that we choose to ignore for the sake
of verification -- for example, the function \ensuretext{\straighttick\ttfamily{}showTree} in \ensuretext{\straighttick\ttfamily{}Data.Set} prints
the internal structure of such a set as an ASCII-art tree. This function is
not used elsewhere in the module. In the interest of a tidier and smaller
output, we skip this function using an edit:
\begingroup\straighttick\begin{hscode}\SaveRestoreHook
\column{B}{@{}>{\hspre}l<{\hspost}@{}}%
\column{E}{@{}>{\hspre}l<{\hspost}@{}}%
\>[B]{}\textbf{skip}{\spacestart}{\nextspace}showTree{}\<[E]%
\ColumnHook
\end{hscode}\resethooks\endgroup
Similarly, we skip functionality related to serialization (the \ensuretext{\straighttick\ttfamily{}Show}
and \ensuretext{\straighttick\ttfamily{}Show1} type classes), deserialization (the \ensuretext{\straighttick\ttfamily{}Read} type class), generic
programming (the \ensuretext{\straighttick\ttfamily{}Data} type class), and overloaded list notation (the \ensuretext{\straighttick\ttfamily{}IsList}
type class).

Furthermore, we skip some operations whose public API is partial.  For example,
evaluating \ensuretext{\straighttick\ttfamily{}findMax{\spacestart}{\nextspace}empty} will raise an exception, as the empty set has no
maximum element.  We cannot model this exception in Coq, so we skip
\ensuretext{\straighttick\ttfamily{}findMax} and similar functions (\ensuretext{\straighttick\ttfamily{}findMin}, \ensuretext{\straighttick\ttfamily{}deleteFindMax}, \ensuretext{\straighttick\ttfamily{}deleteFindMin},
\ensuretext{\straighttick\ttfamily{}findIndex}, \ensuretext{\straighttick\ttfamily{}elemAt} and \ensuretext{\straighttick\ttfamily{}deleteAt}).  This elision is not significant because
the \texttt{containers} API provides total equivalents for many of these functions
(e.g., \ensuretext{\straighttick\ttfamily{}lookupIndex}, which returns \ensuretext{\straighttick\ttfamily{}Nothing} when the index is out of bounds).

Finally, we skip the two functions that use mutual recursion
(\ensuretext{\straighttick\ttfamily{}Data.IntSet.fromAscList} and \ensuretext{\straighttick\ttfamily{}Data.IntSet.fromDistinctAscList}) as this is not
yet supported by \texttt{hs\char45{}to\char45{}coq}.

\subsection{Partiality in total functions}
\label{sec:partial}


In contrast to the skipped functions above, some functions use partiality in
their implementation in ways that cannot be triggered by a user of the
public API. In particular, they may use calls to Haskell’s \ensuretext{\straighttick\ttfamily{}error} function when an invariant is violated.

For example, the central balancing functions for \ensuretext{\straighttick\ttfamily{}Set}s, \ensuretext{\straighttick\ttfamily{}balanceL} and
\ensuretext{\straighttick\ttfamily{}balanceR}, may call \ensuretext{\straighttick\ttfamily{}error} when passed an ill-formed \ensuretext{\straighttick\ttfamily{}Set}. Because our proofs
only reason about well-formed sets, this code is actually dead. It does not
matter how we translate \ensuretext{\straighttick\ttfamily{}error} -- any term that is accepted by Coq is good enough.
However, \ensuretext{\straighttick\ttfamily{}error} in Haskell has the type
\begingroup\straighttick\begin{hscode}\SaveRestoreHook
\column{B}{@{}>{\hspre}l<{\hspost}@{}}%
\column{E}{@{}>{\hspre}l<{\hspost}@{}}%
\>[B]{}error{\spacestart}{\nextspace}::{\spacestart}{\nextspace}String{\spacestart}{\nextspace}->{\spacestart}{\nextspace}a{}\<[E]%
\ColumnHook
\end{hscode}\resethooks\endgroup
which means that a call to error can inhabit \emph{any} type.  We cannot define
such a function in Coq, and adding it as an axiom would be glaringly unsound.


Therefore, we extended \texttt{hs\char45{}to\char45{}coq} to use the following definition for
\ensuretext{\straighttick\ttfamily{}error}:
\begingroup\straighttick\begin{hscode}\SaveRestoreHook
\column{B}{@{}>{\hspre}l<{\hspost}@{}}%
\column{E}{@{}>{\hspre}l<{\hspost}@{}}%
\>[B]{}\textbf{Class}{\spacestart}{\nextspace}Default{\spacestart}{\nextspace}(a{\spacestart}{\nextspace}:\textbf{Type}){\spacestart}{\nextspace}:={\spacestart}{\nextspace}{\char123}{\spacestart}{\nextspace}default{\spacestart}{\nextspace}:{\spacestart}{\nextspace}a{\spacestart}{\nextspace}{\char125}.{\spacestart}{\nextspace}{}\<[E]%
\\[-0.3ex]%
\>[B]{}\textbf{Definition}{\spacestart}{\nextspace}error{\spacestart}{\nextspace}{\char123}a{\char125}{\spacestart}{\nextspace}{\textasciigrave}{\char123}Default{\spacestart}{\nextspace}a{\char125}{\spacestart}{\nextspace}:{\spacestart}{\nextspace}String{\spacestart}{\nextspace}->{\spacestart}{\nextspace}a.{\spacestart}{\nextspace}{}\<[E]%
\\[-0.3ex]%
\>[B]{}\textbf{Proof}.{\spacestart}{\nextspace}exact{\spacestart}{\nextspace}(\textbf{fun}{\spacestart}{\nextspace}\char95 {\spacestart}{\nextspace}=>{\spacestart}{\nextspace}default).{\spacestart}{\nextspace}\textbf{Qed}.{}\<[E]%
\ColumnHook
\end{hscode}\resethooks\endgroup
The type class enforces that we use \ensuretext{\straighttick\ttfamily{}error} only at non-empty types, ensuring
logical consistency. Yet we will notice that something is wrong when we have to
prove something about it. Just as with \ensuretext{\straighttick\ttfamily{}ptrEq} (\cref{sec:ptrEq}), by making the
definition of \ensuretext{\straighttick\ttfamily{}error} opaque using \ensuretext{\straighttick\ttfamily{}\textbf{Qed}}, we are prevented from accidentally or
intentionally using the concrete \ensuretext{\straighttick\ttfamily{}default} value of a given type in a proof
about \ensuretext{\straighttick\ttfamily{}error}. Furthermore, when we extract the Coq code back to Haskell for
testing, we translate this definition back to Haskell's \ensuretext{\straighttick\ttfamily{}error}
function, preserving the original semantics.

This encoding is inspired by Isabelle, where all types are inhabited and there is a polymorphic term \ensuretext{\straighttick\ttfamily{}undefined{\spacestart}{\nextspace}::{\spacestart}{\nextspace}a} that denotes an unspecified element of any type.

\subsection{Translating the \ensuretext{\straighttick\ttfamily{}Int} in \ensuretext{\straighttick\ttfamily{}IntSet}}\label{sec:number-types}

As discussed in \cref{sec:specs-comments}, we map Haskell’s finite-width integer
type \ensuretext{\straighttick\ttfamily{}Int} to Coq’s unbounded integer type \ensuretext{\straighttick\ttfamily{}Z} in the translation of \ensuretext{\straighttick\ttfamily{}Data.Set}
in order to match the specification that integer overflow is outside the scope
of the specified behavior.

For \ensuretext{\straighttick\ttfamily{}IntSet}, however, this choice would cause problems.  Big-endian Patricia
trees require that two different elements have a highest differing bit. This is
not the case for \ensuretext{\straighttick\ttfamily{}Z}, where negative numbers have an infinite number of bits
set to \texttt{1}; for instance, \texttt{\char45{}1} is effectively an infinite sequence of set
bits. Fortunately, \texttt{hs\char45{}to\char45{}coq} is flexible enough to allow us to make a
different choice when translating \ensuretext{\straighttick\ttfamily{}IntSet}; we can pick any suitable type
where all elements have a finite number of bits set, such as the natural numbers
(\ensuretext{\straighttick\ttfamily{}N}) or a fixed width integer type.

Given that Coq’s standard library provides a fairly comprehensive library of
lemmas about \ensuretext{\straighttick\ttfamily{}N} and decision procedures (\ensuretext{\straighttick\ttfamily{}omega} and \ensuretext{\straighttick\ttfamily{}lia}) that work with it,
we chose to use \ensuretext{\straighttick\ttfamily{}N} for now, with the intention to eventually switch to a 64-bit
integer type.  This is the appropriate generalization of \ensuretext{\straighttick\ttfamily{}IntSet} to an infinite
domain. In Haskell, the domain is 64-bit words, which happen to be
interpretable as negative numbers.  When we generalize to an infinite domain, we
generalize to bit strings of unbounded but finite length, which we can most simply interpret as nonnegative.

The \ensuretext{\straighttick\ttfamily{}IntSet} code uses bit-level operations, like \ensuretext{\straighttick\ttfamily{}complement} and \ensuretext{\straighttick\ttfamily{}negate},
that do not exist for \ensuretext{\straighttick\ttfamily{}N}. To deal with this we extended \texttt{hs\char45{}to\char45{}coq} with
support for \emph{rewrite edits} like
\begingroup\straighttick\begin{hscode}\SaveRestoreHook
\column{B}{@{}>{\hspre}l<{\hspost}@{}}%
\column{E}{@{}>{\hspre}l<{\hspost}@{}}%
\>[B]{}\textbf{rewrite}{\spacestart}{\nextspace}\textbf{forall}{\spacestart}{\nextspace}x{\spacestart}{\nextspace}y,{\spacestart}{\nextspace}(x{\spacestart}{\nextspace}.\&.{\spacestart}{\nextspace}complement{\spacestart}{\nextspace}y){\spacestart}{\nextspace}={\spacestart}{\nextspace}(xor{\spacestart}{\nextspace}x{\spacestart}{\nextspace}(x{\spacestart}{\nextspace}.\&.{\spacestart}{\nextspace}y)){}\<[E]%
\ColumnHook
\end{hscode}\resethooks\endgroup
which instruct it to replace any expression that matches the left-hand side by
the right hand side. For signed or bounded integer types, both sides are
equivalent. For unbounded unsigned types, like Coq’s type \ensuretext{\straighttick\ttfamily{}N}, the left hand
side is undefined (values in \ensuretext{\straighttick\ttfamily{}N} have no complement in \ensuretext{\straighttick\ttfamily{}N}), while the right
hand side is perfectly fine. When we switch to bounded integers in the \ensuretext{\straighttick\ttfamily{}IntSet}
code, we can remove these edits.

\subsection{Low-level bit twiddling}

The \texttt{containers} library uses highly tuned bit-twiddling algorithms to operate
on \ensuretext{\straighttick\ttfamily{}IntSet}s.  For example, the function \ensuretext{\straighttick\ttfamily{}revNat} reverses the order of the bits
in a 64-bit number:
\begingroup\straighttick\begin{hscode}\SaveRestoreHook
\column{B}{@{}>{\hspre}l<{\hspost}@{}}%
\column{14}{@{}>{\hspre}l<{\hspost}@{}}%
\column{20}{@{}>{\hspre}l<{\hspost}@{}}%
\column{E}{@{}>{\hspre}l<{\hspost}@{}}%
\>[B]{}revNat{\spacestart}{\nextspace}::{\spacestart}{\nextspace}Nat{\spacestart}{\nextspace}->{\spacestart}{\nextspace}Nat{\spacestart}{\nextspace}{}\<[E]%
\\[-0.3ex]%
\>[B]{}revNat{\spacestart}{\nextspace}x1{\spacestart}{\nextspace}={\spacestart}{\nextspace}{\nextspace}{}\<[14]%
\>[14]{}\textbf{case}{\spacestart}{\nextspace}{\nextspace}{}\<[20]%
\>[20]{}((x1{\spacestart}{\nextspace}{\textasciigrave}shiftRL{\textasciigrave}{\spacestart}{\nextspace}1){\spacestart}{\nextspace}.\&.{\spacestart}{\nextspace}5555555555555555){\spacestart}{\nextspace}.|.{\spacestart}{\nextspace}{}\<[E]%
\\[-0.3ex]%
\>[B]{}{\spacestart}{\nextspace}{\nextspace}{\nextspace}{\nextspace}{\nextspace}{\nextspace}{\nextspace}{\nextspace}{\nextspace}{\nextspace}{\nextspace}{\nextspace}{\nextspace}{\nextspace}{\nextspace}{\nextspace}{\nextspace}{\nextspace}{\nextspace}{}\<[20]%
\>[20]{}((x1{\spacestart}{\nextspace}.\&.{\spacestart}{\nextspace}5555555555555555){\spacestart}{\nextspace}{\textasciigrave}shiftLL{\textasciigrave}{\spacestart}{\nextspace}1){\spacestart}{\nextspace}\textbf{of}{\spacestart}{\nextspace}x2{\spacestart}{\nextspace}->{\spacestart}{\nextspace}{}\<[E]%
\\[-0.3ex]%
\>[B]{}{\spacestart}{\nextspace}{\nextspace}{\nextspace}{\nextspace}{\nextspace}{\nextspace}{\nextspace}{\nextspace}{\nextspace}{\nextspace}{\nextspace}{\nextspace}{\nextspace}{}\<[14]%
\>[14]{}\textbf{case}{\spacestart}{\nextspace}{\nextspace}{}\<[20]%
\>[20]{}((x2{\spacestart}{\nextspace}{\textasciigrave}shiftRL{\textasciigrave}{\spacestart}{\nextspace}2){\spacestart}{\nextspace}.\&.{\spacestart}{\nextspace}3333333333333333){\spacestart}{\nextspace}.|.{\spacestart}{\nextspace}{}\<[E]%
\\[-0.3ex]%
\>[B]{}{\spacestart}{\nextspace}{\nextspace}{\nextspace}{\nextspace}{\nextspace}{\nextspace}{\nextspace}{\nextspace}{\nextspace}{\nextspace}{\nextspace}{\nextspace}{\nextspace}{\nextspace}{\nextspace}{\nextspace}{\nextspace}{\nextspace}{\nextspace}{}\<[20]%
\>[20]{}((x2{\spacestart}{\nextspace}.\&.{\spacestart}{\nextspace}3333333333333333){\spacestart}{\nextspace}{\textasciigrave}shiftLL{\textasciigrave}{\spacestart}{\nextspace}2){\spacestart}{\nextspace}\textbf{of}{\spacestart}{\nextspace}x3{\spacestart}{\nextspace}->{\spacestart}{\nextspace}{}\<[E]%
\\[-0.3ex]%
\>[B]{}{\spacestart}{\nextspace}{\nextspace}{\nextspace}{\nextspace}{\nextspace}{\nextspace}{\nextspace}{\nextspace}{\nextspace}{\nextspace}{\nextspace}{\nextspace}{\nextspace}{}\<[14]%
\>[14]{}\textbf{case}{\spacestart}{\nextspace}{\nextspace}{}\<[20]%
\>[20]{}((x3{\spacestart}{\nextspace}{\textasciigrave}shiftRL{\textasciigrave}{\spacestart}{\nextspace}4){\spacestart}{\nextspace}.\&.{\spacestart}{\nextspace}0F0F0F0F0F0F0F0F){\spacestart}{\nextspace}.|.{\spacestart}{\nextspace}{}\<[E]%
\\[-0.3ex]%
\>[B]{}{\spacestart}{\nextspace}{\nextspace}{\nextspace}{\nextspace}{\nextspace}{\nextspace}{\nextspace}{\nextspace}{\nextspace}{\nextspace}{\nextspace}{\nextspace}{\nextspace}{\nextspace}{\nextspace}{\nextspace}{\nextspace}{\nextspace}{\nextspace}{}\<[20]%
\>[20]{}((x3{\spacestart}{\nextspace}.\&.{\spacestart}{\nextspace}0F0F0F0F0F0F0F0F){\spacestart}{\nextspace}{\textasciigrave}shiftLL{\textasciigrave}{\spacestart}{\nextspace}4){\spacestart}{\nextspace}\textbf{of}{\spacestart}{\nextspace}x4{\spacestart}{\nextspace}->{\spacestart}{\nextspace}{}\<[E]%
\\[-0.3ex]%
\>[B]{}{\spacestart}{\nextspace}{\nextspace}{\nextspace}{\nextspace}{\nextspace}{\nextspace}{\nextspace}{\nextspace}{\nextspace}{\nextspace}{\nextspace}{\nextspace}{\nextspace}{}\<[14]%
\>[14]{}\textbf{case}{\spacestart}{\nextspace}{\nextspace}{}\<[20]%
\>[20]{}((x4{\spacestart}{\nextspace}{\textasciigrave}shiftRL{\textasciigrave}{\spacestart}{\nextspace}8){\spacestart}{\nextspace}.\&.{\spacestart}{\nextspace}00FF00FF00FF00FF){\spacestart}{\nextspace}.|.{\spacestart}{\nextspace}{}\<[E]%
\\[-0.3ex]%
\>[B]{}{\spacestart}{\nextspace}{\nextspace}{\nextspace}{\nextspace}{\nextspace}{\nextspace}{\nextspace}{\nextspace}{\nextspace}{\nextspace}{\nextspace}{\nextspace}{\nextspace}{\nextspace}{\nextspace}{\nextspace}{\nextspace}{\nextspace}{\nextspace}{}\<[20]%
\>[20]{}((x4{\spacestart}{\nextspace}.\&.{\spacestart}{\nextspace}00FF00FF00FF00FF){\spacestart}{\nextspace}{\textasciigrave}shiftLL{\textasciigrave}{\spacestart}{\nextspace}8){\spacestart}{\nextspace}\textbf{of}{\spacestart}{\nextspace}x5{\spacestart}{\nextspace}->{\spacestart}{\nextspace}{}\<[E]%
\\[-0.3ex]%
\>[B]{}{\spacestart}{\nextspace}{\nextspace}{\nextspace}{\nextspace}{\nextspace}{\nextspace}{\nextspace}{\nextspace}{\nextspace}{\nextspace}{\nextspace}{\nextspace}{\nextspace}{}\<[14]%
\>[14]{}\textbf{case}{\spacestart}{\nextspace}{\nextspace}{}\<[20]%
\>[20]{}((x5{\spacestart}{\nextspace}{\textasciigrave}shiftRL{\textasciigrave}{\spacestart}{\nextspace}16){\spacestart}{\nextspace}.\&.{\spacestart}{\nextspace}0000FFFF0000FFFF){\spacestart}{\nextspace}.|.{\spacestart}{\nextspace}{}\<[E]%
\\[-0.3ex]%
\>[B]{}{\spacestart}{\nextspace}{\nextspace}{\nextspace}{\nextspace}{\nextspace}{\nextspace}{\nextspace}{\nextspace}{\nextspace}{\nextspace}{\nextspace}{\nextspace}{\nextspace}{\nextspace}{\nextspace}{\nextspace}{\nextspace}{\nextspace}{\nextspace}{}\<[20]%
\>[20]{}((x5{\spacestart}{\nextspace}.\&.{\spacestart}{\nextspace}0000FFFF0000FFFF){\spacestart}{\nextspace}{\textasciigrave}shiftLL{\textasciigrave}{\spacestart}{\nextspace}16){\spacestart}{\nextspace}\textbf{of}{\spacestart}{\nextspace}x6{\spacestart}{\nextspace}->{\spacestart}{\nextspace}{}\<[E]%
\\[-0.3ex]%
\>[B]{}{\spacestart}{\nextspace}{\nextspace}{\nextspace}{\nextspace}{\nextspace}{\nextspace}{\nextspace}{\nextspace}{\nextspace}{\nextspace}{\nextspace}{\nextspace}{\nextspace}{}\<[14]%
\>[14]{}(x6{\spacestart}{\nextspace}{\textasciigrave}shiftRL{\textasciigrave}{\spacestart}{\nextspace}32){\spacestart}{\nextspace}.|.{\spacestart}{\nextspace}(x6{\spacestart}{\nextspace}{\textasciigrave}shiftLL{\textasciigrave}{\spacestart}{\nextspace}32){}\<[E]%
\ColumnHook
\end{hscode}\resethooks\endgroup
Though complicated, this code is within the scope of what \texttt{hs\char45{}to\char45{}coq} can
translate, and we can verify its correctness.

\goodbreak

However, we can't keep up with \emph{all} their tricks.  For
example, \ensuretext{\straighttick\ttfamily{}indexOfTheOnlyBit}, which was contributed by Edward Kmett,\containerCommit{e076b33f} takes a number with exactly one bit set and
calculates the index of said bit.  It does so by unboxing the input, multiplying
it by a magic constant, and using the upper~6 bits of the product as an index
into a table stored in an unboxed array literal.  This manifests as the
following scary-looking code:
\begingroup\straighttick\begin{hscode}\SaveRestoreHook
\column{B}{@{}>{\hspre}l<{\hspost}@{}}%
\column{6}{@{}>{\hspre}l<{\hspost}@{}}%
\column{13}{@{}>{\hspre}l<{\hspost}@{}}%
\column{30}{@{}>{\hspre}l<{\hspost}@{}}%
\column{E}{@{}>{\hspre}l<{\hspost}@{}}%
\>[B]{}indexOfTheOnlyBit{\spacestart}{\nextspace}::{\spacestart}{\nextspace}Nat{\spacestart}{\nextspace}->{\spacestart}{\nextspace}Int{\spacestart}{\nextspace}{}\<[E]%
\\[-0.3ex]%
\>[B]{}indexOfTheOnlyBit{\spacestart}{\nextspace}bitmask{\spacestart}{\nextspace}={\spacestart}{\nextspace}I\#{\spacestart}{\nextspace}(lsbArray{\spacestart}{\nextspace}{\textasciigrave}indexInt8OffAddr\#{\textasciigrave}{\spacestart}{\nextspace}unboxInt{\spacestart}{\nextspace}{}\<[E]%
\\[-0.3ex]%
\>[B]{}{\spacestart}{\nextspace}{\nextspace}{\nextspace}{\nextspace}{\nextspace}{\nextspace}{\nextspace}{\nextspace}{\nextspace}{\nextspace}{\nextspace}{\nextspace}{}\<[13]%
\>[13]{}(intFromNat{\spacestart}{\nextspace}((bitmask{\spacestart}{\nextspace}*{\spacestart}{\nextspace}magic){\spacestart}{\nextspace}{\textasciigrave}shiftRL{\textasciigrave}{\spacestart}{\nextspace}offset))){\spacestart}{\nextspace}{}\<[E]%
\\[-0.3ex]%
\>[B]{}~~{\spacestart}{\nextspace}{\nextspace}{}\<[6]%
\>[6]{}\textbf{where}{\spacestart}{\nextspace}{}\<[E]%
\\[-0.3ex]%
\>[B]{}{\spacestart}{\nextspace}{\nextspace}{\nextspace}{\nextspace}{\nextspace}{}\<[6]%
\>[6]{}~~{\spacestart}{\nextspace}{\nextspace}{\nextspace}{\nextspace}{}\<[13]%
\>[13]{}unboxInt{\spacestart}{\nextspace}(I\#{\spacestart}{\nextspace}i){\spacestart}{\nextspace}{\nextspace}{}\<[30]%
\>[30]{}={\spacestart}{\nextspace}i{\spacestart}{\nextspace}{}\<[E]%
\\[-0.3ex]%
\>[B]{}{\spacestart}{\nextspace}{\nextspace}{\nextspace}{\nextspace}{\nextspace}{\nextspace}{\nextspace}{\nextspace}{\nextspace}{\nextspace}{\nextspace}{\nextspace}{}\<[13]%
\>[13]{}magic{\spacestart}{\nextspace}{\nextspace}{\nextspace}{\nextspace}{\nextspace}{\nextspace}{\nextspace}{\nextspace}{\nextspace}{\nextspace}{\nextspace}{\nextspace}{}\<[30]%
\>[30]{}={\spacestart}{\nextspace}0x07EDD5E59A4E28C2{\spacestart}{\nextspace}{}\<[E]%
\\[-0.3ex]%
\>[B]{}{\spacestart}{\nextspace}{\nextspace}{\nextspace}{\nextspace}{\nextspace}{\nextspace}{\nextspace}{\nextspace}{\nextspace}{\nextspace}{\nextspace}{\nextspace}{}\<[13]%
\>[13]{}offset{\spacestart}{\nextspace}{\nextspace}{\nextspace}{\nextspace}{\nextspace}{\nextspace}{\nextspace}{\nextspace}{\nextspace}{\nextspace}{\nextspace}{}\<[30]%
\>[30]{}={\spacestart}{\nextspace}58{\spacestart}{\nextspace}{}\<[E]%
\\[-0.3ex]%
\>[B]{}{\spacestart}{\nextspace}{\nextspace}{\nextspace}{\nextspace}{\nextspace}{\nextspace}{\nextspace}{\nextspace}{\nextspace}{\nextspace}{\nextspace}{\nextspace}{}\<[13]%
\>[13]{}!lsbArray{\spacestart}{\nextspace}{\nextspace}{\nextspace}{\nextspace}{\nextspace}{\nextspace}{\nextspace}{\nextspace}{}\<[30]%
\>[30]{}={\spacestart}{\nextspace}\textquotedbl\char92 63\char92 0\char92 58\char92 1\char92 5\textrm{\ldots}15\char92 8\char92 23\char92 7\char92 6\char92 5{\textquotedbl\#}{}\<[E]%
\ColumnHook
\end{hscode}\resethooks\endgroup

We currently cannot translate this code because \texttt{hs\char45{}to\char45{}coq} does not yet
support unboxed arrays or unboxed integers. We therefore replace it with a simple
definition based on a integer logarithm function provided by Coq’s standard
library:
\begingroup\straighttick\begin{hscode}\SaveRestoreHook
\column{B}{@{}>{\hspre}l<{\hspost}@{}}%
\column{E}{@{}>{\hspre}l<{\hspost}@{}}%
\>[B]{}\textbf{redefine}{\spacestart}{\nextspace}\textbf{Definition}{\spacestart}{\nextspace}indexOfTheOnlyBit{\spacestart}{\nextspace}:={\spacestart}{\nextspace}\textbf{fun}{\spacestart}{\nextspace}x{\spacestart}{\nextspace}=>{\spacestart}{\nextspace}N.log2{\spacestart}{\nextspace}x.{}\<[E]%
\ColumnHook
\end{hscode}\resethooks\endgroup
Similarly, we provide simpler definitions for the low-level bit-twiddling functions
\ensuretext{\straighttick\ttfamily{}branchMask}, \ensuretext{\straighttick\ttfamily{}mask}, \ensuretext{\straighttick\ttfamily{}zero} and \ensuretext{\straighttick\ttfamily{}suffixBitMask}.

\subsection{Non-trivial recursion}\label{sec:recursion}

\begin{figure}
\abovedisplayskip=0pt
\belowdisplayskip=0pt
\raggedright
~
\begingroup\straighttick\begin{hscode}\SaveRestoreHook
\column{B}{@{}>{\hspre}l<{\hspost}@{}}%
\column{6}{@{}>{\hspre}l<{\hspost}@{}}%
\column{15}{@{}>{\hspre}l<{\hspost}@{}}%
\column{29}{@{}>{\hspre}l<{\hspost}@{}}%
\column{E}{@{}>{\hspre}l<{\hspost}@{}}%
\>[B]{}\mbox{\onelinecomment  Less obvious structural recursion}{\spacestart}{\nextspace}{}\<[E]%
\\[-0.3ex]%
\>[B]{}link{\spacestart}{\nextspace}::{\spacestart}{\nextspace}a{\spacestart}{\nextspace}->{\spacestart}{\nextspace}Set{\spacestart}{\nextspace}a{\spacestart}{\nextspace}->{\spacestart}{\nextspace}Set{\spacestart}{\nextspace}a{\spacestart}{\nextspace}->{\spacestart}{\nextspace}Set{\spacestart}{\nextspace}a{\spacestart}{\nextspace}{}\<[E]%
\\[-0.3ex]%
\>[B]{}link{\spacestart}{\nextspace}x{\spacestart}{\nextspace}Tip{\spacestart}{\nextspace}r{\spacestart}{\nextspace}{\nextspace}{}\<[15]%
\>[15]{}={\spacestart}{\nextspace}insertMin{\spacestart}{\nextspace}x{\spacestart}{\nextspace}r{\spacestart}{\nextspace}{}\<[E]%
\\[-0.3ex]%
\>[B]{}link{\spacestart}{\nextspace}x{\spacestart}{\nextspace}l{\spacestart}{\nextspace}Tip{\spacestart}{\nextspace}{\nextspace}{}\<[15]%
\>[15]{}={\spacestart}{\nextspace}insertMax{\spacestart}{\nextspace}x{\spacestart}{\nextspace}l{\spacestart}{\nextspace}{}\<[E]%
\\[-0.3ex]%
\>[B]{}link{\spacestart}{\nextspace}x{\spacestart}{\nextspace}l@(Bin{\spacestart}{\nextspace}sizeL{\spacestart}{\nextspace}y{\spacestart}{\nextspace}ly{\spacestart}{\nextspace}ry){\spacestart}{\nextspace}r@(Bin{\spacestart}{\nextspace}sizeR{\spacestart}{\nextspace}z{\spacestart}{\nextspace}lz{\spacestart}{\nextspace}rz){\spacestart}{\nextspace}{}\<[E]%
\\[-0.3ex]%
\>[B]{}~~{\spacestart}{\nextspace}{\nextspace}{}\<[6]%
\>[6]{}|{\spacestart}{\nextspace}delta*sizeL{\spacestart}{\nextspace}<{\spacestart}{\nextspace}sizeR{\spacestart}{\nextspace}{\nextspace}{}\<[29]%
\>[29]{}={\spacestart}{\nextspace}balanceL{\spacestart}{\nextspace}z{\spacestart}{\nextspace}(link{\spacestart}{\nextspace}x{\spacestart}{\nextspace}l{\spacestart}{\nextspace}lz){\spacestart}{\nextspace}rz{\spacestart}{\nextspace}{}\<[E]%
\\[-0.3ex]%
\>[B]{}{\spacestart}{\nextspace}{\nextspace}{\nextspace}{\nextspace}{\nextspace}{}\<[6]%
\>[6]{}|{\spacestart}{\nextspace}delta*sizeR{\spacestart}{\nextspace}<{\spacestart}{\nextspace}sizeL{\spacestart}{\nextspace}{\nextspace}{}\<[29]%
\>[29]{}={\spacestart}{\nextspace}balanceR{\spacestart}{\nextspace}y{\spacestart}{\nextspace}ly{\spacestart}{\nextspace}(link{\spacestart}{\nextspace}x{\spacestart}{\nextspace}ry{\spacestart}{\nextspace}r){\spacestart}{\nextspace}{}\<[E]%
\\[-0.3ex]%
\>[B]{}{\spacestart}{\nextspace}{\nextspace}{\nextspace}{\nextspace}{\nextspace}{}\<[6]%
\>[6]{}|{\spacestart}{\nextspace}otherwise{\spacestart}{\nextspace}{\nextspace}{\nextspace}{\nextspace}{\nextspace}{\nextspace}{\nextspace}{\nextspace}{\nextspace}{\nextspace}{\nextspace}{\nextspace}{}\<[29]%
\>[29]{}={\spacestart}{\nextspace}bin{\spacestart}{\nextspace}x{\spacestart}{\nextspace}l{\spacestart}{\nextspace}r{}\<[E]%
\ColumnHook
\end{hscode}\resethooks\endgroup
\begingroup\straighttick\begin{hscode}\SaveRestoreHook
\column{B}{@{}>{\hspre}l<{\hspost}@{}}%
\column{6}{@{}>{\hspre}l<{\hspost}@{}}%
\column{13}{@{}>{\hspre}l<{\hspost}@{}}%
\column{17}{@{}>{\hspre}l<{\hspost}@{}}%
\column{21}{@{}>{\hspre}l<{\hspost}@{}}%
\column{24}{@{}>{\hspre}l<{\hspost}@{}}%
\column{26}{@{}>{\hspre}l<{\hspost}@{}}%
\column{E}{@{}>{\hspre}l<{\hspost}@{}}%
\>[B]{}\mbox{\onelinecomment  Well-founded recursion}{\spacestart}{\nextspace}{}\<[E]%
\\[-0.3ex]%
\>[B]{}foldlBits{\spacestart}{\nextspace}::{\spacestart}{\nextspace}Int{\spacestart}{\nextspace}->{\spacestart}{\nextspace}(a{\spacestart}{\nextspace}->{\spacestart}{\nextspace}Int{\spacestart}{\nextspace}->{\spacestart}{\nextspace}a){\spacestart}{\nextspace}->{\spacestart}{\nextspace}a{\spacestart}{\nextspace}->{\spacestart}{\nextspace}Nat{\spacestart}{\nextspace}->{\spacestart}{\nextspace}a{\spacestart}{\nextspace}{}\<[E]%
\\[-0.3ex]%
\>[B]{}foldlBits{\spacestart}{\nextspace}prefix{\spacestart}{\nextspace}f{\spacestart}{\nextspace}z{\spacestart}{\nextspace}bitmap{\spacestart}{\nextspace}={\spacestart}{\nextspace}go{\spacestart}{\nextspace}bitmap{\spacestart}{\nextspace}z{\spacestart}{\nextspace}{}\<[E]%
\\[-0.3ex]%
\>[B]{}~~{\spacestart}{\nextspace}{\nextspace}{}\<[6]%
\>[6]{}\textbf{where}{\spacestart}{\nextspace}{\nextspace}{}\<[13]%
\>[13]{}go{\spacestart}{\nextspace}{\nextspace}{}\<[17]%
\>[17]{}0{\spacestart}{\nextspace}{\nextspace}{\nextspace}{}\<[21]%
\>[21]{}acc{\spacestart}{\nextspace}{\nextspace}{}\<[26]%
\>[26]{}={\spacestart}{\nextspace}acc{\spacestart}{\nextspace}{}\<[E]%
\\[-0.3ex]%
\>[B]{}{\spacestart}{\nextspace}{\nextspace}{\nextspace}{\nextspace}{\nextspace}{\nextspace}{\nextspace}{\nextspace}{\nextspace}{\nextspace}{\nextspace}{\nextspace}{}\<[13]%
\>[13]{}go{\spacestart}{\nextspace}{\nextspace}{}\<[17]%
\>[17]{}bm{\spacestart}{\nextspace}{\nextspace}{}\<[21]%
\>[21]{}acc{\spacestart}{\nextspace}{\nextspace}{}\<[26]%
\>[26]{}={\spacestart}{\nextspace}go{\spacestart}{\nextspace}(xor{\spacestart}{\nextspace}bm{\spacestart}{\nextspace}bitmask){\spacestart}{\nextspace}((f{\spacestart}{\nextspace}acc){\spacestart}{\nextspace}\$!{\spacestart}{\nextspace}(prefix+bi)){\spacestart}{\nextspace}{}\<[E]%
\\[-0.3ex]%
\>[B]{}{\spacestart}{\nextspace}{\nextspace}{\nextspace}{\nextspace}{\nextspace}{\nextspace}{\nextspace}{\nextspace}{\nextspace}{\nextspace}{\nextspace}{\nextspace}{}\<[13]%
\>[13]{}~~{\spacestart}{\nextspace}\textbf{where}{\spacestart}{\nextspace}{\nextspace}{}\<[24]%
\>[24]{}!bitmask{\spacestart}{\nextspace}={\spacestart}{\nextspace}lowestBitMask{\spacestart}{\nextspace}bm{\spacestart}{\nextspace}{}\<[E]%
\\[-0.3ex]%
\>[B]{}{\spacestart}{\nextspace}{\nextspace}{\nextspace}{\nextspace}{\nextspace}{\nextspace}{\nextspace}{\nextspace}{\nextspace}{\nextspace}{\nextspace}{\nextspace}{\nextspace}{\nextspace}{\nextspace}{\nextspace}{\nextspace}{\nextspace}{\nextspace}{\nextspace}{\nextspace}{\nextspace}{\nextspace}{}\<[24]%
\>[24]{}!bi{\spacestart}{\nextspace}={\spacestart}{\nextspace}indexOfTheOnlyBit{\spacestart}{\nextspace}bitmask{}\<[E]%
\ColumnHook
\end{hscode}\resethooks\endgroup
\begingroup\straighttick\begin{hscode}\SaveRestoreHook
\column{B}{@{}>{\hspre}l<{\hspost}@{}}%
\column{6}{@{}>{\hspre}l<{\hspost}@{}}%
\column{13}{@{}>{\hspre}l<{\hspost}@{}}%
\column{17}{@{}>{\hspre}c<{\hspost}@{}}%
\column{17E}{@{}l@{}}%
\column{18}{@{}>{\hspre}l<{\hspost}@{}}%
\column{21}{@{}>{\hspre}l<{\hspost}@{}}%
\column{22}{@{}>{\hspre}c<{\hspost}@{}}%
\column{22E}{@{}l@{}}%
\column{24}{@{}>{\hspre}l<{\hspost}@{}}%
\column{26}{@{}>{\hspre}l<{\hspost}@{}}%
\column{31}{@{}>{\hspre}c<{\hspost}@{}}%
\column{31E}{@{}l@{}}%
\column{34}{@{}>{\hspre}c<{\hspost}@{}}%
\column{34E}{@{}l@{}}%
\column{35}{@{}>{\hspre}l<{\hspost}@{}}%
\column{37}{@{}>{\hspre}l<{\hspost}@{}}%
\column{40}{@{}>{\hspre}l<{\hspost}@{}}%
\column{E}{@{}>{\hspre}l<{\hspost}@{}}%
\>[B]{}\mbox{\onelinecomment  Deferred recursion}{\spacestart}{\nextspace}{}\<[E]%
\\[-0.3ex]%
\>[B]{}fromDistinctAscList{\spacestart}{\nextspace}::{\spacestart}{\nextspace}[a]{\spacestart}{\nextspace}->{\spacestart}{\nextspace}Set{\spacestart}{\nextspace}a{\spacestart}{\nextspace}{}\<[E]%
\\[-0.3ex]%
\>[B]{}fromDistinctAscList{\spacestart}{\nextspace}[]{\spacestart}{\nextspace}={\spacestart}{\nextspace}Tip{\spacestart}{\nextspace}{}\<[E]%
\\[-0.3ex]%
\>[B]{}fromDistinctAscList{\spacestart}{\nextspace}(x0{\spacestart}{\nextspace}:{\spacestart}{\nextspace}xs0){\spacestart}{\nextspace}={\spacestart}{\nextspace}go{\spacestart}{\nextspace}(1::Int){\spacestart}{\nextspace}(Bin{\spacestart}{\nextspace}1{\spacestart}{\nextspace}x0{\spacestart}{\nextspace}Tip{\spacestart}{\nextspace}Tip){\spacestart}{\nextspace}xs0{\spacestart}{\nextspace}{}\<[E]%
\\[-0.3ex]%
\>[B]{}~~{\spacestart}{\nextspace}{\nextspace}{}\<[6]%
\>[6]{}\textbf{where}{\spacestart}{\nextspace}{\nextspace}{}\<[13]%
\>[13]{}go{\spacestart}{\nextspace}{\nextspace}{}\<[17]%
\>[17]{}!\char95 {\spacestart}{\nextspace}{\nextspace}{}\<[17E]%
\>[21]{}t{\spacestart}{\nextspace}{\nextspace}{}\<[24]%
\>[24]{}[]{\spacestart}{\nextspace}{\nextspace}{\nextspace}{\nextspace}{\nextspace}{\nextspace}{\nextspace}{\nextspace}{}\<[34]%
\>[34]{}={\spacestart}{\nextspace}{\nextspace}{}\<[34E]%
\>[37]{}t{\spacestart}{\nextspace}{}\<[E]%
\\[-0.3ex]%
\>[B]{}{\spacestart}{\nextspace}{\nextspace}{\nextspace}{\nextspace}{\nextspace}{\nextspace}{\nextspace}{\nextspace}{\nextspace}{\nextspace}{\nextspace}{\nextspace}{}\<[13]%
\>[13]{}go{\spacestart}{\nextspace}{\nextspace}{}\<[17]%
\>[17]{}s{\spacestart}{\nextspace}{\nextspace}{\nextspace}{}\<[17E]%
\>[21]{}l{\spacestart}{\nextspace}{\nextspace}{}\<[24]%
\>[24]{}(x{\spacestart}{\nextspace}:{\spacestart}{\nextspace}xs){\spacestart}{\nextspace}{\nextspace}{}\<[34]%
\>[34]{}={\spacestart}{\nextspace}{\nextspace}{}\<[34E]%
\>[37]{}\textbf{case}{\spacestart}{\nextspace}create{\spacestart}{\nextspace}s{\spacestart}{\nextspace}xs{\spacestart}{\nextspace}\textbf{of}{\spacestart}{\nextspace}(r{\spacestart}{\nextspace}:*:{\spacestart}{\nextspace}ys){\spacestart}{\nextspace}->{\spacestart}{\nextspace}{}\<[E]%
\\[-0.3ex]%
\>[B]{}{\spacestart}{\nextspace}{\nextspace}{\nextspace}{\nextspace}{\nextspace}{\nextspace}{\nextspace}{\nextspace}{\nextspace}{\nextspace}{\nextspace}{\nextspace}{\nextspace}{\nextspace}{\nextspace}{\nextspace}{\nextspace}{\nextspace}{\nextspace}{\nextspace}{\nextspace}{\nextspace}{\nextspace}{\nextspace}{\nextspace}{\nextspace}{\nextspace}{\nextspace}{\nextspace}{\nextspace}{\nextspace}{\nextspace}{\nextspace}{\nextspace}{\nextspace}{\nextspace}{}\<[37]%
\>[37]{}\textbf{let}{\spacestart}{\nextspace}!t'{\spacestart}{\nextspace}={\spacestart}{\nextspace}link{\spacestart}{\nextspace}x{\spacestart}{\nextspace}l{\spacestart}{\nextspace}r{\spacestart}{\nextspace}\textbf{in}{\spacestart}{\nextspace}go{\spacestart}{\nextspace}(s{\spacestart}{\nextspace}{\textasciigrave}shiftL{\textasciigrave}{\spacestart}{\nextspace}1){\spacestart}{\nextspace}t'{\spacestart}{\nextspace}ys{\spacestart}{\nextspace}{}\<[E]%
\\[\blanklineskip]%
\>[B]{}{\spacestart}{\nextspace}{\nextspace}{\nextspace}{\nextspace}{\nextspace}{\nextspace}{\nextspace}{\nextspace}{\nextspace}{\nextspace}{\nextspace}{\nextspace}{}\<[13]%
\>[13]{}create{\spacestart}{\nextspace}{\nextspace}{\nextspace}{}\<[22]%
\>[22]{}!\char95 {\spacestart}{\nextspace}{\nextspace}{}\<[22E]%
\>[26]{}[]{\spacestart}{\nextspace}={\spacestart}{\nextspace}(Tip{\spacestart}{\nextspace}:*:{\spacestart}{\nextspace}[]){\spacestart}{\nextspace}{}\<[E]%
\\[-0.3ex]%
\>[B]{}{\spacestart}{\nextspace}{\nextspace}{\nextspace}{\nextspace}{\nextspace}{\nextspace}{\nextspace}{\nextspace}{\nextspace}{\nextspace}{\nextspace}{\nextspace}{}\<[13]%
\>[13]{}create{\spacestart}{\nextspace}{\nextspace}{\nextspace}{}\<[22]%
\>[22]{}s{\spacestart}{\nextspace}{\nextspace}{\nextspace}{}\<[22E]%
\>[26]{}xs@(x{\spacestart}{\nextspace}:{\spacestart}{\nextspace}xs'){\spacestart}{\nextspace}{}\<[E]%
\\[-0.3ex]%
\>[B]{}{\spacestart}{\nextspace}{\nextspace}{\nextspace}{\nextspace}{\nextspace}{\nextspace}{\nextspace}{\nextspace}{\nextspace}{\nextspace}{\nextspace}{\nextspace}{}\<[13]%
\>[13]{}~~{\spacestart}{\nextspace}{\nextspace}{}\<[18]%
\>[18]{}|{\spacestart}{\nextspace}s{\spacestart}{\nextspace}=={\spacestart}{\nextspace}1{\spacestart}{\nextspace}{\nextspace}{\nextspace}{\nextspace}{\nextspace}{}\<[31]%
\>[31]{}={\spacestart}{\nextspace}{\nextspace}{\nextspace}{}\<[31E]%
\>[35]{}(Bin{\spacestart}{\nextspace}1{\spacestart}{\nextspace}x{\spacestart}{\nextspace}Tip{\spacestart}{\nextspace}Tip{\spacestart}{\nextspace}:*:{\spacestart}{\nextspace}xs'){\spacestart}{\nextspace}{}\<[E]%
\\[-0.3ex]%
\>[B]{}{\spacestart}{\nextspace}{\nextspace}{\nextspace}{\nextspace}{\nextspace}{\nextspace}{\nextspace}{\nextspace}{\nextspace}{\nextspace}{\nextspace}{\nextspace}{}\<[13]%
\>[13]{}~~{\spacestart}{\nextspace}{\nextspace}{}\<[18]%
\>[18]{}|{\spacestart}{\nextspace}otherwise{\spacestart}{\nextspace}{\nextspace}{}\<[31]%
\>[31]{}={\spacestart}{\nextspace}{\nextspace}{\nextspace}{}\<[31E]%
\>[35]{}\textbf{case}{\spacestart}{\nextspace}create{\spacestart}{\nextspace}(s{\spacestart}{\nextspace}{\textasciigrave}shiftR{\textasciigrave}{\spacestart}{\nextspace}1){\spacestart}{\nextspace}xs{\spacestart}{\nextspace}\textbf{of}{\spacestart}{\nextspace}{}\<[E]%
\\[-0.3ex]%
\>[B]{}{\spacestart}{\nextspace}{\nextspace}{\nextspace}{\nextspace}{\nextspace}{\nextspace}{\nextspace}{\nextspace}{\nextspace}{\nextspace}{\nextspace}{\nextspace}{\nextspace}{\nextspace}{\nextspace}{\nextspace}{\nextspace}{\nextspace}{\nextspace}{\nextspace}{\nextspace}{\nextspace}{\nextspace}{\nextspace}{\nextspace}{\nextspace}{\nextspace}{\nextspace}{\nextspace}{\nextspace}{\nextspace}{\nextspace}{\nextspace}{\nextspace}{}\<[35]%
\>[35]{}~~{\spacestart}{\nextspace}{\nextspace}{}\<[40]%
\>[40]{}res@(\char95 {\spacestart}{\nextspace}:*:{\spacestart}{\nextspace}[]){\spacestart}{\nextspace}->{\spacestart}{\nextspace}res{\spacestart}{\nextspace}{}\<[E]%
\\[-0.3ex]%
\>[B]{}{\spacestart}{\nextspace}{\nextspace}{\nextspace}{\nextspace}{\nextspace}{\nextspace}{\nextspace}{\nextspace}{\nextspace}{\nextspace}{\nextspace}{\nextspace}{\nextspace}{\nextspace}{\nextspace}{\nextspace}{\nextspace}{\nextspace}{\nextspace}{\nextspace}{\nextspace}{\nextspace}{\nextspace}{\nextspace}{\nextspace}{\nextspace}{\nextspace}{\nextspace}{\nextspace}{\nextspace}{\nextspace}{\nextspace}{\nextspace}{\nextspace}{\nextspace}{\nextspace}{\nextspace}{\nextspace}{\nextspace}{}\<[40]%
\>[40]{}(l{\spacestart}{\nextspace}:*:{\spacestart}{\nextspace}(y:ys)){\spacestart}{\nextspace}->{\spacestart}{\nextspace}\textbf{case}{\spacestart}{\nextspace}create{\spacestart}{\nextspace}(s{\spacestart}{\nextspace}{\textasciigrave}shiftR{\textasciigrave}{\spacestart}{\nextspace}1){\spacestart}{\nextspace}ys{\spacestart}{\nextspace}\textbf{of}{\spacestart}{\nextspace}{}\<[E]%
\\[-0.3ex]%
\>[B]{}{\spacestart}{\nextspace}{\nextspace}{\nextspace}{\nextspace}{\nextspace}{\nextspace}{\nextspace}{\nextspace}{\nextspace}{\nextspace}{\nextspace}{\nextspace}{\nextspace}{\nextspace}{\nextspace}{\nextspace}{\nextspace}{\nextspace}{\nextspace}{\nextspace}{\nextspace}{\nextspace}{\nextspace}{\nextspace}{\nextspace}{\nextspace}{\nextspace}{\nextspace}{\nextspace}{\nextspace}{\nextspace}{\nextspace}{\nextspace}{\nextspace}{\nextspace}{\nextspace}{\nextspace}{\nextspace}{\nextspace}{}\<[40]%
\>[40]{}~~{\spacestart}{\nextspace}(r{\spacestart}{\nextspace}:*:{\spacestart}{\nextspace}zs){\spacestart}{\nextspace}->{\spacestart}{\nextspace}(link{\spacestart}{\nextspace}y{\spacestart}{\nextspace}l{\spacestart}{\nextspace}r{\spacestart}{\nextspace}:*:{\spacestart}{\nextspace}zs){}\<[E]%
\ColumnHook
\end{hscode}\resethooks\endgroup
\caption{Recursion styles}
\label{fig:recursion-patterns}
\end{figure}

In order to prove the correctness of \ensuretext{\straighttick\ttfamily{}Set} and \ensuretext{\straighttick\ttfamily{}IntSet}, we must deal with
termination.  There are two reasons for this.  First, we intrinsically want to
prove that none of the functions provided by \texttt{containers} go into an infinite
loop. Second, Coq requires that all defined functions are terminating, as
unrestricted recursion would lead to logical inconsistencies.
Depending on how involved the termination argument for a given function is, we
use one of the following four approaches.

\subsubsection{Obvious structural recursion}
By default, \texttt{hs\char45{}to\char45{}coq} implements recursive functions directly using Coq’s
\ensuretext{\straighttick\ttfamily{}\textbf{fix}} keyword.  This works smoothly for primitive structural recursion; indeed,
a majority of the recursive functions that we encountered, such as \ensuretext{\straighttick\ttfamily{}member} in
\cref{fig:setdef}, were of this form and required no further attention.

\subsubsection{Almost-structural recursion}

Another common recursion pattern can be found in binary operations such as
\ensuretext{\straighttick\ttfamily{}link} in \cref{fig:recursion-patterns}. Here, every recursive call shrinks
\emph{either or both} of its arguments to immediate subterms of the originals,
leaving the others unchanged.  This almost-structural recursion is already
beyond the capabilities of Coq’s termination checker. We therefore instruct
\texttt{hs\char45{}to\char45{}coq} to use Coq's \ensuretext{\straighttick\ttfamily{}\textbf{Program}{\spacestart}{\nextspace}\textbf{Fixpoint}} command to translate these functions
in terms of well-founded recursion by adding the edits
\begingroup\straighttick\begin{hscode}\SaveRestoreHook
\column{B}{@{}>{\hspre}l<{\hspost}@{}}%
\column{E}{@{}>{\hspre}l<{\hspost}@{}}%
\>[B]{}\textbf{termination}{\spacestart}{\nextspace}link{\spacestart}{\nextspace}{\char123}measure{\spacestart}{\nextspace}(size\char95 nat{\spacestart}{\nextspace}arg\char95 0\char95 \char95 {\spacestart}{\nextspace}+{\spacestart}{\nextspace}size\char95 nat{\spacestart}{\nextspace}arg\char95 1\char95 \char95 ){\char125}{\spacestart}{\nextspace}{}\<[E]%
\\[-0.3ex]%
\>[B]{}\textbf{obligations}{\spacestart}{\nextspace}link{\spacestart}{\nextspace}termination\char95 by\char95 omega{}\<[E]%
\ColumnHook
\end{hscode}\resethooks\endgroup
This specifies both:
\begin{compactenum}
\item The termination measure, which is the sum of the sizes of the
  arguments (we defined the function \ensuretext{\straighttick\ttfamily{}size\char95 nat{\spacestart}{\nextspace}:{\spacestart}{\nextspace}IntSet{\spacestart}{\nextspace}->{\spacestart}{\nextspace}nat}).
\item The termination proof that the measure decreases on every call.  This is
  represented as the Coq tactic \ensuretext{\straighttick\ttfamily{}termination\char95 by\char95 omega}, which is a thin wrapper
  we defined around \ensuretext{\straighttick\ttfamily{}omega}, a Coq tactic to decide linear integer arithmetic.
\end{compactenum}
We can use these edits (with \ensuretext{\straighttick\ttfamily{}size\char95 nat} and \ensuretext{\straighttick\ttfamily{}termination\char95 by\char95 omega}) to get to
Coq accepts such recursive definitions without the need for any further proofs
or manual intervention.

\subsubsection{Well-founded recursion}
A small number of functions recurse in a non-structural way, such as \ensuretext{\straighttick\ttfamily{}foldlBits}
in \cref{fig:recursion-patterns}, which recurses on the input after clearing the
least-significant set bit (\ensuretext{\straighttick\ttfamily{}bm{\spacestart}{\nextspace}{\textasciigrave}xor{\textasciigrave}{\spacestart}{\nextspace}(lowestBitMask{\spacestart}{\nextspace}bm)}).  We can handle this
sort of logic using the same machinery as before, but now we have to write a
\emph{specialized} termination tactic and declare it in the \texttt{obligations} hint.
To do so, we need to prove necessary lemmas \emph{before} translating the
Haskell module in question.
\iflong
In particular, our lemmas cannot mention functions
that are defined in the translated Haskell module. We can do that by exploiting
the fact that Coq’s value definitions are not generative: since the function
\ensuretext{\straighttick\ttfamily{}lowestBitMask} is defined elsewhere to be %
\ensuretext{\straighttick\ttfamily{}lowestBitMask{\spacestart}{\nextspace}(bm{\spacestart}{\nextspace}:{\spacestart}{\nextspace}N){\spacestart}{\nextspace}={\spacestart}{\nextspace}2{\spacestart}{\nextspace}\char94 {\spacestart}{\nextspace}N\char95 ctz{\spacestart}{\nextspace}bm}, we can still use lemmas such as
\ensuretext{\straighttick\ttfamily{}foldlBits\char95 proof} to reason about \ensuretext{\straighttick\ttfamily{}lowestBitMask}:
\begingroup\straighttick\begin{hscode}\SaveRestoreHook
\column{B}{@{}>{\hspre}l<{\hspost}@{}}%
\column{6}{@{}>{\hspre}l<{\hspost}@{}}%
\column{E}{@{}>{\hspre}l<{\hspost}@{}}%
\>[B]{}\textbf{Lemma}{\spacestart}{\nextspace}foldlBits\char95 proof:{\spacestart}{\nextspace}\textbf{forall}{\spacestart}{\nextspace}a,{\spacestart}{\nextspace}{}\<[E]%
\\[-0.3ex]%
\>[B]{}~~{\spacestart}{\nextspace}{\nextspace}{}\<[6]%
\>[6]{}N.eqb{\spacestart}{\nextspace}a{\spacestart}{\nextspace}0{\spacestart}{\nextspace}={\spacestart}{\nextspace}false{\spacestart}{\nextspace}->{\spacestart}{\nextspace}N.to\char95 nat{\spacestart}{\nextspace}(N.lxor{\spacestart}{\nextspace}a{\spacestart}{\nextspace}(2{\spacestart}{\nextspace}\char94 {\spacestart}{\nextspace}N\char95 ctz{\spacestart}{\nextspace}a)){\spacestart}{\nextspace}<{\spacestart}{\nextspace}N.to\char95 nat{\spacestart}{\nextspace}a.{}\<[E]%
\ColumnHook
\end{hscode}\resethooks\endgroup
\fi

Coq’s \ensuretext{\straighttick\ttfamily{}\textbf{Program}{\spacestart}{\nextspace}\textbf{Fixpoint}} only supports top-level functions, but we frequently
encounter local recursive functions -- the \ensuretext{\straighttick\ttfamily{}go} idiom, as seen here.  To support
this, we extended \texttt{hs\char45{}to\char45{}coq} to offer some of the convenience provided by
\ensuretext{\straighttick\ttfamily{}\textbf{Program}{\spacestart}{\nextspace}\textbf{Fixpoint}} by translating local recursive functions using the same
well-founded-recursion-based fixed-point combinator as \ensuretext{\straighttick\ttfamily{}\textbf{Program}{\spacestart}{\nextspace}\textbf{Fixpoint}}.

\subsubsection{Deferred recursion}
\label{sec:deferredFix}

Finally, we encounter some functions that require elaborate termination
arguments, such as \ensuretext{\straighttick\ttfamily{}fromDistinctAscList} in \cref{fig:recursion-patterns}. It
has two local recursive functions, \ensuretext{\straighttick\ttfamily{}go} and \ensuretext{\straighttick\ttfamily{}create}, and to convince
ourselves that \ensuretext{\straighttick\ttfamily{}fromDistinctAscList} is indeed terminating, we have to
reason as follows:

\begin{quote}
  The function \ensuretext{\straighttick\ttfamily{}create} bitshifts its first argument to the right upon each
  recursive call, until the argument is \ensuretext{\straighttick\ttfamily{}1}. Ergo, it is terminating, but only
  for positive input -- it clearly loops if \ensuretext{\straighttick\ttfamily{}x} is \ensuretext{\straighttick\ttfamily{}0}.  The function \ensuretext{\straighttick\ttfamily{}go}
  recurses on smaller lists as its third argument, but to see that, we first
  have to convince ourself that the list in the tuple returned by \ensuretext{\straighttick\ttfamily{}create} is
  never larger than the list passed to \ensuretext{\straighttick\ttfamily{}create}. Also, \ensuretext{\straighttick\ttfamily{}go} calls \ensuretext{\straighttick\ttfamily{}create}, so
  we need to ensure that the \ensuretext{\straighttick\ttfamily{}s} passed to it is positive.  The \ensuretext{\straighttick\ttfamily{}go} function
  bitshifts \ensuretext{\straighttick\ttfamily{}s} to the left at every call, so if \ensuretext{\straighttick\ttfamily{}go} is called with a positive
  \ensuretext{\straighttick\ttfamily{}s}, then \ensuretext{\straighttick\ttfamily{}s} will remain positive in recursive calls. Finally, we see that
  \ensuretext{\straighttick\ttfamily{}fromDistinctAscList} calls \ensuretext{\straighttick\ttfamily{}go} with \ensuretext{\straighttick\ttfamily{}s} equal to \ensuretext{\straighttick\ttfamily{}1}, which is positive, so
  we can conclude that \ensuretext{\straighttick\ttfamily{}fromDistinctAscList} terminates.
\end{quote}

If we wanted to convince Coq of this termination pattern, we would have to turn
\ensuretext{\straighttick\ttfamily{}create} and \ensuretext{\straighttick\ttfamily{}go} into top-level definitions, change their types to rule out
invalid (non-positive) inputs, define \ensuretext{\straighttick\ttfamily{}create} using \ensuretext{\straighttick\ttfamily{}\textbf{Program}{\spacestart}{\nextspace}\textbf{Fixpoint}}, and
provide an explicit termination argument by well-founded recursion. Then we
could prove that \ensuretext{\straighttick\ttfamily{}create} preserves the list lengths, which we need to define
\ensuretext{\straighttick\ttfamily{}go}, again using \ensuretext{\straighttick\ttfamily{}\textbf{Program}{\spacestart}{\nextspace}\textbf{Fixpoint}}. This is certainly possible, but it is not
simple, especially in an automatic translation.

For hard cases like these we resort to \emph{deferred termination checking}, a
feature that we added to \texttt{hs\char45{}to\char45{}coq}. We can instruct it  to use the following
axiom as a very permissive
fixed-point combinator and translate the code of \ensuretext{\straighttick\ttfamily{}fromDistinctAscList}
essentially unchanged:
\begingroup\straighttick\begin{hscode}\SaveRestoreHook
\column{B}{@{}>{\hspre}l<{\hspost}@{}}%
\column{E}{@{}>{\hspre}l<{\hspost}@{}}%
\>[B]{}\textbf{Axiom}{\spacestart}{\nextspace}deferredFix:{\spacestart}{\nextspace}\textbf{forall}{\spacestart}{\nextspace}{\char123}a{\spacestart}{\nextspace}r{\char125}{\spacestart}{\nextspace}{\textasciigrave}{\char123}Default{\spacestart}{\nextspace}r{\char125},{\spacestart}{\nextspace}((a{\spacestart}{\nextspace}->{\spacestart}{\nextspace}r){\spacestart}{\nextspace}->{\spacestart}{\nextspace}(a{\spacestart}{\nextspace}->{\spacestart}{\nextspace}r)){\spacestart}{\nextspace}->{\spacestart}{\nextspace}a{\spacestart}{\nextspace}->{\spacestart}{\nextspace}r.{}\<[E]%
\ColumnHook
\end{hscode}\resethooks\endgroup
On its own, \ensuretext{\straighttick\ttfamily{}deferredFix} does not do anything; it merely sits in the translated
code applied to the original function body.
It is consistent, since its type could be implemented by a function that 
always returns \ensuretext{\straighttick\ttfamily{}default} (see  \cref{sec:partial}).
And it does not prevent the user from running extracted
code -- we can extract this axiom to the target language’s unrestricted fixpoint
operator (e.g., \ensuretext{\straighttick\ttfamily{}Data.Function.fix} in Haskell),
although this costs us the guarantee that the extracted code is terminating.

When we come to verifying something about a function that is defined using
\ensuretext{\straighttick\ttfamily{}deferredFix}, we need to give it meaning. We do so using a second axiom,
\ensuretext{\straighttick\ttfamily{}deferredFix\char95 eq\char95 on}, which states that for any well-founded relation \ensuretext{\straighttick\ttfamily{}R}
(\ensuretext{\straighttick\ttfamily{}well\char95 founded{\spacestart}{\nextspace}R}), if the recursive calls in \ensuretext{\straighttick\ttfamily{}f} are always at values that are
strictly \ensuretext{\straighttick\ttfamily{}R}-smaller than the input (\ensuretext{\straighttick\ttfamily{}recurses\char95 on{\spacestart}{\nextspace}R}), then we may unroll
the fixpoint of \ensuretext{\straighttick\ttfamily{}f}:
\begingroup\straighttick\begin{hscode}\SaveRestoreHook
\column{B}{@{}>{\hspre}l<{\hspost}@{}}%
\column{6}{@{}>{\hspre}l<{\hspost}@{}}%
\column{E}{@{}>{\hspre}l<{\hspost}@{}}%
\>[B]{}\textbf{Definition}{\spacestart}{\nextspace}recurses\char95 on{\spacestart}{\nextspace}{\char123}a{\spacestart}{\nextspace}b{\char125}{\spacestart}{\nextspace}{}\<[E]%
\\[-0.3ex]%
\>[B]{}~~{\spacestart}{\nextspace}{\nextspace}{}\<[6]%
\>[6]{}(P{\spacestart}{\nextspace}:{\spacestart}{\nextspace}a{\spacestart}{\nextspace}->{\spacestart}{\nextspace}\textbf{Prop}){\spacestart}{\nextspace}(R{\spacestart}{\nextspace}:{\spacestart}{\nextspace}a{\spacestart}{\nextspace}->{\spacestart}{\nextspace}a{\spacestart}{\nextspace}->{\spacestart}{\nextspace}\textbf{Prop}){\spacestart}{\nextspace}(f{\spacestart}{\nextspace}:{\spacestart}{\nextspace}(a{\spacestart}{\nextspace}->{\spacestart}{\nextspace}b){\spacestart}{\nextspace}->{\spacestart}{\nextspace}(a{\spacestart}{\nextspace}->{\spacestart}{\nextspace}b)){\spacestart}{\nextspace}{}\<[E]%
\\[-0.3ex]%
\>[B]{}{\spacestart}{\nextspace}{\nextspace}{\nextspace}{\nextspace}{\nextspace}{}\<[6]%
\>[6]{}:={\spacestart}{\nextspace}\textbf{forall}{\spacestart}{\nextspace}g{\spacestart}{\nextspace}h{\spacestart}{\nextspace}x,{\spacestart}{\nextspace}P{\spacestart}{\nextspace}x{\spacestart}{\nextspace}->{\spacestart}{\nextspace}(\textbf{forall}{\spacestart}{\nextspace}y,{\spacestart}{\nextspace}P{\spacestart}{\nextspace}y{\spacestart}{\nextspace}->{\spacestart}{\nextspace}R{\spacestart}{\nextspace}y{\spacestart}{\nextspace}x{\spacestart}{\nextspace}->{\spacestart}{\nextspace}g{\spacestart}{\nextspace}y{\spacestart}{\nextspace}={\spacestart}{\nextspace}h{\spacestart}{\nextspace}y){\spacestart}{\nextspace}->{\spacestart}{\nextspace}f{\spacestart}{\nextspace}g{\spacestart}{\nextspace}x{\spacestart}{\nextspace}={\spacestart}{\nextspace}f{\spacestart}{\nextspace}h{\spacestart}{\nextspace}x.{\spacestart}{\nextspace}{}\<[E]%
\\[\blanklineskip]%
\>[B]{}\textbf{Axiom}{\spacestart}{\nextspace}deferredFix\char95 eq\char95 on:{\spacestart}{\nextspace}\textbf{forall}{\spacestart}{\nextspace}{\char123}a{\spacestart}{\nextspace}b{\char125}{\spacestart}{\nextspace}{}\<[E]%
\\[-0.3ex]%
\>[B]{}~~{\spacestart}{\nextspace}{\textasciigrave}{\char123}Default{\spacestart}{\nextspace}b{\char125}{\spacestart}{\nextspace}(f{\spacestart}{\nextspace}:{\spacestart}{\nextspace}(a{\spacestart}{\nextspace}->{\spacestart}{\nextspace}b){\spacestart}{\nextspace}->{\spacestart}{\nextspace}(a{\spacestart}{\nextspace}->{\spacestart}{\nextspace}b)){\spacestart}{\nextspace}(P{\spacestart}{\nextspace}:{\spacestart}{\nextspace}a{\spacestart}{\nextspace}->{\spacestart}{\nextspace}\textbf{Prop}){\spacestart}{\nextspace}(R{\spacestart}{\nextspace}:{\spacestart}{\nextspace}a{\spacestart}{\nextspace}->{\spacestart}{\nextspace}a{\spacestart}{\nextspace}->{\spacestart}{\nextspace}\textbf{Prop}),{\spacestart}{\nextspace}{}\<[E]%
\\[-0.3ex]%
\>[B]{}{\spacestart}{\nextspace}{\nextspace}{\nextspace}{\nextspace}{\nextspace}{}\<[6]%
\>[6]{}well\char95 founded{\spacestart}{\nextspace}R{\spacestart}{\nextspace}->{\spacestart}{\nextspace}recurses\char95 on{\spacestart}{\nextspace}P{\spacestart}{\nextspace}R{\spacestart}{\nextspace}f{\spacestart}{\nextspace}->{\spacestart}{\nextspace}{}\<[E]%
\\[-0.3ex]%
\>[B]{}{\spacestart}{\nextspace}{\nextspace}{\nextspace}{\nextspace}{\nextspace}{}\<[6]%
\>[6]{}\textbf{forall}{\spacestart}{\nextspace}x,{\spacestart}{\nextspace}P{\spacestart}{\nextspace}x{\spacestart}{\nextspace}->{\spacestart}{\nextspace}deferredFix{\spacestart}{\nextspace}f{\spacestart}{\nextspace}x{\spacestart}{\nextspace}={\spacestart}{\nextspace}f{\spacestart}{\nextspace}(deferredFix{\spacestart}{\nextspace}f){\spacestart}{\nextspace}x.{}\<[E]%
\ColumnHook
\end{hscode}\resethooks\endgroup
The predicate \ensuretext{\straighttick\ttfamily{}P{\spacestart}{\nextspace}:{\spacestart}{\nextspace}a{\spacestart}{\nextspace}->{\spacestart}{\nextspace}\textbf{Prop}} allows us to restrict the domain to inputs for which the
function is actually terminating -- crucial for \ensuretext{\straighttick\ttfamily{}go} and \ensuretext{\straighttick\ttfamily{}create}.

The predicate \ensuretext{\straighttick\ttfamily{}recurses\char95 on{\spacestart}{\nextspace}P{\spacestart}{\nextspace}R{\spacestart}{\nextspace}f} characterizes the recursion pattern of \ensuretext{\straighttick\ttfamily{}f}, but does
so in a very extensional way and only considers recursive calls that can
actually affect the result of the function.  For instance, it would consider a
recursive function defined by \ensuretext{\straighttick\ttfamily{}f{\spacestart}{\nextspace}n{\spacestart}{\nextspace}={\spacestart}{\nextspace}\textbf{if}{\spacestart}{\nextspace}f{\spacestart}{\nextspace}n{\spacestart}{\nextspace}\textbf{then}{\spacestart}{\nextspace}true{\spacestart}{\nextspace}\textbf{else}{\spacestart}{\nextspace}true} to be terminating.

We can use this non-trivial axiom without losing too much sleep, because we
can actually implement \ensuretext{\straighttick\ttfamily{}deferredFix} and \ensuretext{\straighttick\ttfamily{}deferredFix\char95 eq\char95 on} in terms of
classical logic and the axiom of choice (as provided by the Coq module
\ensuretext{\straighttick\ttfamily{}Coq.Logic.Epsilon}).%
\footnote{In the file \fileHsToCoq{base/GHC/DeferredFixImpl.v}}
This means that both axioms are consistent with plain
Coq. We do not know if
\ensuretext{\straighttick\ttfamily{}deferredFix\char95 eq\char95 on} is strictly weaker than classical choice, so users of
\texttt{hs\char45{}to\char45{}coq} who want to combine the output of \texttt{hs\char45{}to\char45{}coq} with developments
known to be inconsistent with classical choice (e.g.,\ homotopy type theory)
should be cautious.

Pragmatically, working with \ensuretext{\straighttick\ttfamily{}deferredFix} is quite convenient, as we can prove
termination together with the other specifications about these functions. The
actual termination proofs themselves are not fundamentally different from the
proof obligations that \ensuretext{\straighttick\ttfamily{}\textbf{Program}{\spacestart}{\nextspace}\textbf{Fixpoint}} would generate for us and -- although
not needed in this example -- can be carried out even for nested recursion
through higher-order functions like \ensuretext{\straighttick\ttfamily{}map}.

This approach to defining recursive functions extensionality was inspired by Isabelle’s function package \citep{isabelle-function}.

\subsection{Translating Haskell tests to Coq types}
\label{sec:translating-quickcheck}

When we translate code, we usually want to preserve the semantics of the code as
much as possible. Things are very different when we translate the QuickCheck
tests defined in the \texttt{containers} test suite, as we discuss in
\cref{sec:specs-tests}: whereas the semantics of the test suite in Haskell is a
program that creates random input to use as input test executable properties, we
want to re-interpret these properties as logical propositions in Coq.  Put
differently, we are turning executable code into \emph{types}.

QuickCheck’s API provides types and type classes for writing property-based
tests. In particular, it defines an opaque type \ensuretext{\straighttick\ttfamily{}Property} that describes
properties that can checked using randomized testing, a type class \ensuretext{\straighttick\ttfamily{}Testable} that
converts various testable types into a \ensuretext{\straighttick\ttfamily{}Property}, and a type constructor \ensuretext{\straighttick\ttfamily{}Gen}
that describes how to generate a random value. These are combined, for instance,
in the QuickCheck combinator
\begingroup\straighttick\begin{hscode}\SaveRestoreHook
\column{B}{@{}>{\hspre}l<{\hspost}@{}}%
\column{E}{@{}>{\hspre}l<{\hspost}@{}}%
\>[B]{}forAll{\spacestart}{\nextspace}::{\spacestart}{\nextspace}(Show{\spacestart}{\nextspace}a,{\spacestart}{\nextspace}Testable{\spacestart}{\nextspace}prop){\spacestart}{\nextspace}=>{\spacestart}{\nextspace}Gen{\spacestart}{\nextspace}a{\spacestart}{\nextspace}->{\spacestart}{\nextspace}(a{\spacestart}{\nextspace}->{\spacestart}{\nextspace}prop){\spacestart}{\nextspace}->{\spacestart}{\nextspace}Property{}\<[E]%
\ColumnHook
\end{hscode}\resethooks\endgroup
which tests the result of a function on inputs generated by the given generator.

Since we want to \emph{prove}, not \emph{test}, these properties, we do not
convert the QuickCheck implementation to Coq. Instead, we write a small Coq
module that provides the necessary pieces of the interface of \ensuretext{\straighttick\ttfamily{}Test.QuickCheck},
but interprets these types and functions in terms of Coq propositions. In
particular:
\begin{compactitem}
\item We use Coq’s non-computational type of propositions, \ensuretext{\straighttick\ttfamily{}\textbf{Prop}}, instead of QuickCheck’s
  computational \ensuretext{\straighttick\ttfamily{}Property};
\item \ensuretext{\straighttick\ttfamily{}Gen{\spacestart}{\nextspace}a} is simply a wrapper around a logical predicate on \ensuretext{\straighttick\ttfamily{}a}; and
\item \ensuretext{\straighttick\ttfamily{}forAll} quantifies (using Coq's \ensuretext{\straighttick\ttfamily{}\textbf{forall}}) over the type \ensuretext{\straighttick\ttfamily{}a}, and ensures
  that the given function -- now a predicate -- holds for all members of \ensuretext{\straighttick\ttfamily{}a}
  that satisfy the given “generator”.
\end{compactitem}
Concretely, this leads to the following adapted Coq code:
\begingroup\straighttick\begin{hscode}\SaveRestoreHook
\column{B}{@{}>{\hspre}l<{\hspost}@{}}%
\column{6}{@{}>{\hspre}l<{\hspost}@{}}%
\column{E}{@{}>{\hspre}l<{\hspost}@{}}%
\>[B]{}\textbf{Record}{\spacestart}{\nextspace}Gen{\spacestart}{\nextspace}a{\spacestart}{\nextspace}:={\spacestart}{\nextspace}MkGen{\spacestart}{\nextspace}{\char123}{\spacestart}{\nextspace}unGen{\spacestart}{\nextspace}:{\spacestart}{\nextspace}a{\spacestart}{\nextspace}->{\spacestart}{\nextspace}\textbf{Prop}{\spacestart}{\nextspace}{\char125}.{\spacestart}{\nextspace}{}\<[E]%
\\[-0.3ex]%
\>[B]{}\textbf{Class}{\spacestart}{\nextspace}Testable{\spacestart}{\nextspace}(a{\spacestart}{\nextspace}:{\spacestart}{\nextspace}\textbf{Type}){\spacestart}{\nextspace}:={\spacestart}{\nextspace}{\char123}{\spacestart}{\nextspace}toProp{\spacestart}{\nextspace}:{\spacestart}{\nextspace}a{\spacestart}{\nextspace}->{\spacestart}{\nextspace}\textbf{Prop}{\spacestart}{\nextspace}{\char125}.{\spacestart}{\nextspace}{}\<[E]%
\\[-0.3ex]%
\>[B]{}\textbf{Definition}{\spacestart}{\nextspace}forAll{\spacestart}{\nextspace}{\char123}a{\spacestart}{\nextspace}prop{\char125}{\spacestart}{\nextspace}{\textasciigrave}{\char123}Testable{\spacestart}{\nextspace}prop{\char125}{\spacestart}{\nextspace}(g{\spacestart}{\nextspace}:{\spacestart}{\nextspace}Gen{\spacestart}{\nextspace}a){\spacestart}{\nextspace}(p{\spacestart}{\nextspace}:{\spacestart}{\nextspace}a{\spacestart}{\nextspace}->{\spacestart}{\nextspace}prop){\spacestart}{\nextspace}:{\spacestart}{\nextspace}\textbf{Prop}{\spacestart}{\nextspace}:={\spacestart}{\nextspace}{}\<[E]%
\\[-0.3ex]%
\>[B]{}~~{\spacestart}{\nextspace}{\nextspace}{}\<[6]%
\>[6]{}\textbf{forall}{\spacestart}{\nextspace}(x{\spacestart}{\nextspace}:{\spacestart}{\nextspace}a),{\spacestart}{\nextspace}unGen{\spacestart}{\nextspace}g{\spacestart}{\nextspace}x{\spacestart}{\nextspace}->{\spacestart}{\nextspace}toProp{\spacestart}{\nextspace}(p{\spacestart}{\nextspace}x).{}\<[E]%
\ColumnHook
\end{hscode}\resethooks\endgroup
We provide similar translations for QuickCheck’s operators \ensuretext{\straighttick\ttfamily{}===}, \ensuretext{\straighttick\ttfamily{}==>}, \ensuretext{\straighttick\ttfamily{}.\&\&.}
and \ensuretext{\straighttick\ttfamily{}.||.}, and we replace generators such as
\ensuretext{\straighttick\ttfamily{}choose{\spacestart}{\nextspace}::{\spacestart}{\nextspace}Random{\spacestart}{\nextspace}a{\spacestart}{\nextspace}=>{\spacestart}{\nextspace}(a,{\spacestart}{\nextspace}a){\spacestart}{\nextspace}->{\spacestart}{\nextspace}Gen{\spacestart}{\nextspace}a} with their corresponding
predicates.

With this module in place, \texttt{hs\char45{}to\char45{}coq} translates the test suite into a “proof
suite”.  As we saw in \cref{sec:specs-tests}, a test like \ensuretext{\straighttick\ttfamily{}prop\char95 UnionAssoc}
is now a definition of a Coq proposition, that is to say a type, and can be
used as the type of a theorem:
\begingroup\straighttick\begin{hscode}\SaveRestoreHook
\column{B}{@{}>{\hspre}l<{\hspost}@{}}%
\column{E}{@{}>{\hspre}l<{\hspost}@{}}%
\>[B]{}\textbf{Theorem}{\spacestart}{\nextspace}thm\char95 UnionAssoc{\spacestart}{\nextspace}:{\spacestart}{\nextspace}toProp{\spacestart}{\nextspace}prop\char95 UnionAssoc.{}\<[E]%
\ColumnHook
\end{hscode}\resethooks\endgroup

\section{Contributions to \LITcontainers: Theory and Practice}
\label{sec:additional}

We chose the \texttt{Set} and \texttt{IntSet} modules of the \texttt{containers} library as our
target because they nicely represent the kind of Haskell code that we want to
see verified in practice.
Nevertheless, the deep understanding required to verify these modules led
to new insights into the algorithms themselves and to improvements to their
Haskell implementation.

\subsection{New insight into the verification of weight-balanced trees}
\label{sec:balanceL}

\newcommand{\bal}{\operatorname{bal}}
\newcommand{\balNR}{\operatorname{bal}_{\text{NR}}}
\newcommand{\balA}{\operatorname{bal}}
\newcommand{\balAins}{\operatorname{bal}^\ast}

The data structure underlying \ensuretext{\straighttick\ttfamily{}Set} and \ensuretext{\straighttick\ttfamily{}Map} was originally presented by
\citet{nievergelt}.
It is a binary search tree with the invariant that if $s_1$ is the
size of the left subtree and $s_2$ the size of the right subtree of a node, then
\[
\balNR(s_1,s_2) \coloneqq (s_1 + 1)\le \delta \cdot (s_2+1) \wedge (s_2+1) \le \delta \cdot (s_1 + 1)
\]
holds for a balancing tuning parameter $\delta$. In \citeyear{adams-tr}, \citeauthor{adams-tr} suggested a variant of the balancing condition, namely
\[
\balA(s_1,s_2) \coloneqq s_1 + s_2 \le 1 \vee (s_1 \le \delta \cdot s_2 \wedge s_2 \le \delta \cdot s_1).
\]
The conditions are very similar, but not equivalent: the former allows, for example, $\delta = 3$, $s_1 = 2$ and $s_2 = 0$, which the latter rejects.%

Initially, the \texttt{containers} library used \citeauthor{adams-tr}'s
balancing condition with the
parameters $\delta = 4$ (for sets) or $\delta = 5$ (for maps).
\citet{campbell} found that these parameters are
actually invalid and exhibited a sequence of insertions and deletions that
produced an unbalanced tree. Subsequently, the containers
library switched to $\delta=3$ in both modules.  Inspired by this bug report,
\citet{hirai-wbt} thoroughly analyzed this data structure with the help of a Coq
formalization, and identified the valid range for the balancing parameter,
albeit only for \citeauthor{nievergelt}'s variant --
our proof seems to be the first mechanical verification of Adam’s variant.

Given such thorough analysis of the algorithms, we did not expect to learn
anything new about this data structure, and for the most part, this was true.
Our proofs are free of any manual calculations about  tree sizes and the
balancing condition. We just state the proper preconditions for each lemma,
and Coq’s automation for linear integer arithmetic, \ensuretext{\straighttick\ttfamily{}lia}~\citep{lia}, takes
care of the rest.

One exception was the crucial function \ensuretext{\straighttick\ttfamily{}balanceL} which is used, according to the documentation “when the left subtree might have been inserted to or when the right subtree might have been deleted from”. This suggests the precondition
\[
(\balA(s_1-1,s_2) \wedge 0 < s_1) \vee \balA(s_1,s_2) \vee \balA(s_1,s_2+1)
\]
corresponding to the three cases: left tree inserted to, no change, and right tree deleted from. This is also what Hirai and Yamamoto used in their formalization of the original variant. And indeed, this precondition is strong enough to verify that the output of \ensuretext{\straighttick\ttfamily{}balanceL} is balanced -- but we found the precondition too strong.
The \ensuretext{\straighttick\ttfamily{}link} operation, shown in \cref{fig:recursion-patterns}, is supposed to balance two arbitrary trees using \ensuretext{\straighttick\ttfamily{}balanceL}. In the verification of \ensuretext{\straighttick\ttfamily{}link} we were unable to satisfy this precondition.

We found another precondition that is both strong enough for the verification of \ensuretext{\straighttick\ttfamily{}balanceL} and weak enough to allow the verification of \ensuretext{\straighttick\ttfamily{}link}, namely
\begin{gather*}
(\balAins(s_1-1,s_2) \wedge 0 < s_1) \vee \balA(s_1,s_2) \vee \balA(s_1,s_2+1)
\shortintertext{where}
\balAins(s_1,s_2) \coloneqq
\delta \cdot s_1 \le \delta^2\cdot s_2 + \delta\cdot s_2 + s_2  \wedge s_2 \le s_1.
\end{gather*}

Unfortunately we cannot give much insight into why this is the right
inequality -- this is the price we pay for relying completely on proof
automation when verifying the balancing properties. For us, this inequality
fell out of the proof for \ensuretext{\straighttick\ttfamily{}link}, where we call \ensuretext{\straighttick\ttfamily{}balanceL}
with trees with sizes $s_1$ and $s_2$, and what we know about these trees can be expressed as
\begin{align*}
\exists\, s\, s_l\,  s_r,\,
    s_1 = 1 + s + s_l \wedge s_2 = s_r \wedge
    \balA(s_l,s_r) \wedge \delta \cdot s < 1 + s_l + s_r \wedge
    1 \le s \wedge 0 \le s_l \wedge 0 \le s_r
\end{align*}
As \ensuretext{\straighttick\ttfamily{}lia} works best when the formulas are quantifier-free, we manually
eliminated the existential quantifiers and simplified this to arrive at $\balAins(s_1-1,s_2)$.

\subsection{Improvements to \LITcontainers}\label{sec:improvements}

Although our verification did not find bugs in the code of the library, we were
able to improve \texttt{containers}: during the
verification of \ensuretext{\straighttick\ttfamily{}Data.Set.union}, we noticed that it was using a nested pattern
match to check if an argument is a singleton set, when it could be testing if
the size was \texttt{1} directly.  This
change turned out to provide a 4\% speedup to \ensuretext{\straighttick\ttfamily{}union} and will be
present in the next version of the \texttt{containers} library.%
\containerCommit{b1a05c3a2}%

Additionally, as we mention in \cref{sec:specs-valid}, our well-formedness
property for \ensuretext{\straighttick\ttfamily{}IntSet} uncovered a weakness in the \ensuretext{\straighttick\ttfamily{}valid} function for \ensuretext{\straighttick\ttfamily{}IntSet}
and \ensuretext{\straighttick\ttfamily{}IntMap}, which was used in
the test suite.  The \ensuretext{\straighttick\ttfamily{}valid} function failed to ensure that some of the
invariants hold recursively in the tree structure, which was necessary
for the proof.  We notified the \texttt{containers} maintainers,\containerIssue{522}
 who then made the \ensuretext{\straighttick\ttfamily{}valid} function complete.

We also provide a package,
\texttt{containers\char45{}verified},\footnote{\url{https://hackage.haskell.org/package/containers-verified}}
which re-exports the types and definitions we have verified from the precise
version of \texttt{containers} we are working with.  This way, a developer who wants to
use only the verified portion of the implementation can replace their dependency
on \texttt{containers} with a dependency on \texttt{containers\char45{}verified}.

\section{Assumptions and Limitations}
\label{sec:limitations}

We have shown a way to make mechanically checked, formal statements about
existing Haskell code, and have applied this technique to verify parts of the
\texttt{containers} library. But are the theorems that we prove actually true? And
if they are, how useful is this method?

\subsection{The formalization gap}

As always, when a theorem is stated about a object that does not purely exist
within mathematics, its validity depends on a number of assumption.
\begin{itemize}
\item First and foremost, we have to assume that Coq behaves as documented and
  does not allow us to prove false theorems. This is of particular relevance
  because we rely on fine details of Coq’s machinery (e.g.\ the behavior of
  \ensuretext{\straighttick\ttfamily{}\textbf{Qed}}, \cref{sec:partial}) and optionally add consistent axioms (see
  \cref{sec:deferredFix}).

\item The biggest assumption is that the semantics of a Gallina program, as
  defined by the Calculus of Construction~\citep{coc}, models the the behavior of
  a running Haskell program in a meaningful way. At this time, we cannot even
  attempt to close this gap, as there are neither formal semantics nor verified
  compilers for Haskell.

 We know that “Fast and Loose Reasoning is Morally
 Correct”~\citep{fast-and-loose}, which says that theorems about the total
 fragment of a non-total language carry over to the full language. But since we
 relate two very different languages, this argument alone is not enough to
 bridge the gap.

  We can gain additional confidence by \emph{testing} this connection: we
  extracted our Gallina versions of \ensuretext{\straighttick\ttfamily{}Set} and \ensuretext{\straighttick\ttfamily{}IntSet} back to Haskell and
  successfully ran the original test suite of \texttt{containers} against it.
Every successful test indicates that the corresponding theorem (see \cref{sec:specs-tests}) is indeed a theorem about the Haskell program.

\item Furthermore, we rely on \texttt{hs\char45{}to\char45{}coq} translating Haskell code into the correct
Gallina code. The translator itself is a sizable piece of code, unverified (and
such verification is currently out of our reach, not least because there is no
formal semantics of Haskell) and therefore surely
not free of bugs. We get some confidence into the tool from manually inspecting
its output and observing that it is indeed what we would consider the “right”
translation from Haskell into Gallina, and additional confidence from the fact
that we were actually able to prove the specifications, which would not be
possible if the translated code behaved differently than intended. Moreover,
extracting the translated code back to Haskell and running the test suite also
stress-tests the translation.

\item The translation was not completely automatic and required manual edits. With each edit, we add another assumption to the formalization gap: Does our underspecification of pointer equality encompass the actual behavior of GHC’s \ensuretext{\straighttick\ttfamily{}reallyUnsafePtrEqualty\#}? Given our choice of using unbounded integers, does the size field in a \ensuretext{\straighttick\ttfamily{}Set} really never overflow in practice? Are our manually written versions of low-level bit twiddling functions correct? We list and justify our manual interventions in \cref{sec:translating}.
\end{itemize}

The formalization gap of our work is relatively large compared to, say, the gap
for the verification of programs written in Gallina in the first
place. But for the purpose of ensuring the correctness of the Haskell code, that
is less critical.  Even if one of our assumptions is flawed, it is much more
likely that the flaw will get in the way of concluding the proofs, rather than
allowing us to conclude the proofs without noticing a bug. Incomplete proofs can uncover bugs, too.

\subsection{Limitations of our approach}

We proved multiple specifications about a large part of the code, but there are limits to what theorems we can prove, and what we can prove them about.

Since we work with a shallow embedding of Haskell in Coq, we cannot make statements about the performance of the Haskell code -- which is a pity, given that the \texttt{containers} library contains highly optimized code and provides clear specifications of the algorithmic complexity of their operations. Similarly, we cannot verify that the operations are as strict or lazy as documented.

\label{sec:untranslated}
We also cannot translate and verify all code in the \texttt{containers} library, because
some functions use language features not yet supported by
\texttt{hs\char45{}to\char45{}coq}, such as mutual recursion, unboxed values and generic programming.
When this affects a crucial utility function we can provide a manual
translation. This widens the formalization gap, but enables verification of
code that depends on it. When it affects less central code, e.g.\ the \ensuretext{\straighttick\ttfamily{}Data}
instances, we can simply skip the translation (\cref{sec:unwanted-code}).

Partiality is a particular interesting limit. Coq only allows total functions,
but practical Haskell code uses partiality, often in a benign way: Calls to
\ensuretext{\straighttick\ttfamily{}error} in code paths that are unreachable as long as invariants are
maintained, or recursive functions that terminate on the actual arguments they
are called with, but may diverge on other inputs. Whereas
\citet{hs-to-coq-cpp} considered such code out of scope,
we have found ways to deal with “internal partiality”
(see \cref{sec:partial,sec:deferredFix}).

Taking a step back, it might seem that our approach may see limited adoption
in the Haskell community because it requires expertise in Coq. But though tied
to a Haskell artifact, verification is isolated from the codebase. Haskell
programmers can focus on their domain, trying to get the best performance out
of the code and without having to know about verification, while proof
engineers can work solely within Coq and do not have to be fluent in Haskell.

In this paper we verify a specific version of the Haskell code, and do not
discuss ongoing maintenance of such a verification. It remains to be seen how
resilient the proofs are against changes in the Haskell code. Changes of
syntactic nature, or changes to a function’s strictness, might be swallowed by \texttt{hs\char45{}to\char45{}coq}. Other changes might affect the translated code, but still allow the proofs to go through, if our proof tactics are flexible enough. In general, though, we expect that changes in the code require changes in the proofs. Since Coq is an interactive theorem prover, it will at least clearly point out which parts of our development need to be updated.

\section{Related work}
\label{sec:related}
%
%
%

\subsection{Verification of purely functional data structures}

Purely functional data structures, such as those found in \citeauthor{okasaki}'s book \citeyearpar{okasaki}
are frequent targets of mechanical verification.
That said, we believe that we are the first to verify the Patricia tree
algorithms that underlie \ensuretext{\straighttick\ttfamily{}Data.IntSet}.

\paragraph*{Verifying weight-balanced trees in Haskell}
%
Similar to \texttt{hs\char45{}to\char45{}coq}, LiquidHaskell~\citep{Vazou:2014:RTH:2628136.2628161} can
be used to verify existing Haskell code.  Users of this tool annotate their
Haskell source files with refinement types and other annotations. LiquidHaskell
then uses an SMT solver to automatically discharge proof obligations described
by the refinements.  This means that LiquidHaskell provides more automation than
\texttt{hs\char45{}to\char45{}coq}; however, the language of Coq is higher-order and more expressive
than the language of SMT solvers.  \Citet{Vazou:2017:TTP:3122955.3122963}
compared the experiences of using LiquidHaskell vs.\ plain Coq, and found that
both have advantages.

\Citet{liquidhaskell2013} also described the use of LiquidHaskell to verify the
\ensuretext{\straighttick\ttfamily{}Data.Map} module of finite maps from \texttt{containers}.  Although not the same as
\ensuretext{\straighttick\ttfamily{}Data.Set}, this code shares the same underlying data structure (weight-balanced
trees) and the implementations of the two are similar.
Their verification has
similarities with our work; they also use unbounded integers as the number
representation and left functions like \ensuretext{\straighttick\ttfamily{}showTree} unspecified.
However, we develop a richer specification, which includes a semantic
description of each operation we verified, constraints about the tree balance,
and the ordering of the elements in the tree. In contrast,
\citeauthor{liquidhaskell2013} limit their specifications to ordering only. For
example, in addition to showing that the insertion operation preserves the order
of elements of the tree, our work also shows that: (1)~insertion preserves the
balancing conditions of the weight-balanced tree, (2)~the \ensuretext{\straighttick\ttfamily{}size} field at each
node in the tree is maintained correctly~(i.e.,~the \ensuretext{\straighttick\ttfamily{}size} equals to the number
of all its descendants), and (3)~the tree returned by this operation contains
the inserted element, and all elements in the original tree, but nothing more.
Although it might be possible to replicate our specifications by using theories
of finite sets and maps in SMT~\citep{smt-set} to encode these properties using
refinement types in LiquidHaskell, this approach has not been explored.

Furthermore, our specification also includes type class laws, and we
are able to verify that \ensuretext{\straighttick\ttfamily{}Set} and \ensuretext{\straighttick\ttfamily{}IntSet} have lawful instances of the \ensuretext{\straighttick\ttfamily{}Eq},
\ensuretext{\straighttick\ttfamily{}Ord}, \ensuretext{\straighttick\ttfamily{}Semigroup}, and \ensuretext{\straighttick\ttfamily{}Monoid} type classes.  When
\citeauthor{liquidhaskell2013}'s original work was developed in
\citeyear{liquidhaskell2013}, LiquidHaskell did not have the
capability to state and prove these properties.  Since then,
there have been new developments in LiquidHaskell, particularly
refinement reflection~\citep{refinement-reflection}, which could make
it possible to specify and prove type class laws.

Both LiquidHaskell and \texttt{hs\char45{}to\char45{}coq} check for termination of Haskell
functions. In LiquidHaskell, the termination check is an option that can be
deactivated, allowing the sound verification of nonstrict, non-terminating
functions~\citep{Vazou:2014:RTH:2628136.2628161}. In contrast, a proof of
termination is a requirement for verifying functions using
\texttt{hs\char45{}to\char45{}coq}~\citep{hs-to-coq-cpp}.  However, \texttt{hs\char45{}to\char45{}coq} is able to take
advantage of many options available in Coq for proving termination of
non-trivial recursion, including structural recursion, \ensuretext{\straighttick\ttfamily{}\textbf{Program}{\spacestart}{\nextspace}\textbf{Fixpoint}} and
our own approach based on \ensuretext{\straighttick\ttfamily{}deferredFix}.  This latter approach alowed us to
reason about \ensuretext{\straighttick\ttfamily{}fromDistinctAscList} (see \cref{sec:deferredFix}) and prove that
it is indeed terminating; on the other hand, \citeauthor{liquidhaskell2013}
deactivate the termination check for this function.

\paragraph*{Verifying weight-balanced trees in other languages}
\Citet{hirai-wbt} implemented a weight-balanced tree similar to Haskell’s
\ensuretext{\straighttick\ttfamily{}Data.Set} library (albeit using the balancing condition of
\citet{nievergelt}) and mechanically verified its balancing properties
in Coq. More recently, \citet{tobias-wbt} extended this work to formalize
similar weight-balanced trees in Isabelle and further verified the functional
correctness of insertions and deletions.

We verify more functions in the \ensuretext{\straighttick\ttfamily{}Data.Set} library than prior work.  Operations
that are unique to our development include \ensuretext{\straighttick\ttfamily{}foldl}, \ensuretext{\straighttick\ttfamily{}isSubsetOf}, and
\ensuretext{\straighttick\ttfamily{}fromDistinctAscList}.  The code verified both by \citeauthor{hirai-wbt} and by
\citeauthor{tobias-wbt} is also different from the latest \texttt{containers} library;
for example, it does not use of pointer equality (\cref{sec:ptrEq}). Another
difference is that \citeauthor{hirai-wbt} defined the \ensuretext{\straighttick\ttfamily{}union}, \ensuretext{\straighttick\ttfamily{}intersection},
and \ensuretext{\straighttick\ttfamily{}difference} functions based on the ``hedge union'' algorithm, but
\texttt{containers} has since changed to use the ``divide and conquer'' algorithm.

\Citeauthor{hirai-wbt} specify only the balancing constraints, whereas we
develop a richer specification that also includes a semantic description of each
operation we verified and the ordering of the elements in the tree.  We also
gained new insights about the balancing conditions of the weight-balanced tree
through our verification effort~(see \cref{sec:balanceL}).
\Citeauthor{tobias-wbt}'s specification contains the same properties as ours,
but they only specify the behavior of \ensuretext{\straighttick\ttfamily{}insert} and \ensuretext{\straighttick\ttfamily{}delete}; we have verified a
significantly larger set of functions (see \cref{fig:verifiedAPI}).




\paragraph*{Other verifications of balanced trees}

There are many existing works on mechanically verifying purely functional
balanced trees. We briefly mention a few here.  \citet{Filliatre04}
implemented AVL trees in Coq, and verified their functional correctness as
well as their balancing conditions.  \citet{appel-rbt} did the same
thing for red-black trees. These implementations have now become parts of Coq
standard library.  \citet{chargueraud-10-cfml} translated OCaml
implementations of \citet{okasaki}'s functional data structures to
characteristic formulae expressed as Coq axioms. \Citet{licata} lectured on verifying red-black trees in Agda at the Oregon Programming Languages Summer School.
\citet{McBride-order} showed how to represent the ordering relationships in
Agda for general data structures, not just binary search
trees. \citet{Nipkow-ITP16} showed how to automatically verify the ordering
properties of eight different binary search tree structures, by specifying
each in terms of the sorted list of their elements, a method he used again in his verification of weight-balanced trees \citep{tobias-wbt}.
\citet{Ralston:2009} verified
AVL trees in ACL2.



\subsection{Verification tools for Haskell}

Previous work using \texttt{hs\char45{}to\char45{}coq} has only applied it to small examples. 
\citet{hs-to-coq-cpp} describe three case studies, two of
which require less than~20 lines of Haskell. The longest example (the \texttt{Bag}
module taken directly from the GHC compiler) is~247
lines of code. Furthermore, none of the reasoning required for these examples
is particularly deep. Our work provides experience with more complex, 
externally-sourced, industrial-strength examples.

The \texttt{coq-haskell} library~\citep{coq-haskell} is a
handwritten Coq library designed to make it easier for Haskell programmers to
work in Coq.
In addition to enabling easier Coq programming, it
also provides support for extracting Coq programs to Haskell.

The prototype contract checker \texttt{halo}~\citep{Vytiniotis:2013}
takes a Haskell program, uses GHC to desugar it into the intermediate language
Core, and translates the Core program into a first-order logic formula.  It
then invokes an SMT solver to prove this formula; a successful proof implies
that the original program is crash-free.

\emph{Haskabelle} was \texttt{hs-to-coq}'s counterpart in Isabelle. 
It translated total Haskell code into equivalent Isabelle function
definitions. Similar to \texttt{hs-to-coq}, it parsed Haskell, desugared
syntactic constructs, and configurably adapted basic types and functions to their
counterparts in Isabelle’s standard library. It used to be bundled with the
Isabelle release, but it has not been updated recently and was dropped from
the distribution.





Haskell has been used as a prototyping language for mechanically verified systems
in the past. The \emph{seL4 verified microkernel} started with a Haskell prototype
that was semi-automatically translated to
Isabelle/HOL~\citep{sel4,Derrin:2006:RMA}. 
%
The authors found that the availability of the Haskell prototype provided a
machine-checkable formal executable specification of the system. They used
this prototype to refine their designs via testing, allowing them to make corrections
before full verification.

The \emph{Programmatica project}~\citep{Hallgren04anoverview} included a tool that
translates Haskell code into the Alfa proof editor. Their tool only produces
valid proofs for total functions over finite data structures. The logic of the
Alfa proof assistant is based on dependent type theory, but without as many
features as Coq. In particular, the Programmatica tool compiles away type
classes and nested pattern matching; both of these features are retained by
\texttt{hs-to-coq}.

\citeauthor{DybjerHT04} developed a tool for automatically
translating Haskell programs to the \emph{Agda/Alfa proof assistant}~\citeyearpar{DybjerHT04}.
They explicitly note the
interplay between testing and theorem proving and show how to verify a tautology checker.
%
\citet{Abel:2005} translate Haskell expressions into the logic of the
Agda 2 proof assistant. Their tool works later in the GHC pipeline than \texttt{hs\char45{}to\char45{}coq}
 and translates Core expressions.
%

\subsection{Translating other higher-order functional languages}

There are many large and successful verification projects that demonstrate 
that functional languages are well suited for verification. 
In contrast to our work, they require re-implementing the code either in a new 
functional language, as is the case for Cogent~\citep{OConnor_CRALMNSK_16,Amani_HCRCOBNLSTKMKH_16} and F*~\citep{Swamy_HKRDAFBFSPKZZ_16}, or in a proof assistant, 
such as HOL4 in the case of CakeML~\citep{Myreen_Owens_14,Kumar_MNO_14}. 
The CakeML and Cogent projects have a different focus than ours, and they provide a higher assurance in their verified code. 
CakeML~\citep{Kumar_MNO_14} has a verified compiler and Cogent has a certifying compiler~\citep{OConnor_CRALMNSK_16,Rizkallah_LNSCOMKK_16}. Both tools provide mechanically 
checked proofs that their shallow embeddings correspond to the functional code being verified.

\emph{Cogent} is a restricted higher-order functional language~\citep{OConnor_CRALMNSK_16} that was used to verify
filesystems~\citep{Amani_HCRCOBNLSTKMKH_16}. Its compiler produces C code, a
high-level specification in Isabelle/HOL, and an Isabelle/HOL refinement proof
linking the two~\citep{OConnor_CRALMNSK_16,Rizkallah_LNSCOMKK_16}.
\citet{Chen_OKKH_17} integrated property-based testing into the Cogent
framework. The
authors claim that property-based testing enables an incremental approach to a
fully verified system, as it allows for the replacement of tests of properties
stated in the specification by formal proofs.  Our work substantiates this
claim, as indeed one of the ways in which we obtain specifications is through
the QuickCheck properties provided by Haskell as discussed in \cref{sec:specs-tests}.

\emph{CakeML} is a large subset of ML with a verified compiler and runtime
system~\citep{Kumar_MNO_14}. Users of CakeML can write pure
functional programs in the HOL4 proof assistant and verify them in HOL4. They
can then extract an equivalent correct by construction CakeML program.  This
method has been used to verify several data structures including red black
trees, crypto protocols, and a CakeML version of the HOL light theorem
prover~\citep{Myreen_Owens_14}.

\emph{F*} is a general-purpose functional
language that allows for a mixture of proving and general-purpose programming~\citep{Swamy_HKRDAFBFSPKZZ_16}. Programs
can be specified using dependent and refinement types and automatically verified using one
of F*'s backend SMT solvers.
Its subset Low*~\citep{Protzenko_ZRRWZDHBFS_17} has been used to verify high-assurance optimized cryptographic libraries.

\subsection{Extraction}
The semantic proximity of Haskell and Coq, which we rely on, is also
used in the other direction by Coq’s support for code extraction to
Haskell~\citep{code-extraction}. Several projects use this feature to verify
Haskell code~\citep{Chen:2015,Megacz}.  However, since extraction starts with
Coq code and generates Haskell, it cannot be used to verify pre-existing
Haskell.  Although in a certain sense \texttt{hs\char45{}to\char45{}coq} and extraction are
inverses, round-tripping does not produce syntactically equivalent output in
either direction. In one direction, \texttt{hs\char45{}to\char45{}coq} extensively annotates the
resulting Coq code; in the other, extraction ignores many Haskell features and
inserts unsafe type coercions. In this work, we use testing to verify that
round-tripping produces operationally equivalent output; this provides 
assurance about the correctness of both \texttt{hs-to-coq} and extraction.

CertiCoq~\citep{CertiCoq_17} and Œuf~\citep{Mullen_PWJTG_2018} are verified compilers
from Gallina to assembly. CertiCoq can compile any Gallina program, while Œuf can
only compile a limited subset of Gallina. However, Œuf provides stronger guarantees
about the Gallina code that is in the limited subset it translates.


\section{Conclusions and Future Work}
\label{sec:conclusion}

We verified the two finite set modules that are part of the widely used and
highly-optimized \texttt{containers} library. Our efforts provide the deepest
specification and verification of this code to date, covering more of the API
and proving stronger, more descriptive properties than prior work.

In future work, there is yet more to verify in \texttt{containers}. For example, we
plan to add a version of \ensuretext{\straighttick\ttfamily{}IntSet} that uses 64-bit ints as the element type in
addition to our current version with unbounded natural numbers. That way users could
choose the treatment of overflow that makes the most sense for their
application. We have also started to verify \ensuretext{\straighttick\ttfamily{}Data.Map} and
\ensuretext{\straighttick\ttfamily{}Data.IntMap}. Because these modules use the same algorithms as \ensuretext{\straighttick\ttfamily{}Data.Set} and
\ensuretext{\straighttick\ttfamily{}Data.IntSet}, we already adapted some of our existing proofs to this
setting.


The fact that we did not find bugs says a lot about the tools that are already
available to Haskell programmers for producing correct code, such as a strong,
expressive type system and a mature property-based testing
infrastructure.  However, few would dare to extrapolate from these results to
say that all Haskell programs are bug free!  Instead, we view verification as a
valuable opportunity for functional programmers and an activity that we hope
will become more commonplace.

\begin{acks}
This material is based upon work supported by the
  \grantsponsor{GS100000001}{National Science
    Foundation}{http://dx.doi.org/10.13039/100000001} under Grant
  No.~\grantnum{GS100000001}{1319880} and Grant
  No.~\grantnum{GS100000001}{1521539}.
\end{acks}

\bibliography{bib}

\appendix

\iflong
\section{Big-endian Patricia trees}

\label{sec:patricia-trees}

In this self-contained section, we explain Patricia trees as they are implemented in the containers library for the \ensuretext{\straighttick\ttfamily{}IntSet} and \ensuretext{\straighttick\ttfamily{}IntMap} data structures and outline their correctness proofs.

\subsection{The implementation}

\begin{figure}
\abovedisplayskip=0pt
\belowdisplayskip=0pt
\raggedright
\begingroup\straighttick\begin{hscode}\SaveRestoreHook
\column{B}{@{}>{\hspre}l<{\hspost}@{}}%
\column{14}{@{}>{\hspre}c<{\hspost}@{}}%
\column{14E}{@{}l@{}}%
\column{17}{@{}>{\hspre}l<{\hspost}@{}}%
\column{E}{@{}>{\hspre}l<{\hspost}@{}}%
\>[B]{}\textbf{data}{\spacestart}{\nextspace}IntSet{\spacestart}{\nextspace}{\nextspace}{}\<[14]%
\>[14]{}={\spacestart}{\nextspace}{\nextspace}{}\<[14E]%
\>[17]{}Bin{\spacestart}{\nextspace}Prefix{\spacestart}{\nextspace}Mask{\spacestart}{\nextspace}IntSet{\spacestart}{\nextspace}IntSet{\spacestart}{\nextspace}{}\<[E]%
\\[-0.3ex]%
\>[B]{}{\spacestart}{\nextspace}{\nextspace}{\nextspace}{\nextspace}{\nextspace}{\nextspace}{\nextspace}{\nextspace}{\nextspace}{\nextspace}{\nextspace}{\nextspace}{\nextspace}{}\<[14]%
\>[14]{}|{\spacestart}{\nextspace}{\nextspace}{}\<[14E]%
\>[17]{}Tip{\spacestart}{\nextspace}Prefix{\spacestart}{\nextspace}BitMap{\spacestart}{\nextspace}{}\<[E]%
\\[-0.3ex]%
\>[B]{}{\spacestart}{\nextspace}{\nextspace}{\nextspace}{\nextspace}{\nextspace}{\nextspace}{\nextspace}{\nextspace}{\nextspace}{\nextspace}{\nextspace}{\nextspace}{\nextspace}{}\<[14]%
\>[14]{}|{\spacestart}{\nextspace}{\nextspace}{}\<[14E]%
\>[17]{}Nil{\spacestart}{\nextspace}{}\<[E]%
\\[-0.3ex]%
\>[B]{}\textbf{type}{\spacestart}{\nextspace}Prefix{\spacestart}{\nextspace}{\nextspace}{}\<[14]%
\>[14]{}={\spacestart}{\nextspace}{\nextspace}{}\<[14E]%
\>[17]{}Int{\spacestart}{\nextspace}{}\<[E]%
\\[-0.3ex]%
\>[B]{}\textbf{type}{\spacestart}{\nextspace}Mask{\spacestart}{\nextspace}{\nextspace}{\nextspace}{\nextspace}{}\<[14]%
\>[14]{}={\spacestart}{\nextspace}{\nextspace}{}\<[14E]%
\>[17]{}Int{\spacestart}{\nextspace}{}\<[E]%
\\[-0.3ex]%
\>[B]{}\textbf{type}{\spacestart}{\nextspace}BitMap{\spacestart}{\nextspace}{\nextspace}{}\<[14]%
\>[14]{}={\spacestart}{\nextspace}{\nextspace}{}\<[14E]%
\>[17]{}Word{\spacestart}{\nextspace}{}\<[E]%
\\[-0.3ex]%
\>[B]{}\textbf{type}{\spacestart}{\nextspace}Key{\spacestart}{\nextspace}{\nextspace}{\nextspace}{\nextspace}{\nextspace}{}\<[14]%
\>[14]{}={\spacestart}{\nextspace}{\nextspace}{}\<[14E]%
\>[17]{}Int{}\<[E]%
\ColumnHook
\end{hscode}\resethooks\endgroup
\caption{The \ensuretext{\straighttick\ttfamily{}IntSet} data structure}
\label{fig:intset}
\end{figure}

\begin{figure}
\raggedright
\begingroup\straighttick\begin{hscode}\SaveRestoreHook
\column{B}{@{}>{\hspre}l<{\hspost}@{}}%
\column{3}{@{}>{\hspre}l<{\hspost}@{}}%
\column{10}{@{}>{\hspre}l<{\hspost}@{}}%
\column{12}{@{}>{\hspre}l<{\hspost}@{}}%
\column{29}{@{}>{\hspre}l<{\hspost}@{}}%
\column{51}{@{}>{\hspre}l<{\hspost}@{}}%
\column{E}{@{}>{\hspre}l<{\hspost}@{}}%
\>[B]{}member{\spacestart}{\nextspace}::{\spacestart}{\nextspace}Key{\spacestart}{\nextspace}->{\spacestart}{\nextspace}IntSet{\spacestart}{\nextspace}->{\spacestart}{\nextspace}Bool{\spacestart}{\nextspace}{}\<[E]%
\\[-0.3ex]%
\>[B]{}member{\spacestart}{\nextspace}!x{\spacestart}{\nextspace}={\spacestart}{\nextspace}go{\spacestart}{\nextspace}{}\<[E]%
\\[-0.3ex]%
\>[B]{}{\spacestart}{\nextspace}{\nextspace}{}\<[3]%
\>[3]{}\textbf{where}{\spacestart}{\nextspace}{\nextspace}{}\<[10]%
\>[10]{}go{\spacestart}{\nextspace}(Bin{\spacestart}{\nextspace}p{\spacestart}{\nextspace}m{\spacestart}{\nextspace}l{\spacestart}{\nextspace}r){\spacestart}{\nextspace}{}\<[E]%
\\[-0.3ex]%
\>[B]{}{\spacestart}{\nextspace}{\nextspace}{\nextspace}{\nextspace}{\nextspace}{\nextspace}{\nextspace}{\nextspace}{\nextspace}{\nextspace}{\nextspace}{}\<[12]%
\>[12]{}|{\spacestart}{\nextspace}nomatch{\spacestart}{\nextspace}x{\spacestart}{\nextspace}p{\spacestart}{\nextspace}m{\spacestart}{\nextspace}{\nextspace}{}\<[29]%
\>[29]{}={\spacestart}{\nextspace}False{\spacestart}{\nextspace}{}\<[E]%
\\[-0.3ex]%
\>[B]{}{\spacestart}{\nextspace}{\nextspace}{\nextspace}{\nextspace}{\nextspace}{\nextspace}{\nextspace}{\nextspace}{\nextspace}{\nextspace}{\nextspace}{}\<[12]%
\>[12]{}|{\spacestart}{\nextspace}zero{\spacestart}{\nextspace}x{\spacestart}{\nextspace}m{\spacestart}{\nextspace}{\nextspace}{\nextspace}{\nextspace}{\nextspace}{\nextspace}{\nextspace}{}\<[29]%
\>[29]{}={\spacestart}{\nextspace}go{\spacestart}{\nextspace}l{\spacestart}{\nextspace}{}\<[E]%
\\[-0.3ex]%
\>[B]{}{\spacestart}{\nextspace}{\nextspace}{\nextspace}{\nextspace}{\nextspace}{\nextspace}{\nextspace}{\nextspace}{\nextspace}{\nextspace}{\nextspace}{}\<[12]%
\>[12]{}|{\spacestart}{\nextspace}otherwise{\spacestart}{\nextspace}{\nextspace}{\nextspace}{\nextspace}{\nextspace}{\nextspace}{}\<[29]%
\>[29]{}={\spacestart}{\nextspace}go{\spacestart}{\nextspace}r{\spacestart}{\nextspace}{}\<[E]%
\\[-0.3ex]%
\>[B]{}{\spacestart}{\nextspace}{\nextspace}{\nextspace}{\nextspace}{\nextspace}{\nextspace}{\nextspace}{\nextspace}{\nextspace}{}\<[10]%
\>[10]{}go{\spacestart}{\nextspace}(Tip{\spacestart}{\nextspace}p{\spacestart}{\nextspace}bm){\spacestart}{\nextspace}{\nextspace}{\nextspace}{\nextspace}{\nextspace}{\nextspace}{}\<[29]%
\>[29]{}={\spacestart}{\nextspace}prefixOf{\spacestart}{\nextspace}x{\spacestart}{\nextspace}=={\spacestart}{\nextspace}p{\spacestart}{\nextspace}\&\&{\spacestart}{\nextspace}{\nextspace}{}\<[51]%
\>[51]{}(bitmapOf{\spacestart}{\nextspace}x{\spacestart}{\nextspace}.\&.{\spacestart}{\nextspace}bm){\spacestart}{\nextspace}/={\spacestart}{\nextspace}0{\spacestart}{\nextspace}{}\<[E]%
\\[-0.3ex]%
\>[B]{}{\spacestart}{\nextspace}{\nextspace}{\nextspace}{\nextspace}{\nextspace}{\nextspace}{\nextspace}{\nextspace}{\nextspace}{}\<[10]%
\>[10]{}go{\spacestart}{\nextspace}Nil{\spacestart}{\nextspace}{\nextspace}{\nextspace}{\nextspace}{\nextspace}{\nextspace}{\nextspace}{\nextspace}{\nextspace}{\nextspace}{\nextspace}{\nextspace}{\nextspace}{}\<[29]%
\>[29]{}={\spacestart}{\nextspace}False{}\<[E]%
\ColumnHook
\end{hscode}\resethooks\endgroup
\caption{\ensuretext{\straighttick\ttfamily{}IntSet.member}}
\label{fig:intset-member}
\end{figure}

The \ensuretext{\straighttick\ttfamily{}IntSet} type an algebraic data structure with three constructors (\cref{fig:intset}):
\begin{itemize}
\item An \ensuretext{\straighttick\ttfamily{}IntSet} of the form \ensuretext{\straighttick\ttfamily{}Bin{\spacestart}{\nextspace}p{\spacestart}{\nextspace}m{\spacestart}{\nextspace}l{\spacestart}{\nextspace}r} combines the two sets \ensuretext{\straighttick\ttfamily{}l} and \ensuretext{\straighttick\ttfamily{}r}. In the mask \ensuretext{\straighttick\ttfamily{}m}, a single bit is set. It discriminates the members of \ensuretext{\straighttick\ttfamily{}l} (bit not set) and \ensuretext{\straighttick\ttfamily{}r} (bit set). All elements in the map have the bits above the mask bit in common; the prefix \ensuretext{\straighttick\ttfamily{}p} indicates that.
\item The constructor \ensuretext{\straighttick\ttfamily{}Tip{\spacestart}{\nextspace}p{\spacestart}{\nextspace}bm} stores uses the bits of the bitmap \ensuretext{\straighttick\ttfamily{}bm} of width $w$ (commonly 64) to store a subset of the keys $[\ensuretext{\straighttick\ttfamily{}p},\ldots, (\ensuretext{\straighttick\ttfamily{}p}+2^{w}-1)]$.
\item The constructor \ensuretext{\straighttick\ttfamily{}Nil} describes the empty set.
\end{itemize}
These three constructors are deliberately ordered from most to least frequent, as this improves the efficiency of pattern matching.

To get a better feeling of this data structure, let us walk through the function \ensuretext{\straighttick\ttfamily{}member} in \cref{fig:intset-member}, which checks if a key \ensuretext{\straighttick\ttfamily{}x} is in an \ensuretext{\straighttick\ttfamily{}IntSet}.
If the set is a \ensuretext{\straighttick\ttfamily{}Bin}, it uses the function \ensuretext{\straighttick\ttfamily{}nomatch} to compare the upper bits of \ensuretext{\straighttick\ttfamily{}x} and \ensuretext{\straighttick\ttfamily{}p}; if they differ, then \ensuretext{\straighttick\ttfamily{}x} is certainly not in the map. Otherwise, the function \ensuretext{\straighttick\ttfamily{}zero} checks whether the mask bit \ensuretext{\straighttick\ttfamily{}m} is set and the search continues in \ensuretext{\straighttick\ttfamily{}l} or \ensuretext{\straighttick\ttfamily{}r}.
If the set is a \ensuretext{\straighttick\ttfamily{}Tip}, then \ensuretext{\straighttick\ttfamily{}x} is a member if it is in the range described by the bitmap, and whether the corresponding bit is set. This uses \ensuretext{\straighttick\ttfamily{}prefixOf{\spacestart}{\nextspace}x}, which is \ensuretext{\straighttick\ttfamily{}x} with the lower $\log_2 w$ bits cleared, and \ensuretext{\straighttick\ttfamily{}bitmapOf{\spacestart}{\nextspace}x}, which has one bit set at the index corresponding to the lower $\log_2 w$ bits of \ensuretext{\straighttick\ttfamily{}x}.
Finally, the empty set surely does not contain \ensuretext{\straighttick\ttfamily{}x}.

In the classic literature, including the two sources cited by the IntSet implementation -- Morrison work on PATRICIA~\citep{Morrison:1968} and a paper by Okasaki and Gill~\citep{okasakigill} -- the \ensuretext{\straighttick\ttfamily{}Tip} constructor stores just a single value. The containers library uses a bit map here to improve the space and time performance of sets where members are dense.%
\footnote{Although this feature was contributed in 2011 by the first author, he certainly did this without having an eventual formal verification in mind.}

\subsection{Invariants and semantics of sets}

Functions like \ensuretext{\straighttick\ttfamily{}member} will only yield sensible results on sets that are well-formed and satisfy certain invariants. The first step towards verifying these sets is to specify precisely when a values of type \ensuretext{\straighttick\ttfamily{}IntSet} constitutes a valid set.

We formulate the theory initially only for unsigned unbounded integers, and discuss the treatment of bounded and possibly negative number afterwards.

Every \ensuretext{\straighttick\ttfamily{}IntSet} describes the elements of a set in an interval of the form $[p\cdot 2^b,\ldots,(p+1)\cdot 2^b-1] \eqqcolon d(p,b)$ for some $p\in\mathbb N$ and $b\in \mathbb N $. These \emph{dyadic intervals} have an interesting algebraic structure, which plays a crucial role in the verification of the Patricia trees. In particular:
\begin{itemize}
\item A dyadic interval is non-empty.
\item Two dyadic intervals are either disjoint, or one is contained in the other.
\item A dyadic interval $d(p,b)$ with $b > 0$ is the disjoint union of its two halves, $h_1(d(p,b)) \coloneqq d(p,b-1)$ and $h_2(d(p,b)) \coloneqq d(p+1,b-1)$.
\item In particular, if another dyadic interval $r$ overlaps with with both halves of $d(p,b)$, then  $d(p,b)\subseteq r$.
\item The dyadic intervals, ordered by inclusion, form a join semi-lattice: For two dyadic intervals $r_1, r_2$ there exists a least dyadic interval $r_1 \sqcup r_2$ that contains the two.
\end{itemize}

\newcommand{\WF}[1]{\ensuretext{\straighttick\ttfamily{}WF}~#1}
\newcommand{\Sem}[2]{#1\dblcolon#2}
\newcommand{\Desc}[3]{#1\dblcolon#2\dblcolon#3}
\newcommand{\DescO}[3]{#1\dblcolon_0#2\dblcolon#3}

\begin{figure}
\begin{mathpar}
\inferrule {\Sem{\ensuretext{\straighttick\ttfamily{}s}}D}{\WF{\ensuretext{\straighttick\ttfamily{}s}}}
\and
\inferrule {\Desc{\ensuretext{\straighttick\ttfamily{}s}}rD}{\Sem{\ensuretext{\straighttick\ttfamily{}s}}D}
\and
\inferrule {~}{\Sem{\ensuretext{\straighttick\ttfamily{}Nil}}\emptyset}
\\
\inferrule{%
\Desc{\ensuretext{\straighttick\ttfamily{}s$_\text{1}$}}{r_1}{D_1} \and
\Desc{\ensuretext{\straighttick\ttfamily{}s$_\text{2}$}}{r_2}{D_2}\\
r_1 \subseteq h_1(d(p,b)) \and
r_2 \subseteq h_2(d(p,b)) \\
b > 0 \and \ensuretext{\straighttick\ttfamily{}p'} = p\cdot2^b \and \ensuretext{\straighttick\ttfamily{}m} = 2^{b-1}
}{\Desc{\ensuretext{\straighttick\ttfamily{}Bin{\spacestart}{\nextspace}p'{\spacestart}{\nextspace}m{\spacestart}{\nextspace}s$_\text{1}${\spacestart}{\nextspace}s$_\text{2}$}}{d(p,b)}{D_1 \cup D_2}}
\\
\inferrule{%
\ensuretext{\straighttick\ttfamily{}p'} = p\cdot w \and 0 < \ensuretext{\straighttick\ttfamily{}bm} < 2^w
}{\Desc{\ensuretext{\straighttick\ttfamily{}Tip{\spacestart}{\nextspace}p'{\spacestart}{\nextspace}bm}}{d(p,\log_2 w)}{\{ p\cdot w + j \mid \text{bit $j$ set in \ensuretext{\straighttick\ttfamily{}bm}} \}}}
\end{mathpar}
\caption{Well-formedness and denotation of \ensuretext{\straighttick\ttfamily{}IntSet}}
\label{fig:intset-inv}
\end{figure}

A value \ensuretext{\straighttick\ttfamily{}s} of type \ensuretext{\straighttick\ttfamily{}IntSet} satisfies all the necessary invariants, if the predicate $\WF{s}$ holds. It is defined by the rules in \cref{fig:intset-inv}, using the following two auxiliary relations:
\begin{itemize}
\item The predicate $\Sem{\ensuretext{\straighttick\ttfamily{}s}}D$ expresses that $s$ is well-formed and denotes the (mathematical) set $D \subseteq \mathbb N$.
\item The predicate $\Desc{\ensuretext{\straighttick\ttfamily{}s}}rD$ additionally expresses that the set is non-empty ($D \ne \emptyset$), and that $r$ is the smallest dyadic interval that contains $D$.
\end{itemize}

We can state and prove these characterizations as lemmas, all of which are proven by induction on the assumed relation:

\begin{lemma}
If $\Desc{\ensuretext{\straighttick\ttfamily{}s}}{r}{D}$, then $\emptyset \subset D\subseteq r$.
\end{lemma}

\begin{lemma}
\label{lem:intset-member}%
If $\Sem{\ensuretext{\straighttick\ttfamily{}s}}{D}$, then $\ensuretext{\straighttick\ttfamily{}member{\spacestart}{\nextspace}x{\spacestart}{\nextspace}=={\spacestart}{\nextspace}true}$ iff $x \in D$.
\end{lemma}

In contrast to many other search trees, an \ensuretext{\straighttick\ttfamily{}IntSet} representation is unique:

\begin{lemma}
If $\Desc{\ensuretext{\straighttick\ttfamily{}s$_\text{1}$}}{r_1}{D}$ and $\Desc{\ensuretext{\straighttick\ttfamily{}s$_\text{2}$}}{r_2}{D}$, then $r_1 = r_2$ and $\ensuretext{\straighttick\ttfamily{}s$_\text{1}$} = \ensuretext{\straighttick\ttfamily{}s$_\text{2}$}$.
\end{lemma}

\begin{lemma}
If $\Sem{\ensuretext{\straighttick\ttfamily{}s$_\text{1}$}}{D}$ and $\Sem{\ensuretext{\straighttick\ttfamily{}s$_\text{2}$}}{D}$, then $\ensuretext{\straighttick\ttfamily{}s$_\text{1}$} = \ensuretext{\straighttick\ttfamily{}s$_\text{2}$}$.
\end{lemma}

\subsection{Specifying set operations}

With the notion of a well-formed tree in place, we can proceed to specify the various set operations. \Cref{lem:intset-member} is actually the first of these, relating the \ensuretext{\straighttick\ttfamily{}member} function to denotation $D$ of the set \ensuretext{\straighttick\ttfamily{}s}, the other query operations sport similar lemmas:

\begin{lemma}
If $\Sem{\ensuretext{\straighttick\ttfamily{}s$_\text{1}$}}{D_1}$ and $\Sem{\ensuretext{\straighttick\ttfamily{}s$_\text{2}$}}{D_2}$, then
\begin{itemize}
\item \ensuretext{\straighttick\ttfamily{}null{\spacestart}{\nextspace}s$_\text{1}${\spacestart}{\nextspace}=={\spacestart}{\nextspace}true} iff $D_1 = \emptyset$,
\item $\ensuretext{\straighttick\ttfamily{}size{\spacestart}{\nextspace}s$_\text{1}$} = |D_1|$,
\item $\ensuretext{\straighttick\ttfamily{}toList{\spacestart}{\nextspace}s$_\text{1}${\spacestart}{\nextspace}=={\spacestart}{\nextspace}xs}$ where $\ensuretext{\straighttick\ttfamily{}xs}$ is a duplicate-free list representing $D_1$,
\item $\ensuretext{\straighttick\ttfamily{}equal{\spacestart}{\nextspace}s$_\text{1}${\spacestart}{\nextspace}s$_\text{2}$}$ iff $D_1 = D_2$ and
\item $\ensuretext{\straighttick\ttfamily{}nequal{\spacestart}{\nextspace}s$_\text{1}${\spacestart}{\nextspace}s$_\text{2}$}$ iff $D_1 \ne D_2$ and
\item $\ensuretext{\straighttick\ttfamily{}isSubsetOf{\spacestart}{\nextspace}s$_\text{1}${\spacestart}{\nextspace}s$_\text{2}$}$ iff $D_1 \subseteq D_2$.
\end{itemize}
\end{lemma}

For set-returning operations, the specifications simultaneously express functional correctness and invariant preservation:
\begin{lemma}
If $\Sem{\ensuretext{\straighttick\ttfamily{}s$_\text{1}$}}{D_1}$ and $\Sem{\ensuretext{\straighttick\ttfamily{}s$_\text{2}$}}{D_2}$, then
\begin{itemize}
\item $\Sem{\ensuretext{\straighttick\ttfamily{}singleton{\spacestart}{\nextspace}x}}{\{\ensuretext{\straighttick\ttfamily{}x}\}}$,
\item $\Sem{\ensuretext{\straighttick\ttfamily{}insert{\spacestart}{\nextspace}x{\spacestart}{\nextspace}s$_\text{1}$}}{(D_1 \cup \{\ensuretext{\straighttick\ttfamily{}x}\})}$,
\item $\Sem{\ensuretext{\straighttick\ttfamily{}delete{\spacestart}{\nextspace}x{\spacestart}{\nextspace}s$_\text{1}$}}{(D_1 \setminus \{\ensuretext{\straighttick\ttfamily{}x}\})}$,
\item $\Sem{\ensuretext{\straighttick\ttfamily{}union{\spacestart}{\nextspace}s$_\text{1}${\spacestart}{\nextspace}s$_\text{2}$}}{(D_1 \cup D_2)}$,
\item $\Sem{\ensuretext{\straighttick\ttfamily{}intersection{\spacestart}{\nextspace}s$_\text{1}${\spacestart}{\nextspace}s$_\text{2}$}}{(D_1 \cap D_2)}$,
\item $\Sem{\ensuretext{\straighttick\ttfamily{}difference{\spacestart}{\nextspace}s$_\text{1}${\spacestart}{\nextspace}s$_\text{2}$}}{(D_1 \setminus D_2)}$,
\item $\Sem{\ensuretext{\straighttick\ttfamily{}filter{\spacestart}{\nextspace}f{\spacestart}{\nextspace}s$_\text{1}$}}{\{\ensuretext{\straighttick\ttfamily{}x} \in D_1 \mid \ensuretext{\straighttick\ttfamily{}f{\spacestart}{\nextspace}x{\spacestart}{\nextspace}=={\spacestart}{\nextspace}true}\}}$ and
\item $\Sem{\ensuretext{\straighttick\ttfamily{}fromList{\spacestart}{\nextspace}xs}}{\{\ensuretext{\straighttick\ttfamily{}x} \mid \text{\ensuretext{\straighttick\ttfamily{}x} occurs in the list \ensuretext{\straighttick\ttfamily{}xs}}\}}$.
\end{itemize}
\end{lemma}

The various variants of fold are specified in terms of \ensuretext{\straighttick\ttfamily{}toList}:
\begin{lemma}
If $\WF{\ensuretext{\straighttick\ttfamily{}s}}$, then
\begin{itemize}
\item \ensuretext{\straighttick\ttfamily{}IntSet.foldr{\spacestart}{\nextspace}k{\spacestart}{\nextspace}n{\spacestart}{\nextspace}s{\spacestart}{\nextspace}={\spacestart}{\nextspace}List.foldr{\spacestart}{\nextspace}k{\spacestart}{\nextspace}n{\spacestart}{\nextspace}(toList{\spacestart}{\nextspace}s)} and
\item \ensuretext{\straighttick\ttfamily{}IntSet.foldl{\spacestart}{\nextspace}k{\spacestart}{\nextspace}n{\spacestart}{\nextspace}s{\spacestart}{\nextspace}={\spacestart}{\nextspace}List.foldl{\spacestart}{\nextspace}k{\spacestart}{\nextspace}n{\spacestart}{\nextspace}(toList{\spacestart}{\nextspace}s)}.
\end{itemize}
\end{lemma}

All proofs build on a corresponding lemma that also involves the dyadic interval of the set. For example, the proof for \ensuretext{\straighttick\ttfamily{}union} will build on a lemma that states that for $\Desc{\ensuretext{\straighttick\ttfamily{}s$_\text{1}$}}{r_1}{D_1}$ and $\Desc{\ensuretext{\straighttick\ttfamily{}s$_\text{2}$}}{r_2}{D_2}$ we have
\[
\Desc{\ensuretext{\straighttick\ttfamily{}union{\spacestart}{\nextspace}s$_\text{1}${\spacestart}{\nextspace}s$_\text{2}$}}{r_1 \sqcup r2}{(D_1 \cup D_2)}.
\]
which can be proved by induction on both assumed relations.%

Things are a bit more complicated with operations that may \emph{remove} elements, such as \ensuretext{\straighttick\ttfamily{}delete} or \ensuretext{\straighttick\ttfamily{}intersection}: Because they return a possibility empty set, we cannot describe their result using the relation $\Desc\_\_\_$. Therefore, we introduce
the relation $\DescO{\ensuretext{\straighttick\ttfamily{}s}}r{D}$, where we still have $D \subseteq r$, but $r$ needs not be the smallest such dyadic interval, and $D$ may be empty. More concretely
\begin{mathpar}
\inferrule {~}{\DescO{\ensuretext{\straighttick\ttfamily{}Nil}}r\emptyset}
\and
\inferrule{%
\Desc{\ensuretext{\straighttick\ttfamily{}s}}rD \and r \subseteq r'
}{\DescO{\ensuretext{\straighttick\ttfamily{}s}}{r'}{D}}
\end{mathpar}
With this relation we can express that for $\Desc{\ensuretext{\straighttick\ttfamily{}s$_\text{1}$}}{r_1}{D_1}$ and $\Desc{\ensuretext{\straighttick\ttfamily{}s$_\text{2}$}}{r_2}{D_2}$ we have
\[
\DescO{\ensuretext{\straighttick\ttfamily{}intersection{\spacestart}{\nextspace}s$_\text{1}${\spacestart}{\nextspace}s$_\text{2}$}}{r_1}{(D_1 \cap D_2)},
\]
again by induction on the assumed relations.%

In each of these lemma, the base case of a \ensuretext{\straighttick\ttfamily{}Tip} requires a corresponding lemma about the bitmaps, which usually follows from the corresponding bit-wise operations (AND, OR, etc.)%
\jb{Shall we spell one such proofs out? What else should I say?}

\subsection{Bounded and negative integers}

\label{sec:intset-negative}

This theory does not impose a bound on the size the integers stored in the \ensuretext{\straighttick\ttfamily{}IntSet}. Since the implementation works on the individual bits of the elements, e.g.\ when checking the mask bit in a \ensuretext{\straighttick\ttfamily{}Bin} constructor, the theory immediately applies to bounded unsigned integer types.

Negative numbers are more tricky, as the nice algebraic structure of dyadic intervals breaks down: A dyadic interval is either all negative or all positive. So for the the two dyadic intervals $r_1 = d(-1,1) = \{-2,-1\}$ and $r_2 = d(0,1) = \{0,1\}$ the least upper bound $r_1 \sqcup r_2$ does not exist. So in order to apply our \ensuretext{\straighttick\ttfamily{}IntSet} theory to signed integers, we have to map them to unsigned integers.

For signed integers of bounded bit-width $w$ the usual encoding of negative numbers \emph{two’s complement}, is quite suitable. It represents a negative number $i$ by negating all bits of its absolute value $|i|$ and then adding 1 to the result. This effectively maps the negative numbers $\{-2^{w-1},\ldots,-1\}$ onto the range $\{2^{w-1},\ldots 2^{w}-1\}$. Because the implementation uses only bit-level operations, it works just as well for signed, bounded types, simply by treating them as unsigned.

For signed unbounded types, the two’s complement implementation is unsuitable: It represents $-1$ as a number with infinitely many bits set, which we cannot just interpret as a unsigned number. In order to build a \ensuretext{\straighttick\ttfamily{}IntSet}-like data structure for signed unbounded types, one would either have to chose a different encoding of natural numbers where all numbers are represented with finitely many bit set (such as \emph{signed magnitude representation}), or simply use a pair of sets, one for the negative and one for the positive numbers.


\fi 

\ifediting

\section{More metrics}

\begin{table}
\summarytallytable
\end{table}

\subsection{What have we translated}

This is obviously not for the final paper, but mostly for us.

\paragraph{Verified functions}
\begin{description}
\item[\ensuretext{\straighttick\ttfamily{}Set}:]  \listVerifSet
\item[\ensuretext{\straighttick\ttfamily{}IntSet}:]  \listVerifIntSet
\end{description}

\paragraph{Unverified functions}
\begin{description}
\item[\ensuretext{\straighttick\ttfamily{}Set}:]  \listUnverifSet
\item[\ensuretext{\straighttick\ttfamily{}IntSet}:]  \listUnverifIntSet
\end{description}

\paragraph{Untranslated functions}
\begin{description}
\item[\ensuretext{\straighttick\ttfamily{}Set}:]  \listUntransSet
\item[\ensuretext{\straighttick\ttfamily{}IntSet}:]  \listUntransIntSet
\end{description}

\section{Old text}
\subsection{Proof Automation}

\criz{proofs about math involve little automation, proofs about compilers full automation, we lie somewhere in the middle.}
Writing formal proofs typically add an additional challenge as opposed to pen-and-paper proofs.
There are various styles of writing formal proofs based on the particular theorem being proven.
\begin{itemize}
\item  Theorems about pure mathematics typically benefit from very little automation.
\item with some exceptions such as the four color theorem which relies on a program for it's proof.
\item Moreover, little proof automation can be achieved when verifying
  theorems about abstract code against a formal specification  for example
  Joachim's PhD work, Christine's initial LEDA work.  \criz{this one is pretty
    questionable -- might want to drop/modify that}

\item At the other end lie theorems that certify generated code against generated specifications.
As both the code and the specifications are generated, nothing needs to be guessed in a sense and
the proofs linking them can also be fully automated. As is the case for Cogent, CakeML, as well as some CompCert optimizations.

\item For verifying code written in general purpose languages such as C or Haskell against it's specification, one can still achieve some level of proof automation.
\item However, full static verification of Turing complete languages such as C and Haskell is undecidable~\citep{Rice:53}.
\item For C code, the C-to-Isabelle parser does a straightforward translation of the C code into the Simpl language which is embedded into Isabelle/HOL. Then a tool called AutoCorres  does a verified translation of the Simpl output of the  parser to a monadic representation  which is much easier to reason about. Hence automating part of the verification process.
\item Unlike C, Haskell is a functional language that is in general easier to reason about. The straightforward translation of our hs-to-coq tool  is already convenient for direct reasoning.
\item Many of our theorems have the same form, maintaining an inductive well-formedness predicate. Therefore we were able to write some specialized tactics that helped automate large parts of our proofs.
\item We suspect that others can do the same.

\end{itemize}

\subsection{Lemma forms used in proving the tests}
\jb{The following two subsections should come after we introduce the abstract model. They could be part of a narrative “We use the abstract model for the other specifications, but write these more convenient lemmas first.” Or maybe dropped completely.}
\paragraph*{Theorems about membership} To prove the QuickCheck properties
correct, we needed more lemmas than had already been directly proven by our
other lower-level specifications.  In particular, since the QuickCheck
properties were almost all equalities between various functions from the
\ensuretext{\straighttick\ttfamily{}IntSet} API, it was important to be able to do all our reasoning in terms of
these functions.  A standard approach to this\asz{citation needed?} is to
prove that equality between sets is extensional (i.e., that it just depends on
them having the same members), which we had already proved, and then to prove
theorems about set membership and proceed by rewriting.  Thus, for each \ensuretext{\straighttick\ttfamily{}IntSet}
function we verified, we proved an additional theorem such as
\begingroup\straighttick\begin{hscode}\SaveRestoreHook
\column{B}{@{}>{\hspre}l<{\hspost}@{}}%
\column{3}{@{}>{\hspre}l<{\hspost}@{}}%
\column{E}{@{}>{\hspre}l<{\hspost}@{}}%
\>[B]{}\textbf{Theorem}{\spacestart}{\nextspace}union\char95 member{\spacestart}{\nextspace}(s1{\spacestart}{\nextspace}s2{\spacestart}{\nextspace}:{\spacestart}{\nextspace}IntSet){\spacestart}{\nextspace}:{\spacestart}{\nextspace}{}\<[E]%
\\[-0.3ex]%
\>[B]{}{\spacestart}{\nextspace}{\nextspace}{}\<[3]%
\>[3]{}WF{\spacestart}{\nextspace}s1{\spacestart}{\nextspace}->{\spacestart}{\nextspace}WF{\spacestart}{\nextspace}s2{\spacestart}{\nextspace}->{\spacestart}{\nextspace}{}\<[E]%
\\[-0.3ex]%
\>[B]{}{\spacestart}{\nextspace}{\nextspace}{}\<[3]%
\>[3]{}\textbf{forall}{\spacestart}{\nextspace}k,{\spacestart}{\nextspace}member{\spacestart}{\nextspace}k{\spacestart}{\nextspace}(union{\spacestart}{\nextspace}s1{\spacestart}{\nextspace}s2){\spacestart}{\nextspace}={\spacestart}{\nextspace}member{\spacestart}{\nextspace}k{\spacestart}{\nextspace}s1{\spacestart}{\nextspace}||{\spacestart}{\nextspace}member{\spacestart}{\nextspace}k{\spacestart}{\nextspace}s2.{}\<[E]%
\ColumnHook
\end{hscode}\resethooks\endgroup
Then we could translate the equalities into statements about membership and
proceed by rewriting.  Proving these higher-level membership lemmas was doable
directly from the mathematical model for sets that we used as our lower-level
specification (see \cref{sec:specs-math}), and they proved much easier to work
with directly.

\asz{Is this next subsection necessary?}%
\paragraph*{Theorems about sortedness} The other approach some of the QuickCheck
properties took was to specify the behavior of \ensuretext{\straighttick\ttfamily{}IntSet} functions in terms of
\ensuretext{\straighttick\ttfamily{}toList}, such as [sic]\scw{What does [sic] refer to?}
\begingroup\straighttick\begin{hscode}\SaveRestoreHook
\column{B}{@{}>{\hspre}l<{\hspost}@{}}%
\column{6}{@{}>{\hspre}l<{\hspost}@{}}%
\column{11}{@{}>{\hspre}l<{\hspost}@{}}%
\column{E}{@{}>{\hspre}l<{\hspost}@{}}%
\>[B]{}prop\char95 Int{\spacestart}{\nextspace}::{\spacestart}{\nextspace}[Int]{\spacestart}{\nextspace}->{\spacestart}{\nextspace}[Int]{\spacestart}{\nextspace}->{\spacestart}{\nextspace}Property{\spacestart}{\nextspace}{}\<[E]%
\\[-0.3ex]%
\>[B]{}prop\char95 Int{\spacestart}{\nextspace}xs{\spacestart}{\nextspace}ys{\spacestart}{\nextspace}={\spacestart}{\nextspace}{}\<[E]%
\\[-0.3ex]%
\>[B]{}~~{\spacestart}{\nextspace}{\nextspace}{}\<[6]%
\>[6]{}\textbf{case}{\spacestart}{\nextspace}intersection{\spacestart}{\nextspace}(fromList{\spacestart}{\nextspace}xs){\spacestart}{\nextspace}(fromList{\spacestart}{\nextspace}ys){\spacestart}{\nextspace}\textbf{of}{\spacestart}{\nextspace}t{\spacestart}{\nextspace}->{\spacestart}{\nextspace}{}\<[E]%
\\[-0.3ex]%
\>[B]{}{\spacestart}{\nextspace}{\nextspace}{\nextspace}{\nextspace}{\nextspace}{}\<[6]%
\>[6]{}~~{\spacestart}{\nextspace}{\nextspace}{}\<[11]%
\>[11]{}valid{\spacestart}{\nextspace}t{\spacestart}{\nextspace}.\&\&.{\spacestart}{\nextspace}toAscList{\spacestart}{\nextspace}t{\spacestart}{\nextspace}==={\spacestart}{\nextspace}List.sort{\spacestart}{\nextspace}(nub{\spacestart}{\nextspace}((List.intersect){\spacestart}{\nextspace}(xs){\spacestart}{\nextspace}(ys))){}\<[E]%
\ColumnHook
\end{hscode}\resethooks\endgroup
(The \ensuretext{\straighttick\ttfamily{}valid} function is the condition mentioned in the previous section,
\ensuretext{\straighttick\ttfamily{}(.\&\&.)} is a QuickCheck conjunction connective which we translate to Coq's
\ensuretext{\straighttick\ttfamily{}/\char92 }, and \ensuretext{\straighttick\ttfamily{}(===)} is a QuickCheck equality connective which we translate to
Haskell's \ensuretext{\straighttick\ttfamily{}(==)}.)  This required us to be able to reason about sorted lists
with no duplicate elements (\ensuretext{\straighttick\ttfamily{}nub} removes duplicates), as \ensuretext{\straighttick\ttfamily{}toAscList} produces
an ascending list of the set's elements, which will naturally contain no
duplicates.  In particular, we had to prove that equality between such lists is
extensional.

\begin{figure}
  \centering
  \def\Margin{0.5cm}
  \def\SpecsWidth{\textwidth}
  \def\SpecsHeight{7cm}

  \pgfmathsetlengthmacro{\ConsumedWidth}{\SpecsWidth - 3*\Margin}

  \pgfmathsetlengthmacro{\LeftWidth}  {\ConsumedWidth * 2 / 3}
  \pgfmathsetlengthmacro{\RightWidth} {\ConsumedWidth - \LeftWidth}

  \pgfmathsetlengthmacro{\ConsumedLeftHeight}  {\SpecsHeight - 2*\Margin}
  \pgfmathsetlengthmacro{\ConsumedRightHeight} {\SpecsHeight - 3*\Margin}

  \pgfmathsetlengthmacro{\FullHeight}   {\ConsumedLeftHeight}
  \pgfmathsetlengthmacro{\TopHeight}    {\ConsumedRightHeight / 2}
  \pgfmathsetlengthmacro{\BottomHeight} {\ConsumedRightHeight - \TopHeight}

  \pgfmathsetlengthmacro{\LeftStart}   {\Margin}
  \pgfmathsetlengthmacro{\TopStart}    {\Margin}
  \pgfmathsetlengthmacro{\RightStart}  {\LeftWidth + 2*\Margin}
  \pgfmathsetlengthmacro{\BottomStart} {\TopHeight + 2*\Margin}

  \begin{tikzpicture}[rounded corners, thick]
    \draw (0,0) rectangle ++(\SpecsWidth, -\SpecsHeight) ;

    \draw (\LeftStart,  -\TopStart)    rectangle ++(\LeftWidth,  -\FullHeight)   ;
    \draw (\RightStart, -\TopStart)    rectangle ++(\RightWidth, -\TopHeight)    ;
    \draw (\RightStart, -\BottomStart) rectangle ++(\RightWidth, -\BottomHeight) ;

  \end{tikzpicture}
  \caption{The seven specification sources for \ensuretext{\straighttick\ttfamily{}Set} and \ensuretext{\straighttick\ttfamily{}IntSet}}
  \label{fig:specifications}
\end{figure}

\fi 

\end{document}